\documentclass[final,3p,times,11pt]{elsarticle}

\usepackage[utf8]{inputenc}
\usepackage[english]{babel}
\usepackage[ngerman]{datetime}
\usepackage[
	pdftex,
	bookmarks=true,
	bookmarksnumbered=true,
	bookmarksopen=true,
	colorlinks=true,
	filecolor=black,
	linkcolor=black,
	urlcolor=grey,
	plainpages=false,
	pdfpagelabels,
	citecolor=black,
	unicode=true,
	pdftitle={Kinetic Field Theory: Higher-Order Power Spectra},
	pdfauthor={Stefan Zentarra},
	pdfdisplaydoctitle=true]{hyperref}
\usepackage{amsmath}
\usepackage{amssymb}
\usepackage{amsthm}
\usepackage{stmaryrd}
\usepackage{paralist}
\usepackage{dsfont}
\usepackage{txfonts}
\usepackage{wrapfig}
\usepackage{xspace}
\usepackage{graphicx}

\usepackage{tikz}
\usetikzlibrary{arrows.meta,matrix}
\usetikzlibrary{decorations.pathmorphing}
\usetikzlibrary{shapes}
\tikzset{
	prop/.style={draw,ultra thick},
	vertex/.style={circle,thick,draw=black,fill=black!20,minimum size=18pt,inner sep=3pt},
	empty/.style={circle,thick,draw=black,minimum size=18pt,inner sep=3pt},
	outtraj/.style={rectangle,thick,minimum size=18pt,inner sep=3pt},
	outdens/.style={rectangle,thick,draw=black,fill=black!20,minimum size=18pt,inner sep=3pt},
	outdensempty/.style={rectangle,thick,draw=black,minimum size=18pt,inner sep=3pt},
	subdiagr/.style={star,star points=7,star point ratio=0.8,thick,draw=black,fill=black!20,minimum size=18pt,inner sep=3pt},
	subdiagrempty/.style={star,star points=7,star point ratio=0.8,thick,draw=black,minimum size=18pt,inner sep=3pt}
}

\usepackage{pgfplots}
\pgfplotsset{width=7cm,compat=1.8}

\makeatletter
\def\thm@space@setup{%
  \thm@preskip=\topsep 
  \thm@postskip=\thm@preskip
}
\makeatother

\makeatletter 
\newcommand{\oset}[3][0ex]{%
  \mathrel{\mathop{#3}\limits^{
    \vbox to#1{\kern-2\ex@
    \hbox{$\scriptstyle#2$}\vss}}}}
\makeatother

\newcommand\st{\textsuperscript{st}\xspace}

\newcommand\nth{\textsuperscript{th}\xspace} 
\newcommand{\ce}{\coloneqq}
\newcommand{\ec}{\eqqcolon}
\renewcommand{\c}{\colon}
\renewcommand{\b}[1]{\boldsymbol{#1}}

\newcommand{\tp}{\intercal}

\newcommand{\LRa}{\Leftrightarrow}
\newcommand{\ra}{\rightarrow}

\newcommand{\diff}{\mathrm{d}}
\newcommand{\N}{\mathbb{N}}

\newcommand{\R}{\mathbb{R}}

\renewcommand{\mid}{\;\middle|\;}
\newcommand{\all}{\ \forall~}

\newcommand{\restr}[2]{{
		\left.\kern-\nulldelimiterspace
		#1 
		\vphantom{\big|}\,
		\right|_{#2}
	}}
\newcommand{\abs}[1]{\left\lvert #1 \right\rvert}
\newcommand{\norm}[1]{\left\| #1 \right\|}

\newcommand{\qpart}[2][0]{
		{{#2}\hspace*{-0.11em}{}_{\rule[\the\numexpr #1 / 2 - 3pt]{0.7pt}{\the\numexpr 7 - #1 pt}}}\hspace*{0.1em}{}
	}
\newcommand{\ppart}[2][0]{
		{{}_{\rule[\the\numexpr #1 / 2 - 3pt]{0.7pt}{\the\numexpr 7 - #1 pt}}\hspace*{-0.13em}{#2}}{}
	}

\newdateformat{myformat}
	{\THEDAY{.\ }\monthnamengerman[\THEMONTH] \THEYEAR}
\newdateformat{myformatshort}
	{\THEDAY{\textsuperscript{th} }\monthnameenglish[\THEMONTH]{, }\THEYEAR}

\newtheoremstyle{myThm}
	{.7em} 
	{.7em} 
	{\em} 
	{} 
	{\bfseries} 
	{.} 
	{.5em} 
	{} 
\newtheoremstyle{myThm2}
	{.7em} 
	{.7em} 
	{} 
	{} 
	{\bfseries} 
	{.} 
	{.5em} 
	{} 

\theoremstyle{myThm}

\theoremstyle{myThm2}

\allowdisplaybreaks

\journal{Physical Review D}

\begin{document}
	
	\begin{frontmatter}
		
		
		
		\title{\huge{Kinetic Field Theory:\\Higher-Order Perturbation Theory}}
		
		
		\author[ETH,HD]{Lavinia Heisenberg}
		\ead{l.heisenberg@thphys.uni-heidelberg.de}
		\author[ETH,HD]{Shayan Hemmatyar}
		\ead{hemmatyar@thphys.uni-heidelberg.de}
		\author[ETH]{Stefan Zentarra}
		\ead{szentarra@ethz.ch}
		\address[ETH]{Institute for Theoretical Physics, ETH Zurich, Wolfgang-Pauli-Str.~27, 8093 Zurich, Switzerland}
		\address[HD]{Institute for Theoretical Physics, Heidelberg University, Philosophenweg 16, 69120 Heidelberg, Germany}

		\begin{abstract}
			We give a detailed exposition of the formalism of Kinetic Field Theory (KFT) with emphasis on the perturbative determination of observables. KFT is a statistical non-equilibrium classical field theory based on the path integral formulation of classical mechanics, employing the powerful techniques developed in the context of quantum field theory to describe classical systems. Unlike previous work on KFT, we perform the integration over the probability distribution of initial conditions in the very last step. This significantly improves the clarity of the perturbative treatment and allows for physical interpretation of intermediate results. We give an introduction to the general framework, but focus on the application to interacting $N$-body systems. Specializing the results to cosmic structure formation, we reproduce the linear growth of the cosmic density fluctuation power spectrum on all scales from microscopic, Newtonian particle dynamics alone.
		\end{abstract}
		
		\begin{keyword}
			Kinetic Field Theory (KFT) \sep Perturbation Theory \sep N-Body Systems \sep Cosmic Structure Formation \sep Cosmology
			
			
			
		\end{keyword}
		
	\end{frontmatter}
	
	
	
	\newpage
	
	\setlength{\parskip}{-0.1em}
	\tableofcontents
	\setlength{\parskip}{1em}
	

\section{Introduction}

In a 1988 paper~\cite{Gozzi1988}, E.~Gozzi outlined how the successful path integral formalism of quantum field theory can be used to describe classical systems as well. In the following years, Gozzi further developed and studied this approach in collaboration with M.~Reuter and W.D.~Thacker~\cite{Gozzi+1989,L10} and others~\cite{L10a,L12}. Relatively recently, G.F.~Mazenko and S.P.~Das~\cite{L8,L7} developed a perturbation scheme and applied the formalism to phase transitions in glass. Based on this, the group of M.~Bartelmann began to apply the path integral approach of classical mechanics to the study of cosmic structure formation~\cite{L3}. Since then, the approach has been subject to significant further development~\cite{L3e,L3a,L3b,L3f,L3c} and is now known as \emph{Kinetic Field Theory (KFT)}~\cite{bartelmann2019cosmic}.

Conceptually, KFT mimics the approach of numerical simulations to describe the time evolution of an interacting $N$-particle system. Via use of path integrals over phase space trajectories, one keeps track of the positions and momenta of all particles over time. Mathematically, this information is stored in the generating functional~$Z$ of the theory. Collective quantities like, e.g., the density field can be extracted via functional derivatives of~$Z$. Explicit expressions for the generating functional are obtained either via a perturbative approach~\cite{L3} or through the use of suitable approximations~\cite{L3e}.

There is one significant difference between the formalism of KFT and numerical $N$-body simulations. Namely, the latter always have to start out with some explicit initial conditions, i.e., initial positions and momenta for all particles. A priori this is also true for KFT. However, in an analytic setting we can abstractly integrate over a probability distribution of initial conditions to obtain expectation values. This would be prohibitively expensive for numerical $N$-body simulations which therefore have to resort to other methods of extracting macroscopic quantities. The integration over initial conditions in KFT yields an ensemble average which in many applications is the observable we are interested in.

In this article we revisit to the initial perturbative approach to KFT in~\cite{L3}. We give a self-contained and pedagogical introduction to the framework of KFT in a general setting and then specialize it to interacting $N$-body systems. Unlike the perturbative treatment in~\cite{L3}, we perform the integration over the initial conditions not on the level of the generating functional, but on the level of observables. This allows us to interpret intermediate results describing observables of a system with explicit initial conditions. Upon integrating them over a suitable probability distribution, we obtain statistically averaged expectation values, i.e., macroscopic observables describing the collective system. While ultimately mathematically equivalent, this change in procedure makes the perturbative treatment significantly more transparent and simpler.

Our ultimate aim is to supply a simple and consistent procedure for obtaining observables in cosmic structure formation, in particular the non-linear density fluctuation power spectrum. The initial probability distribution for this particular application is relatively complicated. For the sake of self-containedness and clarity, we repeat the discussion from appendix~$A$ of~\cite{L3}. In particular, we explain how the integration over initial conditions fits together with the perturbative treatment. Having done so, we will be able to obtain analytic parameter-free perturbative expressions for the non-linear density fluctuation power spectrum. These expressions are expanded in both the initial correlations and the interactions. More precisely, on the one hand we expand the density and momentum correlations imprinted in the initial conditions of the $N$-particle system and on the other hand we treat the particle interactions perturbatively by expanding the forces relative to the free particle trajectories.

Surprisingly, we find that the expansion in the initial correlations is significantly more difficult than the expansion in the interactions. Indeed, for higher-order correlations our expressions acquire integrals over wave vectors which are challenging to solve even numerically. Conversely, each interaction order merely supplies a simple time integral. To conclude this article, we calculate the density fluctuation power spectrum in first order (i.e.\ linear) correlations and up to very high interaction order. Combining the contributions for linear correlations and arbitrarily high interaction orders precisely reproduces the linear growth of the power spectrum at all scales. This reveals that any non-linearities in the power spectrum are solely due to higher-order correlations in the initial conditions and thus can be studied completely independently of the linear evolution. Results for higher-order correlations will be presented in a future article~\cite{we2022}.


\section{KFT Framework}
\label{sec:theory}

Let us consider a generic classical physical system. Denote its space of possible states by~$\b{X}$, such that any point~$\b{x} \in \b{X}$ uniquely characterizes a state of this system.\footnote{All bold faced symbols take values in~$\b{X}$, i.e., they are generally some high-dimensional vectors or tensors.} In practice, we usually take~$\b{X}$ to be phase space consisting of all (generalized) coordinates and associated momenta. Any evolution of such a system can be described as a map~$\b{\varphi} \c \R \ra \b{X}$ which associates to any point in time~$t \in \R$ a state of the system~$\b{\varphi}(t) \in \b{X}$. However, the vast majority of such maps describe unphysical evolution. We are only interested in the maps which satisfy the equations of motion which we symbolically write as~$\b{E}(\b{\varphi}(t)) = \b{0}$. In almost all cases, the equations of motion are differential equations which require that the change of the state is determined only by the state of the system at the same instance in time. For classical systems, given a state~$\b{x} \in \b{X}$ at initial time~$t_i$, there is exactly one solution of the equations of motion~$\b{\tilde{\varphi}}(t;\b{x})$ which satisfies~$\b{\tilde{\varphi}}(t_i;\b{x}) = \b{x}$.

In a more general setting, we might not precisely know the initial state~$\b{x} \in \b{X}$, but instead our knowledge of the system at initial time~$t_i$ may be encoded in form of a probability distribution~$P(\b{x})$ on the space~$\b{X}$. In case~$P(\b{y}) = \delta_D(\b{y} - \b{x})$ is given by a Dirac $\delta$-function, we recover the deterministic setting above. However, if the probability distribution is more complicated, the state of the system is stochastic for all times~$t$. This means that for any time~$t$ there is a unique probability distribution~$P(\b{x};t)$ describing the system. In fact, it is simple to give an explicit expression for this probability distribution by using the solution to the equations of motion,
\begin{align}
	P(\b{x};t) = P\left(\b{\tilde{\varphi}}(-t;\b{x})\right) \,,
\end{align}
where on the right-hand side we have the initial probability distribution. Obviously it is $P(\b{x};t_i) = P(\b{x})$.

The probability distribution~$P(\b{x};t)$ allows us to extract expressions for observables at any point in time. As an observable we regard in this context any function~$\mathcal{O} \c \b{X} \ra \R$, i.e., a map which assigns to a state of a system some number (this can easily be generalized to tensor-valued observables). For such an observable, we define its expectation value at time~$t$ as
\begin{align}
\label{eq:theory_averageObs}
	\langle \mathcal{O} \rangle(t) = \int_{\b{X}} \diff\b{x} \, \mathcal{O}(\b{x}) P(\b{x};t) \,.
\end{align}
In the deterministic case~$P(\b{y}) = \delta_D(\b{y} - \b{x})$, we simply have
\begin{align}
\label{eq:theory_observable}
	\langle \mathcal{O} \rangle(t) = \tilde{\mathcal{O}}(t;\b{x}) \ce \mathcal{O}\left(\b{\tilde{\varphi}}(t;\b{x})\right) \,.
\end{align}
Here we defined the notation~$\tilde{\mathcal{O}}(t;\b{x})$ for the observable evaluated along the solution of the equations of motion for initial value~$\b{x}$ at time~$t$.

From the discussion above one may arrive at a seemingly simple recipe to calculate observables: determine the solution~$\b{\tilde{\varphi}}$ and use it to express any quantity of interest. However, as is well known, even for seemingly simple classical systems one can prove that there is no elementary expression for~$\b{\tilde{\varphi}}(t;\b{x})$. In many cases one can determine~$\b{\tilde{\varphi}}$ numerically instead, but this can become computationally costly for systems with a high-dimensional space of states~$\b{X}$. A prime example for this is an interacting $N$-particle system. The formalism of KFT provides an alternative approach for calculating observables by analytical methods which we want to introduce below.

\subsection{Path Integral Formulation of Classical Mechanics}

The formalism of KFT is based on the path integral formulation of classical mechanics which to the best of our knowledge was first proposed by E.~Gozzi~\cite{Gozzi1988}. Readers familiar with the path integral approach to Quantum Field Theory (QFT) will notice many similarities between KFT and QFT. In fact, these similarities are one of the main theoretical motivations to use KFT, as one can use the powerful tools developed in the framework of QFT to describe the evolution of a classical physical system. 

Let us consider a classical physical system which at initial time~$t_i$ is in a state~$\b{x} \in \b{X}$. Denote the solution of the equations of motion~$\b{E}(\b{\varphi}) = \b{0}$ subject to these initial conditions by~$\b{\tilde{\varphi}}(t; \b{x})$. Further, let~$\b{y} \in \b{X}$ be some possible state of the system at a later time~$t > t_i$. Then the transition probability~$T^{fi}(\b{y}, t;\b{x}, t_i)$, i.e., the probability of observing~$\b{y}$ at~$t > t_i$ given initial conditions~$\b{x}$ at~$t = t_i$, is given by
\begin{align}
	T^{fi}(\b{y}, t;\b{x}, t_i) = \int_{\b{\varphi}(t_i) = \b{x}}^{\b{\varphi}(t) = \b{y}} \mathcal{D}\b{\varphi} \, \delta_D\left[ \b{\varphi}(t) - \b{\tilde{\varphi}}(t;\b{x}) \right] \,.
\end{align}
The interpretation of this quantity is rather simple: While the path integral adds contributions from all possible paths in state space~$\b{X}$ from~$\b{x}$ to~$\b{y}$, the $\delta_D$-functional (a generalization of the Dirac $\delta$-function) picks out a single one of these. This is reminiscent of taking the classical limit in the path integral formulation of quantum mechanics, where the classical path supplies the dominant contribution as~$\hbar \ra 0$.

Clearly, the transition probability~$T^{fi}(\b{y}, t;\b{x}, t_i)$ is one if~$\b{y} = \b{\tilde{\varphi}}(t;\b{x})$ and zero otherwise. This is not surprising given that the evolution of classical systems with explicitly specified initial conditions is deterministic. Using the fact that the path selected by the $\delta_D$-functional is the solution of the equations of motion, we can replace condition~$\b{\varphi}(t) = \b{\tilde{\varphi}}(t;\b{x})$ encoded in its argument by the condition~$\b{E}(\b{\varphi}) = \b{0}$ which avoids explicit reference to the solution~$\b{\tilde{\varphi}}$. Generally, such a change in the argument of a $\delta_D$-functional changes the normalization according to
\begin{align}
\label{eq:theory_deltaDidentity}
	\delta_D\left[ \b{E}(\b{\varphi}) \right] = \abs{\det\left( \frac{\delta \b{E}}{\delta \b{\varphi}}\left(\b{\tilde{\varphi}}\right) \right)}^{-1} \, \delta_D\left[ \b{\varphi} - \b{\tilde{\varphi}} \right] \,.
\end{align}
This is a generalization of the identity
\begin{align}
	\delta_D \left( f(x) \right) = \abs{\det\left( \frac{\partial f}{\partial x}\left(\tilde{x}\right) \right)}^{-1} \, \delta_D \left( x - \tilde{x} \right)
\end{align}
for the usual Dirac $\delta$-function which is valid if the function~$f$ has exactly one root~$x = \tilde{x}$. However, as proven in equation~$(3.51)$ of~\cite{Gozzi+1989}, the determinant in equation~\eqref{eq:theory_deltaDidentity} is constant in the case of Hamiltonian equations of motion. This may be understood in terms of Liuville's theorem which states that the Hamiltonian flow in phase space preserves volumes. The numerical value of the determinant is irrelevant and can be absorbed into the normalization of the path integral.

While it is possible to keep the determinant in equation~\eqref{eq:theory_deltaDidentity}, it would unnecessarily complicate our treatment. Therefore, we shall restrict ourselves to the case of Hamiltonian dynamics, i.e., the space of states~$\b{X}$ actually is phase space and the equations of motion~$\b{E}(\b{\varphi}) = \b{0}$ are of Hamiltonian form
\begin{align}
\label{eq:theory_eomH}
	\b{E}(\b{\varphi}) = \partial_t \b{\varphi} - \b{\mathcal{J}} \, \partial_{\b{\varphi}} H(\b{\varphi}) = \b{0} \,.
\end{align}
Here, $\b{\mathcal{J}}$ is the usual symplectic matrix and~$H$ is the Hamiltonian. Using equation~\eqref{eq:theory_deltaDidentity}, the transition probability can be expressed as
\begin{align}
\label{eq:theory_TransitionProb}
	T^{fi}(\b{y}, t; \b{x}, t_i) &= \int_{\b{\varphi}(t_i) = \b{x}}^{\b{\varphi}(t) = \b{y}} \mathcal{D}\b{\varphi} \, \delta_D\left[ \b{E}(\b{\varphi}) \right] \\*
\label{eq:theory_Tfi}
	&= \int_{\b{\varphi}(t_i) = \b{x}}^{\b{\varphi}(t) = \b{y}} \mathcal{D}\b{\varphi} \int \mathcal{D}\b{\chi} \, \exp\left( i \int_{t_i}^{t} \diff t' \, \b{\chi}(t') \cdot \b{E}(\b{\varphi}(t')) \right) \,.
\end{align}
In the second line, we used a standard functional Fourier transform to write the $\delta_D$-functional as an exponential function. The object~$\b{\chi}$ is a purely auxiliary abstract mathematical function and does not have an obvious physical interpretation. In analogy to the path integral approach to QFT, we may regard the exponent as the action of the system under consideration.

Further building on the analogy to QFT, we now define the so-called \emph{generating functional} which is the central mathematical object of KFT. Integrating over the final state~$\b{y}$ of the system (or, equivalently, setting $\b{y} \ce \b{\tilde{\varphi}}(\b{x},t)$) to get rid of the endpoint of the path integral over~$\b{\varphi}$ and introducing a source field~$\b{J}$ for~$\b{\varphi}$, it is
\begin{align}
\label{eq:theory_generatingFunctional}
	Z[\b{J}; \b{x}] = \int_{\b{\varphi}(t_i) = \b{x}} \mathcal{D}\b{\varphi} \int \mathcal{D}\b{\chi} \, \exp\left( i \int_{t_i}^\infty \diff t' \, \b{\chi}(t') \cdot \b{E}(\b{\varphi}(t')) + i \int_{t_i}^\infty \diff t' \, \b{J}(t') \cdot \b{\varphi}(t') \right) \,.
\end{align}
The generating functional contains complete information about the physical system under consideration. Any observable depending on the phase space position at a single time~$t$ can be derived from it by means of functional derivation with respect to~$\b{J}(t)$.\footnote{In principle, it is also possible to construct observables depending on the phase space variables on a time-interval (e.g.\ a backward lightcone). While this may become relevant in the future, here we restrict ourselves to the simpler case.} Indeed, for a given observable~$\mathcal{O} \c \b{X} \ra \R$ it is
\begin{align}
\label{eq:theory_observableFD}
	\tilde{\mathcal{O}}(t;\b{x}) &= \restr{\mathcal{O}\left(\frac{\delta}{i \, \delta \b{J}(t)}\right) \, Z[\b{J}; \b{x}]}{\b{J} = \b{0}} \\
	&= \int_{\b{\varphi}(t_i) = \b{x}} \mathcal{D}\b{\varphi} \int \mathcal{D}\b{\chi} \, \mathcal{O}\left(\b{\varphi}(t)\right) \, \exp\left( i \int_{t_i}^t \diff t' \, \b{\chi}(t') \cdot \b{E}(\b{\varphi}(t')) \right) \\
	&= \int_{\b{\varphi}(t_i) = \b{x}} \mathcal{D}\b{\varphi} \, \mathcal{O}\left(\b{\varphi}(t)\right) \, \delta_D\left[ \b{E}(\b{\varphi}) \right] \\*
	&= \mathcal{O}\left(\b{\tilde{\varphi}}(t;\b{x})\right) \,.
\end{align}
We used the notation~$\tilde{\mathcal{O}}$ introduced in equation~\eqref{eq:theory_observable} to denote the observable as evaluated along the solution of the equations of motion~$\b{\tilde{\varphi}}(t;\b{x})$ for initial value~$\b{x}$ -- which evidently is exactly what we obtain via this differentiation. In fact, this can also be seen by performing the integration over~$\b{\chi}$ directly in the generating functional which yields
\begin{align}
	Z[\b{J}; \b{x}] &= \int_{\b{\varphi}(t_i) = \b{x}} \mathcal{D}\b{\varphi} \, \exp\left( i \int_{t_i}^\infty \diff t' \, \b{J}(t') \cdot \b{\varphi}(t') \right) \, \delta_D\left[ \b{E}(\b{\varphi}) \right] \\*
\label{eq:theory_generatingFunctionaltraj}
	&= \exp\left( i \int_{t_i}^\infty \diff t' \, \b{J}(t') \cdot \b{\tilde{\varphi}}(t';\b{x}) \right) \,.
\end{align}
We remark that we implicitly assume that the observable~$\mathcal{O}$ is an analytic function of the state~$\b{x} \in \b{X}$, i.e., it is possible to write~$\mathcal{O}$ as convergent power series in~$\b{x}$ (to make equation~\eqref{eq:theory_observableFD} mathematically rigorous).

As a simple example for an observable, take the phase space position~$\b{\mathcal{O}}(\b{y}) = \b{y}$ for~$\b{y} \in \b{X}$. It has values in phase space~$\b{X}$ itself (as opposed to being real-valued), but this is no issue at all. Using the prescription~\eqref{eq:theory_observableFD}, we obtain
\begin{align}
	\b{\tilde{\mathcal{O}}}(t;\b{x}) &= \restr{\mathcal{O}\left(\frac{\delta}{i \, \delta \b{J}(t)}\right) \, Z[\b{J}; \b{x}]}{\b{J} = \b{0}} \\*
	&= \restr{\frac{\delta}{i \, \delta \b{J}(t)} \exp\left( i \int_{t_i}^\infty \diff t' \, \b{J}(t') \cdot \b{\tilde{\varphi}}(t';\b{x}) \right)}{\b{J} = \b{0}} \\*
	&= \b{\tilde{\varphi}}(t;\b{x}) .
\end{align}
Thus, the observable~$\b{\mathcal{O}}$ precisely extracts the phase space trajectory which is indeed the phase space position as evaluated along the solution of the equations of motion.

\subsection{Free Generating Functional}

If we are able to find an explicit analytical expression for the generating functional, i.e., express~$Z[\b{J};\b{x}]$ as an elementary functional of~$\b{J}$, we would know everything about the physical system under consideration. Indeed, we could simply obtain the phase space trajectory of the system by functional derivation. Hence, it is expected that in any non-trivial case the path integrals in the generating functional cannot be evaluated analytically without explicitly referring to the solution of the equations of motion. As such, the formalism as developed so far is not all that useful for obtaining explicit results for observables of a classical physical system.

To remedy this, we make the crucial assumption that the equations of motion can be split into a free part and an interaction part, i.e.,
\begin{align}
	\b{E}(\b{\varphi}) = \b{E_0}(\b{\varphi}) + \b{E_I}(\b{\varphi}) = \b{0} \,,
\end{align}
where we demand that the free part~$\b{E_0}(\b{\varphi})$ is given by a linear differential operator acting on~$\b{\varphi}$. A simple example would be~$\b{E_0}(\b{\varphi}) = \partial_t \b{\varphi}$ which shows that this decomposition is always possible for Hamiltonian equations of motion. Ultimately, we will perform a perturbative expansion of the interaction part~$\b{E_I}(\b{\varphi})$, such that it is advantageous to include as many terms as possible into the free part of the equations of motion.

Any linear differential operator admits a Green's function (also known as propagator) which can be used to write down a solution to the associated differential equation. In the specific case of~$\b{E_0}(\b{\varphi}) = \b{0}$, it is
\begin{align}
	\b{\bar{\varphi}}(t;\b{x}) = \b{\mathcal{G}}(t,t_i) \, \b{x} \,,
\end{align}
where we denote the free solution by~$\b{\bar{\varphi}}$ and the Green's function by~$\b{\mathcal{G}}(t,t')$. Note that the free solution usually describing straight paths has been overset by a bar and should not be confused with the phase space trajectory~$\b{\tilde{\varphi}}$ generally describing curved paths and accordingly featuring a tilde. 

Using the decomposition of the equation of motion, we can simplify the generating functional significantly. Suppressing the time dependence for compactness of notation, we find
\begin{align}
	Z[\b{J}; \b{x}] &= \int_{\b{\varphi}(t_i) = \b{x}} \mathcal{D}\b{\varphi} \int \mathcal{D}\b{\chi} \, \exp\left( i \int \diff t' \left( \b{\chi} \cdot \b{E_0}(\b{\varphi}) + \b{\chi} \cdot \b{E_I}(\b{\varphi}) + \b{J} \cdot \b{\varphi} \right) \right) \\
	&= \int_{\b{\varphi}(t_i) = \b{x}} \mathcal{D}\b{\varphi} \int \mathcal{D}\b{\chi} \, \exp\left( i \int \diff t' \, \b{\chi} \cdot \b{E_I}(\b{\varphi}) \right) \restr{\exp\left( i \int \diff t' \left( \b{\chi} \cdot (\b{E_0}(\b{\varphi}) + \b{K}) + \b{J} \cdot \b{\varphi} \right) \right)}{\b{K}=\b{0}} \\
\label{eq:theory_generatingFunctionalwithKprev}
	&= \int_{\b{\varphi}(t_i) = \b{x}} \mathcal{D}\b{\varphi} \int \mathcal{D}\b{\chi} \, \exp\left( i \int \diff t' \, \frac{\delta}{i\,\delta \b{K}} \cdot \b{E_I}\left(\frac{\delta}{i\,\delta \b{J}} \right) \right) \restr{\exp\left( i \int \diff t' \left( \b{\chi} \cdot (\b{E_0}(\b{\varphi}) + \b{K}) + \b{J} \cdot \b{\varphi} \right) \right)}{\b{K}=\b{0}} \\*
\label{eq:theory_generatingFunctionalwithK}
	&= \restr{\exp\left( i \int \diff t' \, \frac{\delta}{i\,\delta \b{K}} \cdot \b{E_I}\left(\frac{\delta}{i\,\delta \b{J}} \right) \right) \, Z_0[\b{J},\b{K}; \b{x}]}{\b{K}=\b{0}} \,.
\end{align}
We introduced another auxiliary function~$\b{K}$ which acts as a source field for~$\b{\chi}$ in order to move the interaction term in front of the path integrals. In the last line we isolated the so-called \emph{free generating functional}
\begin{align}
	Z_0[\b{J}, \b{K}; \b{x}] &= \int_{\b{\varphi}(t_i) = \b{x}} \mathcal{D}\b{\varphi} \int \mathcal{D}\b{\chi} \, \exp\left( i \int \diff t' \left( \b{\chi} \cdot (\b{E_0}(\b{\varphi}) + \b{K}) + \b{J} \cdot \b{\varphi} \right) \right) \\
	&= \int_{\b{\varphi}(t_i) = \b{x}} \mathcal{D}\b{\varphi} \, \delta_D\left[ \b{E_0}(\b{\varphi}) + \b{K} \right] \, \exp\left( i \int \diff t' \, \b{J} \cdot \b{\varphi} \right) \\
	&= \exp \left( i \int_{t_i}^\infty \diff t' \, \b{J}(t') \cdot \b{\bar{\varphi}}(t;\b{x}) - i \int_{t_i}^\infty  \diff t' \int_{t_i}^\infty \diff t'' \, \b{J^\tp}(t') \,\b{\mathcal{G}}(t',t'') \,\b{K}(t'') \right) \,.
\end{align}
The second term in the last line is to be understood as a matrix product and precisely cancels the additional factor~$\b{K}$ in the equation~$\b{E_0}(\b{\varphi}) + \b{K} = \b{0}$. We remark that in performing the integration over~$\b{\varphi}$ we assumed that this equation has symplectic structure such that the associated functional determinant is unity. This is generically the case for Hamiltonian equations of motion.

Using the explicit form obtained for the free generating functional, we can immediately perform the functional derivatives with respect to~$\b{K}$ in equation~\eqref{eq:theory_generatingFunctionalwithK}.\footnote{\label{foot:1}Note that the term~$\b{J^\tp}(t') \, \b{\mathcal{G}}(t',t'')$ obtained from the differentiation can be moved past the functional derivative in~$\b{E_I}\left(\frac{\delta}{i\,\delta \b{J}(t'')}\right)$. Indeed, $\b{\mathcal{G}}(t',t'') = \b{0}$ if~$t' \le t''$ such that the following procedure is possible: We write the exponential as a power series and in each term we sort the products by their time variable~$t''$ (this is accomplished by splitting the integral into time-ordered regions and considering each of them separately). In each product, we move all the functional derivatives with respect to~$\b{K}$ to the right and let them act on the free generating functional. The resulting terms then can be moved back to the original positions of the corresponding functional derivatives, allowing the power series to be converted back into an exponential.} This yields
\begin{align}
\label{eq:theory_Z}
	Z[\b{J}; \b{x}] &= \exp\left( - i \int_{t_i}^\infty  \diff t' \int_{t_i}^\infty \diff t'' \, \b{J^\tp}(t') \, \b{\mathcal{G}}(t',t'') \, \b{E_I}\left(\frac{\delta}{i\,\delta \b{J}(t'')} \right) \right) \, Z_0[\b{J}; \b{x}] \qquad \text{with} \\*
\label{eq:theory_Z0}
	Z_0[\b{J}; \b{x}] &= \exp \left( i \int_{t_i}^\infty \diff t' \, \b{J}(t') \cdot \b{\bar{\varphi}}(t';\b{x}) \right) \,.
\end{align}
Evidently, there are no path integrals left in these expressions and there is no explicit reference to the solution of the equations of motion. Instead, only the free trajectory appears which can be easily calculated using the Green's function. The free generating functional takes exactly the same form as the generating functional~\eqref{eq:theory_generatingFunctionaltraj} with~$\b{\tilde{\varphi}}$ replaced by~$\b{\bar{\varphi}}$.

\subsection{Interactions}

The expression in equations~\eqref{eq:theory_Z} and~\eqref{eq:theory_Z0} for the generating functional strongly suggests a perturbative treatment of the interactions by means of a Taylor expansion. We pursue this strategy momentarily, but as a preparatory step we need to investigate the term~$\b{E_I}\left(\frac{\delta}{i\,\delta \b{J}(t'')} \right)$ in slightly more detail. As usual, the derivative in the argument is to be understood by the virtue of an analytic power series for the function~$\b{E_I}$. As such, we have the identity
\begin{align}
	\b{E_I}\left(\frac{\delta}{i\,\delta \b{J}(t'')} \right) \, \exp \left( i \int \diff t' \, \b{J}(t') \cdot \b{\bar{\varphi}}(t';\b{x}) \right) = \b{E_I}\left( \b{\bar{\varphi}}(t'';\b{x}) \right) \, \exp \left( i \int \diff t' \, \b{J}(t') \cdot \b{\bar{\varphi}}(t';\b{x}) \right) \,.
\end{align}
This comes as no surprise as this is exactly the purpose for which we introduced the functional derivative in equation~\eqref{eq:theory_generatingFunctionalwithKprev} -- it extracts the trajectory from the free generating functional. This identity can be generalized to the case we will be facing in the perturbative treatment. There, in addition to the generating functional, the derivative in~$\b{E_I}$ can also act on single factors of~$\b{J}$. We can use
\begin{align}
	\b{E_I}\left(\frac{\delta}{i\,\delta \b{J}(t'')} \right) (i \, \b{J} \cdot \b{A}) \cdots = \left( (i \, \b{J} \cdot \b{A}) \, \b{E_I}\left(\frac{\delta}{i\,\delta \b{J}(t'')} \right) + (\b{A} \cdot \b{\nabla}) \, \b{E_I}\left(\frac{\delta}{i\,\delta \b{J}(t'')} \right) \right) \cdots \,,
\end{align}
where~$\b{A}$ is a placeholder for terms of the form~$\b{\mathcal{G}}(t',t'') \, \b{E_I}\left(\frac{\delta}{i\,\delta \b{J}(t'')} \right)$ and~$\cdots$ denotes further terms on which the derivatives may act. The appearance of two terms on the right-hand side is due to the product rule. The second term warrants some further remarks as it may seem counter-intuitive that we obtain a derivative acting on~$\b{E_I}$. However, this can once again be understood by virtue of power series. Indeed, for a monomial
\begin{align}
\label{eq:theory_identityderivative}
	(\partial_x)^n (ax) \cdots = \left( (ax) (\partial_x)^n + an \, (\partial_x)^{n-1}\right) \cdots \,,
\end{align}
where the factor~$n$ in the second term accounts for the choice of which of the derivatives acts on the term~$ax$. Observe that resulting expression has the form of the derivative of the monomial. The generalization to a power series is straightforward as the same procedure can be applied term-by-term.

It is possible to proceed in full generality. In higher-orders we obtain higher derivatives of the interaction term~$\b{E_I}$ and it is possible to devise a diagrammatic representation of terms order by order to keep track of how the derivatives (which mathematically are tensors) are contracted with the remaining terms. We are happy to supply such prescription upon request, but refrain from detailing it here. Instead we momentarily specify our physical system to be an interacting $N$-body system in an attempt to sacrifice generality for clarity. Nonetheless, below we give the explicit expressions for the generating functional in full generality up to second order in a Taylor expansion of the exponential in equation~\eqref{eq:theory_Z}:
\begin{align}
	Z_0[\b{J}; \b{x}] &= \exp \left( i \int \diff t' \, \b{J}(t') \cdot \b{\bar{\varphi}}(t';\b{x}) \right) \,, \\
	Z_1[\b{J}; \b{x}] 
		&= -i \iint \diff t_1' \, \diff t_1'' \, \b{J^\tp}(t_1') \, \b{\mathcal{G}}(t_1',t_1'') \, \b{E_I}\left(\b{\bar{\varphi}}(t_1'';\b{x}) \right) Z_0[\b{J}; \b{x}] \,, \\
	Z_2[\b{J}; \b{x}] 
		&= - \iiiint \diff t_1' \, \diff t_1'' \, \diff t_2' \, \diff t_2'' \, \b{J^\tp}(t_1') \, \b{\mathcal{G}}(t_1',t_1'') \, \left[ \b{E_I}\left(\b{\bar{\varphi}}(t_1'';\b{x}) \right) \, \b{J^\tp}(t_2') \, \right. \nonumber \\
		&\hspace{0.1\textwidth} \left. - \, i \, \delta_D(t_1''-t_2') \, \b{\nabla} \b{E_I}\left(\b{\bar{\varphi}}(t_1'';\b{x}) \right) \, \right] \b{\mathcal{G}}(t_2',t_2'') \, \b{E_I}\left(\b{\bar{\varphi}}(t_2'';\b{x}) \right) Z_0[\b{J}; \b{x}] \,.
\end{align}
Note that we denote with~$Z_1$ and~$Z_2$, respectively, the additional contributions from first and second order. The full second order generating functional is given by the sum~$Z_0 + Z_1 + Z_2$. Observables may be obtained via functional derivatives with respect to the source field~$\b{J}$. As an example, the contributions to the phase space trajectory~$\b{\tilde{\varphi}}(t;\b{x})$ up to second order is obtained by acting with~$\frac{\delta}{i \, \delta \b{J}(t)}$ (evaluated at~$\b{J} = \b{0}$) on the expressions above. We obtain
\begin{align}
\label{eq:theory_phi0}
	\b{\tilde{\varphi}}_0(t;\b{x}) &= \frac{\delta}{i \, \delta \b{J}(t)} Z_0[\b{0}; \b{x}] = \b{\bar{\varphi}}(t;\b{x}) = \b{\mathcal{G}}(t,t_i) \, \b{x} \, \\
\label{eq:theory_phi1}
	\b{\tilde{\varphi}}_1(t;\b{x}) 
		&= -\int \diff t_1'' \, \b{\mathcal{G}}(t,t_1'') \, \b{E_I}\left(\b{\bar{\varphi}}(t_1'';\b{x}) \right) \, \\
\label{eq:theory_phi2}
	\b{\tilde{\varphi}}_2(t;\b{x}) 
		&= \iint \diff t_1'' \, \diff t_2'' \, \b{\mathcal{G}}(t,t_1'') \, \b{\nabla} \b{E_I}\left(\b{\bar{\varphi}}(t_1'';\b{x}) \right) \, \b{\mathcal{G}}(t_1'',t_2'') \, \b{E_I}\left(\b{\bar{\varphi}}(t_2'';\b{x}) \right)
\end{align}
Note that the term containing two factors of~$\b{J}$ at second order vanishes for this particular observable because the leftover factor~$\b{J}$ is set to zero (for the same reason, the functional derivative cannot act on~$Z_0$ at any positive order). The first order phase space trajectory, given by~$\b{\tilde{\varphi}}_0(t;\b{x}) + \b{\tilde{\varphi}}_1(t;\b{x})$, coincides with the so-called Born approximation which adds the interactions as evaluated along the free trajectory and is a common ad hoc approximation used, e.g., in scattering problems.\footnote{\label{foot:theory_Born}Note that if we replace the free trajectory~$\b{\bar{\varphi}}$ inside~$\b{E_I}$ on the right-hand side of equation~\eqref{eq:theory_phi1} by the actual phase space trajectory~$\b{\tilde{\varphi}}$, the sum of the resulting term and the free trajectory reproduce~$\b{\tilde{\varphi}}$ exactly. Indeed, we have the well-known integral equation
\begin{align*}
	\b{\tilde{\varphi}}(t;\b{x}) = \b{\bar{\varphi}}(t;\b{x}) - \int_{t_i}^t \diff t_1 \, \b{\mathcal{G}}(t,t_1) \, \b{E_I}\left(\b{\tilde{\varphi}}(t_1;\b{x}) \right) \,.
\end{align*}
The Born approximation is the first step in solving this integral equation iteratively (this process is sometimes referred to as iterative Born approximation). We remark that the result of this iteration in second order (where one inserts the (first order) Born approximation for the trajectory on the right-hand side) does not coincide with the second order KFT trajectory -- infinitely many terms of second order iterative Born approximation are assigned to higher-orders in KFT. We elaborate on this in footnote~\ref{foot:feynman_Born} using the diagrammatic treatment of KFT perturbation theory.} Here it is derived as the first order perturbative approximation in the KFT framework.

\subsection{Choice of Free Motion}

As a last topic for the general treatment, we want to briefly discuss the splitting of the equations into a free and an interaction part. It should be stressed that this is a choice which is to a certain degree arbitrary. Indeed, one can always take the trivial splitting~$\b{E_0} = \b{0}$ and~$\b{E_I} = \b{E}$ regarding the entirety of the equations of motion as interactions. This way, the Green's function is the identity and the free trajectory is constant~$\b{\bar{\varphi}}(t;\b{x}) = \b{x}$. However, this is a poor choice because the interaction part of the equations of motion is only treated perturbatively, while the free part is solved exactly. Hence, it is advantageous to include as many terms of the equations of motion as possible into the free part~$\b{E_0}$. The ability to do so is of course limited by the requirement of the existence of a Green's function for the free motion. Nonetheless, minimizing the interaction part~$\b{E_I}$ generally is expected to improve the accuracy and convergence of the perturbative treatment.

This situation is familiar from many other areas of physics. Fundamentally, what we consider as free (or background) evolution and what we regard as interactions (or perturbations) on top of it, is a choice. As an example, given a density field~$\rho(q)$ we may regard it as the sum of a background component~$\bar{\rho}(q)$ and a perturbation~$\delta \rho(q)$, i.e., $\rho(q) = \bar{\rho}(q) + \delta\rho(q)$. However, there is a large freedom in performing this split and depending on the situation different choices might be advantageous. In the example, the most natural split is obtained by introducing a condition like~$\frac{1}{V} \int \diff q \, \rho(q) = \bar{\rho}$. But if we, e.g., describe the density field of gas in a galaxy, one might want to consider the dark matter halo as part of the background density.

Likewise in our situation, there is a natural choice for the free motion. Newton' first law asserts that the reference motion is uniform along straight lines. His second law can then be used to determine the acceleration relative to this reference motion. Commonly we would claim that this acceleration is due to a force caused by interactions. However, in some situations it may be advantageous to include part of this acceleration into the free motion. For example, on an expanding background, we usually prefer to use comoving coordinates. In doing so, the physical trajectories of the particles are no longer straight lines. While this makes the free motion a bit more complicated, it has the advantage that the forces between particles are no longer sourced by the density field, but only by the density excess over the mean density (cf.~\ref{sec:app_ParticlesExpandingBackground}).

There is an even more sophisticated choice. It is possible to use the Zel'dovich approximation for the free motion and model the interactions relative to it~\cite{L29,L3d}. In this case it is less clear what the interaction potential should be and, in fact, it is not even clear whether a simple two-body interaction can reproduce the forces between the particles. Recent work suggests, however, that this is indeed possible and demonstrates that a Yukawa-like interaction potential describes the force relative to Zel'dovich trajectories well~\cite{bartelmann2020kinetic}. Here, we limit ourselves to the more conventional case of using uniform motion along straight paths in comoving coordinates as our free motion.

We emphasise that the choice of free motion does not have any effect on observables. The trajectory~$\b{\tilde{\varphi}}$ and all other observables are exactly the same for any split of the equations of motion~$\b{E} = \b{E_0} + \b{E_I}$. The free trajectory~$\b{\bar{\varphi}}$ and the interaction part~$\b{E_I}$ can be altered significantly, though, which can have significant effects on computational complexity and the validity of approximations. For example, the perturbative treatment of interactions presented in this work may have very different convergence properties and speed depending on the choice of free motion. It might be an interesting task for future works to formalize and study the symmetry~$\left(\b{E_0},\b{E_I}\right) \mapsto \left(\b{E_0} + \b{\xi}, \b{E_I} - \b{\xi}\right)$ and find optimal choices for~$\b{\xi}$.


\section{Interacting N-body Systems}

The discussion above was completely general, but also rather abstract. The only assumption we made was that the equations of motion are of Hamiltonian form and split into a free and an interaction part. We needed their symplectic structure to omit a functional determinant which otherwise would have been a bit of a nuisance. As such, we have been studying a Hamiltonian system governed by an equation~$\b{E_0}(\b{\varphi}) + \b{E_I}(\b{\varphi}) = \b{0}$ evolving from initial conditions~$\b{x} \in \b{X}$ along a phase space trajectory~$\b{\tilde{\varphi}}(t;\b{x})$.

In this section we specialize our abstract treatment and notation to interacting $N$-body systems. We stress that it is possible, albeit notationally laborious, to keep the entire perturbative expansion abstract and general. In particular, it is possible to define Feynman rules for the general setting which yield the Feynman rules derived below as special cases. We refrain from listing them here or even in a dedicated appendix as this would require to introduce an excessive amount of non-trivial notation or, alternatively, work with equations featuring a zoo of indices.

\subsection{Generating Functional}

Let us consider a system of~$N$ interacting particles in three-dimensional Euclidean space. The state of this physical system can be described by a point in phase space~$\b{x} \in \b{X} \cong \R^{6N}$. Let us take one such point as our initial conditions and define
\begin{align}
	\b{\tilde{\varphi}} \c \R \times \b{X} \ra \b{X} \qquad \text{with} \qquad \b{\tilde{\varphi}}(t;\b{x}) = \left( \tilde{\varphi}_1(t;\b{x}), \tilde{\varphi}_2(t;\b{x}), \ldots, \tilde{\varphi}_N(t;\b{x}) \right)^\tp \in \b{X}
\end{align}
to be the phase space trajectory of the system. Here, $\tilde{\varphi}_j(t;\b{x})$ is a six-dimensional vector describing the position and momentum of the $j\nth$~particle at time~$t$. As we have to be able to access the position and momentum parts of this and other objects frequently below, let us introduce the following notation: For any point~$x$ in single-particle phase space~$X \cong \R^6$ we write~${x}^{(q)}$ for its position and~${x}^{(p)}$ for its momentum, i.e., it is~$x = \left({x}^{(q)},{x}^{(p)}\right)^\tp \in X$. Similarly, the $3N$-dimensional vector containing all the positions is given by~$\b{{x}^{(q)}} = \left({x}_1^{(q)}, {x}_2^{(q)}, \ldots, {x}_N^{(q)}\right)^\tp$.

Using this notation and particularizing to an interacting $N$-body system, the equations of motion for the $j\nth$~particle are given by
\begin{align}
\label{eq:Nbody_eom}
	\partial_t \varphi_j(t) - \begin{pmatrix} \frac{1}{m} {\varphi}_j^{(p)}(t) \\ - m \, \partial_{{\varphi}_j^{(q)}} V\left(t; \b{\varphi^{(q)}} \right) \end{pmatrix} = 0 \,.
\end{align}
The potential~$V$ is given by
\begin{align}
\label{eq:Nbody_potential}
	V\left(t; \b{\varphi^{(q)}}\right) = \frac{1}{2} \sum_j \sum\limits_{l \neq j} v\left({\varphi}_j^{(q)}(t) - {\varphi}_l^{(q)}(t), t \right)
\end{align}
featuring the two-particle interaction potential~$v$. We allow for an explicit time-dependence of this potential for later convenience. It is useful to perform a Fourier transform resulting in
\begin{align}
	V\left(t; \b{\varphi^{(q)}}\right) = \frac{1}{2} \sum_j \sum\limits_{l \neq j} \int \frac{\diff^3 k'}{(2\pi)^3} \, v(k', t) \, \exp\left( i k' \left( {\varphi}_j^{(q)}(t) - {\varphi}_l^{(q)}(t) \right) \right) \,,
\end{align}
where we use the symbol~$v$ for the Fourier transform of the two-particle interaction potential, too. We abuse notation in this way throughout and distinguish functions and their Fourier transforms only via their arguments. For Newtonian gravitational interactions in Newtonian gauge we have~$v(k') \propto - \frac{1}{\norm{k'}^2}$ and the splitting of the equations of motion is given by
\begin{align}
	(E_0)_j &= \partial_t \varphi_j(t) - \begin{pmatrix} \frac{1}{m} {\varphi}_j^{(p)}(t) \\ 0 \end{pmatrix} \,, \\*
	(E_I)_j &= \begin{pmatrix} 0 \\ \sum\limits_{l \neq j} \int \frac{\diff^3 k'}{(2\pi)^3} \, i \, m \, k' \, v(k',t) \, \exp\left( i k' \left( {\varphi}_j^{(q)}(t) - {\varphi}_l^{(q)}(t) \right) \right) \end{pmatrix} \,.
\end{align}

A Green's function for the free equations of motion is given by
\begin{align}
\label{eq:Nbody_GreensFunction}
	\mathcal{G}(t,t') = \begin{pmatrix} \theta(t-t') \, \mathcal{I}_3 & g_{qp}(t,t') \, \mathcal{I}_3 \\ 0 & g_{pp}(t,t') \, \mathcal{I}_3 \end{pmatrix} 
\end{align}
with~$g_{qp}(t,t') = \frac{1}{m} (t-t') \, \theta(t-t')$ and~$g_{qp}(t,t') = \theta(t-t')$ in Newtonian gauge on a static background. We keep the propagators~$g_{qp}(t,t')$ and $g_{pp}(t,t')$ as well as the interaction potential~$v(k',t')$ general below to allow for gauge changes in the final results. Moreover, this leaves sufficient freedom to transfer our results to an expanding background necessary for the cosmological case (cf.~\ref{sec:app_ParticlesExpandingBackground}).

Using the explicit forms for the free and interaction parts of the equations of motion, we can simplify the expressions in equations~\eqref{eq:theory_Z} and~\eqref{eq:theory_Z0} for the generating functional. Indeed, we obtain
\begin{align}
	\label{eq:Nbody_Z}
	Z[\b{{J}^{(q)}}; \b{x}] &= \exp\left( \sum\limits_{j} \sum\limits_{l \neq j} \iint \diff t' \, \diff t'' \int \frac{\diff^3 k'}{(2\pi)^3} \, m \, g_{qp}(t',t'') \, v(k',t'') \right. \nonumber\\*
	&\hspace{0.1\textwidth} \left. {J}_j^{(q)}(t') \cdot k' \, \exp\left( i k' \left( \frac{\delta}{i \,\delta {J}_j^{(q)}(t'')} - \frac{\delta}{i\,\delta {J}_l^{(q)}(t'')} \right) \right) \right) \, Z_0[\b{{J}}^{(q)}; \b{x}] \,, \\
	\label{eq:Nbody_Z0}
	Z_0[\b{{J}^{(q)}}; \b{x}] &= \exp \left( i \sum_j \int \diff t' \, {J}_j^{(q)}(t') \cdot \bar{{\varphi}}{}_j^{(q)}(t';\b{x}) \right) \qquad \text{with} \\*
	\bar{{\varphi}}{}_j^{(q)}(t';\b{x}) &= {x}_j^{(q)} + g_{qp}(t',t_i) \, {x}_j^{(p)} \,.
\end{align}
Note that we have set the momentum part of the source field~$\b{{J}^{(p)}}$ to zero as we are exclusively interested in observables featuring positions of particles. We remark for later reference that this avoids any appearance of the function~$g_{pp}(t,t')$ here and below.

\subsection{Density Correlation Functions}
\label{subsec:DensityCorrelationFunctions}

Aside from the trajectories themselves, the density and its $r$-point correlation functions are the most interesting observables for an interacting $N$-body system. In fact, the collective information encoded in the density correlation functions tends to be much more useful in practice compared to the microscopic information encoded in the specific positions of the individual particles. In our applications we integrate the observables over a probability distribution of initial conditions making the specific realization of the system irrelevant. Instead, we are interested in statistical quantities like the density fluctuation power spectrum. In this section we explain how density $r$-point functions can be obtained from the generating functional.

The particle number density at position~$q$, given the positions of particles~$\b{{\varphi}^{(q)}}(t)$ at time~$t$, is
\begin{align}
\label{eq:NBody_densityDefinition}
	\rho\left(q;\b{{\varphi}^{(q)}}(t)\right) = \sum_j \delta_D\left(q - {\varphi}_j^{(q)}(t)\right) \,
\end{align}
Performing a Fourier transformation which introduces a wave vector~$k$ conjugate to~$q$, the density takes the form
\begin{align}
	\rho\left(k;\b{{\varphi}^{(q)}}(t)\right) = \sum_{j} \exp\left( - i \, k \cdot {\varphi}_j^{(q)}(t) \right) \,.
\end{align}
We remark that these equations are valid both for the comoving and the Eulerian densities (cf.~\ref{sec:app_ParticlesExpandingBackground}) because the normalization of the Dirac $\delta$-function, resp.\ the wave vectors, compensate tacitly the factors of the scale factor. Thus, the equations written down here and below are valid both for a static and an expanding background as long as we regard the wave vector as comoving in the expanding case.

One of the advantages of working in Fourier space is that it is very easy to construct the density correlation functions. Indeed, the (Fourier space) density $r$-point correlation function simply is
\begin{align}
	G_{\underset{r~\text{times}}{\underbrace{\rho \cdots \rho}}}(k_1,\ldots,k_r;\b{{\varphi}^{(q)}}(t)) = \prod_{s=1}^r \rho(k_s;\b{\varphi}(t)) \,.
\end{align}
Written out, this yields
\begin{align}
\label{eq:NBody_rpointDensDef}
	G_{\rho \cdots \rho}(k_1,\ldots,k_r;\b{{\varphi}^{(q)}}(t)) = \sum_{\{j_1,\ldots,j_r\}} \exp\left( -i \, \sum_{s=1}^r k_s \cdot {\varphi}_{j_s}^{(q)}(t) \right) \,,
\end{align}
where the sum in front of the exponential runs over all $r$-tuples of particles indices $1 \le j_s \le N$.

The observables defined above are all given in terms of the positions of the ensemble of particles~$\b{{\varphi}^{(q)}}(t)$. In order to calculate these observables for the specific system under consideration, we ought to replace these positions by the solution of the equations of motion~$\b{\tilde{\varphi}^{(q)}}(t;\b{x})$ for intial values~$\b{x}$. However, we do not attempt to find this solution explicitly, but instead use the identity in equation~\eqref{eq:theory_observableFD}, i.e.,
\begin{align}
	\tilde{\mathcal{O}}(t;\b{x}) = \mathcal{O}\left(\b{\tilde{\varphi}}(t;\b{x})\right) = \restr{\mathcal{O}\left(\frac{\delta}{i \, \delta \b{J}(t)}\right) \, Z[\b{J}; \b{x}]}{\b{J} = \b{0}} \,.
\end{align}
Explicitly, we replace the positions~$\b{{\varphi}^{(q)}}(t)$ by a functional derivative with respect to~$\b{{J}^{(q)}}(t)$ in the expressions for the density and the $r$-point correlation function above. This yields the operator
\begin{align}
\label{eq:NBody_rpointDensOp}
	\hat{G}_{\rho \cdots \rho}(k_1,\ldots,k_r,t) = \sum_{\{j_1,\ldots,j_r\}} \exp\left( -i \, \sum_{s=1}^r k_s \cdot \frac{\delta}{i\,\delta {J}_{j_s}^{(q)}(t)} \right)
\end{align}
with which we can act on the generating functional~$Z[\b{J}; \b{x}]$ to obtain the r-point correlation function. The density operator~$\hat{\rho}(k,t)$ is obtained for the special choice~$r = 1$.


\section{KFT Perturbation Scheme}
\label{sec:feynman}

In this section we present a perturbation scheme for obtaining the density correlation functions for an interacting $N$-body system. We stress again that the entire procedure can be done for any Hamiltonian system allowing for a separation of free motion and interactions. For clarity, concreteness and simplicity of notation we focus on interacting $N$-body systems here which are also our main application.

Above we already hinted at the possibility of performing a Taylor expansion of the exponential containing the interaction part of the equations of motion in the generating functional~\eqref{eq:theory_Z}. In doing so, we obtain a polynomial expression containing in each term factors of the source field~$\b{J}$ and functional derivatives with respect to it. To the very right-hand side of the expression is the free generating functional which also contains~$\b{J}$. Evidently, in each term a certain amount of combinatorics is required to keep track of all the ways the derivatives could act on the various factors of~$\b{J}$.

These combinatorics can be easily done algorithmically via successive product rules, but this is neither enlightening, nor efficient. Instead we take inspiration from the path integral formulation of QFT which arrives at very similar expressions for observables. There, the method of diagrammatically representing mathematical expressions has been applied with great success. While our expressions are somewhat unusual from a QFT point of view -- most of our functional derivatives come inside Fourier phase factors -- it is nonetheless possible to obtain a diagrammatic representation in our case, too. Below we derive our Feynman rules and use them to find perturbative expressions for the observables we are interested in.

\subsection{Taylor Expansion}

Let us consider the generating functional for interacting $N$-body systems given in equation~\eqref{eq:Nbody_Z}. Replacing the exponential by its Taylor expansion we have the expression
\begin{align}
\label{eq:pert_ZTaylor}
	Z[\b{{J}^{(q)}}; \b{x}] &= \sum_{n=0}^{\infty} \frac{1}{n!} \left[ \sum\limits_{j} \sum\limits_{l \neq j} \iint \diff t' \, \diff t'' \int \frac{\diff^3 k'}{(2\pi)^3}\, m \,g_{qp}(t',t'') \, v(k',t'') \right. \nonumber\\*
	&\hspace{0.1\textwidth} \left. \vphantom{\sum\limits_{l \neq j}} {J}_j^{(q)}(t') \cdot k' \, \exp\left( i \, k' \cdot \left( \frac{\delta}{i \,\delta {J}_j^{(q)}(t'')} - \frac{\delta}{i\,\delta {J}_l^{(q)}(t'')} \right) \right) \right]^n \, Z_0[\b{{J}^{(q)}}; \b{x}]
\end{align}
with the free generating functional as given in equation~\eqref{eq:Nbody_Z0}. It is worthwhile to point out that up to this point the entire treatment has been exact. Perturbatively, the $n\nth$~order contribution, denoted below by~$Z_n$, is simply given by the $n\nth$~term in this sum. The full expression for the generating functional to $n\nth$~order is the sum of the first $n$~terms in the sum. Evidently, this truncation constitutes an approximation which may have a limited validity. Taylor's theorem comes with estimates of the truncation error, but it is not trivial to apply them to our case of a derivative operator present in the expansion. We intend to investigate this in future works.

Just like one would do in QFT, we intend to derive Feynman rules to diagrammatically represent the perturbative expressions for observables. The class of observables we want to focus on are density correlation functions as introduced in section~\ref{subsec:DensityCorrelationFunctions}. The usefulness of the Fourier transform becomes evident now, as the phase factors coming from the interaction are similar to the ones coming from the observables. This allows for a unified and hence simplified treatment. We remark that similar Feynman rules can be obtained for, e.g., the trajectories of particles yielding a general recipe for the expressions in equations~\eqref{eq:theory_phi0} to~\eqref{eq:theory_phi2}. 

Let us write down the expression for the $n\nth$~order density $r$-point correlation function. It is obtained by acting with the operator~$\hat{G}_{\rho \cdots \rho}$ given in equation~\eqref{eq:NBody_rpointDensOp} on the $n\nth$~term of the sum in equation~\eqref{eq:pert_ZTaylor}. We obtain
\begin{align}
\label{eq:feynman_denscorrFull}
	\left(G_{\rho \cdots \rho}\right)_n(k_1,\ldots,k_r,t;\b{x}) &= \frac{1}{n!} \sum_{\{m_1,\ldots,m_r\}} \exp\left( -i \, \sum_{s=1}^r k_s \cdot \frac{\delta}{i\,\delta {J}_{m_s}^{(q)}(t)} \right) \nonumber\\*
	&\qquad\left[ \sum\limits_{j} \sum\limits_{l \neq j} \iint \diff t' \, \diff t'' \int \frac{\diff^3 k'}{(2\pi)^3}\, m \, g_{qp}(t',t'') \, v(k',t'') \, {J}_j^{(q)}(t') \cdot k' \right. \nonumber\\*
	&\qquad \left. \vphantom{\sum\limits_{l \neq j}} \exp\left( i \, k' \cdot \left( \frac{\delta}{i \,\delta {J}_j^{(q)}(t'')} - \frac{\delta}{i\,\delta {J}_l^{(q)}(t'')} \right) \right) \right]^n \, \restr{\vphantom{\sum\limits_{l \neq j}} Z_0[\b{{J}^{(q)}}; \b{x}]}{\b{{J}^{(q)}}=\b{0}} \,.
\end{align}
Despite the seeming complexity, this expression has actually quite a pleasant structure. The key mathematical identity to use is
\begin{align}
	&\exp\left( k \cdot \frac{\delta}{\delta {J}_m^{(q)}(t)} \right) \, {J}_j^{(q)}(t') \cdot k' \, \cdots \, Z_0[\b{{J}^{(q)}}; \b{x}] \nonumber \\*
	=~&\left( {J}_j^{(q)}(t') \cdot k' + \delta_{jm} \, \delta_D\left(t-t'\right) \, k \cdot k' \right) \, \exp\left( k \cdot \frac{\delta}{\delta {J}_m^{(q)}(t)} \right) \, \cdots  \,Z_0[\b{{J}^{(q)}}; \b{x}],
\end{align}
which can be seen as a special case of the discussion around equation~\eqref{eq:theory_identityderivative}. Again the dots~$\cdots$ denote further terms on which the derivative may act. Given that the phase factors are not modified, they ultimately act in this form on the free generating functional~$Z_0$. Hence, in the final expression we can replace all the functional derivatives by the free trajectory~$\b{\bar{\varphi}^{(q)}}$. The only difficulty is to keep track of which of the factors of~${J}_j^{(q)}$ is acted upon by which of the phase factors. Note that this action happens exactly once as any remaining factors of~${J}_j^{(q)}$ would cause the entire term to vanish upon setting the source field to zero.

\subsection{Feynman Rules}
\label{sec:feynman_rules}

We want to find a diagrammatic representation of equation~\eqref{eq:feynman_denscorrFull}. The first step in achieving this is to understand the term in the square bracket which is taken to the $n\nth$~power. It contains both factors of the source field~$\b{{J}^{(q)}}$ as well as derivatives with respect to it. Hence, the ordering of the individual factors is crucial. Back in equation~\eqref{eq:theory_generatingFunctionalwithK} the ordering was irrelevant because the two derivatives commute. But as soon as we performed the functional derivative with respect to~$\b{K}$ in equation~\eqref{eq:theory_Z} we had to be careful with this issue. In fact, in footnote~\ref{foot:1}, we explained in detail how it is even possible to perform these derivatives -- by time ordering all terms and exploiting the fact that the Green's function is causal. This allowed us to move the factors of~$\b{J}$ back to the left past some of the functional derivatives.

We can postpone the step in which we move the factors of~$\b{J}$ back past the functional derivatives for the moment. Then all these factors -- here in the form of~${J}_j^{(q)}$ -- are towards the very right of the expression just next to the free generating functional from which they were obtained via differentiation. All Fourier phase factors containing the functional derivatives are left of it. Hence, a priori the situation is such that any phase factor could act on any~${J}_j^{(q)}$.

We choose this as the starting point for formulating our Feynman rules. Let us use the phase factors as the vertices of our diagrams:
\begin{align}
	\label{rule:dens_vertex}
	\begin{tikzpicture}[baseline={([yshift=-.5ex]current bounding box.center)}]
		\node[outdens] (G1) at (0,0) {$k_s$, $t$};
	\end{tikzpicture}\ &\ce \sum_{m_s} \exp\left( -i \, k_s \cdot {\bar{\varphi}}_{m_s}^{(q)}(t;\b{x}) \right) \,,\\*
	\label{rule:int_vertex}
	\begin{tikzpicture}[baseline={([yshift=-.5ex]current bounding box.center)}]
		\node[vertex] (G1) at (0,0) {$a$};
	\end{tikzpicture}\ &\ce \sum\limits_{j_a} \sum\limits_{l_a \neq j_a} \int_{t_i}^\infty \diff t_a' \int \frac{\diff^3 k_a'}{(2\pi)^3} \, m \, v(k_a',t_a') \, \exp\left( i \, k_a' \cdot \left( {\bar{\varphi}}_{j_a}^{(q)}(t_a';\b{x}) - {\bar{\varphi}}_{l_a}^{(q)}(t_a';\b{x}) \right) \right) \,.
\end{align}
We call the first of them \emph{density vertex} and the second \emph{interaction vertex}. Note that we replaced the functional derivatives with the free trajectory in accordance with our earlier comments. The remaining parts of the expression in equation~\eqref{eq:feynman_denscorrFull} are encoded in the edges of the diagrams. These are
\begin{align}
	\label{rule:internal_edge}
	\begin{tikzpicture}[baseline={([yshift=-.5ex]current bounding box.center)}]
		\node[empty] (A) at (-2,0) {$a$};
		\node[empty] (B) at (0,0) {$b$};
		\draw[prop] (A) -- (B);
	\end{tikzpicture}\ &\ce g_{qp}(t_b',t_a') \, k_a' \cdot k_b' \, \left( \delta_{j_aj_b} - \delta_{j_al_b} \right) \,,\\*
	\label{rule:dens_edge}
	\begin{tikzpicture}[baseline={([yshift=-.5ex]current bounding box.center)}]
		\node[empty] (A) at (-2,0) {$a$};
		\node[outdensempty] (B) at (0,0) {$k_s$, $t$};
		\draw[prop] (A) -- (B);
	\end{tikzpicture}\ &\ce - g_{qp}(t,t_a') \, k_a' \cdot k_s \, \delta_{j_am_s} \,,
\end{align}
where the uncoloured vertices mean that the corresponding vertex factors are not yet included.

We claim the following:\footnote{We use graph-theoretical language here: A tree is a connected and acyclic graph, i.e., a connected diagram without loops. A forest is the disjoint union of trees.} The expression~\eqref{eq:feynman_denscorrFull} is given by the sum over all possible forests formed from the vertices and edges defined above. Each forest consists of $r$~trees of which each has one root given by a density vertex. All other vertices are interaction vertices. We comment on our notion of equivalence of diagrams in section~\ref{sec:feynman_symmetryFactors} below. The mathematical expression encoded by a diagram is the product of all vertices times the product of all edges (this way all the factors coming from the edges appear under the sums and integrals as they should). The $n\nth$~order contribution to the $r$-point density correlation function is the set of diagrams with exactly $n$~interaction vertices and $r$~density vertices.

The crucial ingredient for proving this claim is the time ordering due to causality. Mathematically this is encoded in the Heaviside function~$\theta(t-t')$ inside the propagator~$g_{qp}(t,t')$ which ensures that functional derivatives may only act on factors~${J}_j^{(q)}$ with earlier time argument. This prevents loops and is also the reason why the density vertices (which come with the highest possible time coordinate~$t$) are the roots of the trees. Upon time ordering the diagram, all interaction vertices have exactly one outgoing edge due to the fact that each factor~${J}_j^{(q)}$ is acted upon exactly once -- yet another way of seeing that we obtain collections of trees with density vertices as roots. As a final remark, the lower bound for the time integration is due to our initial conditions~$\b{x}$ being set at~$t = t_i$.\footnote{We arguably have been somewhat cavalier with integration boundaries above which we excuse with the aim of keeping equations compact. In equation~\eqref{eq:theory_Tfi} a lower bound~$t \ge t_i$ applies to the time integration which carries through. In fact, there is technically also an upper bound which we simply denoted by~$t$ in equation~\eqref{eq:theory_Tfi} and which needs to be at least as large as the time coordinate at which we calculate observables. Here we have simply send it to infinity.}

For concreteness, we list the diagrams and expressions for the density, i.e., the case $r = 1$, up to first order below. In zeroth order we have zero interaction vertices and therefore simply obtain the free density
\begin{align}
	\label{eq:feynman_dens0}
	\tilde{\rho}_0(k,t;\b{x}) &= \
	\begin{tikzpicture}[baseline={([yshift=-.5ex]current bounding box.center)}]
		\node[outdens] (G1) at (0,0) {$k$, $t$};
	\end{tikzpicture}\ = \sum_{m_1} \exp\left( - i \, k \cdot {\bar{\varphi}}_{m_1}^{(q)}(t;\b{x}) \right) \,.
\intertext{As this corresponds to the interactions being completely absent, it is no surprise that we evaluate the (Fourier transform) of the density function along the free trajectory. The first order correction is given by the diagram}
	\tilde{\rho}_1(k,t;\b{x}) &= \
	\begin{tikzpicture}[baseline={([yshift=-.5ex]current bounding box.center)}]
		\node[vertex] (I1) at (-2,0) {$1$};
		\node[outdens] (D) at (0,0) {$k$, $t$};
		\draw[prop] (I1) -- (D);
	\end{tikzpicture}\\*
		&= - \sum_{m_1} \exp\left( - i \, k \cdot {\bar{\varphi}}_{m_1}^{(q)}(t;\b{x}) \right) \sum\limits_{l_1 \neq m_1} \int_{t_i}^\infty \diff t_1' \, g_{qp}(t,t_1') \nonumber \\*
		&\hspace{0.1\textwidth} \int \frac{\diff^3 k_1'}{(2\pi)^3} \, m \, v(k_1',t_1') \, k_1' \cdot k \, \exp\left( i \, k_1' \cdot \left( {\bar{\varphi}}_{m_1}^{(q)}(t_1';\b{x}) - {\bar{\varphi}}_{l_1}^{(q)}(t_1';\b{x}) \right) \right) \,.
\end{align}
In addition, let us list the diagrams which appear in second and third order. The corresponding expressions follow from the Feynman rules and become rather lengthy quite quickly.
\begin{align}
	\label{eq:feynman_dens2}
	\tilde{\rho}_2(k,t;\b{x}) &= \
	\begin{tikzpicture}[baseline={([yshift=-.5ex]current bounding box.center)}]
		\node[vertex] (I1) at (-2,0.4) {$1$};
		\node[vertex] (I2) at (-1.5,-0.4) {$2$};
		\node[outdens] (D) at (0,0) {$k$, $t$};
		\draw[prop] (I1) -- (D);
		\draw[prop] (I2) -- (D);
	\end{tikzpicture}\ +\
	\begin{tikzpicture}[baseline={([yshift=-.5ex]current bounding box.center)}]
		\node[vertex] (I1) at (-3.0,0) {$1$};
		\node[vertex] (I2) at (-1.5,0) {$2$};
		\node[outdens] (D) at (0,0) {$k$, $t$};
		\draw[prop] (I1) -- (I2);
		\draw[prop] (I2) -- (D);
	\end{tikzpicture}\ \,,\\[2ex]
	\label{eq:feynman_dens3}
	\tilde{\rho}_3(k,t;\b{x}) &= \
	\begin{tikzpicture}[baseline={([yshift=-.5ex]current bounding box.center)}]
		\node[vertex] (I1) at (-2.5,0.66) {$1$};
		\node[vertex] (I2) at (-2,0) {$2$};
		\node[vertex] (I3) at (-1.5,-0.66) {$3$};
		\node[outdens] (D) at (0,0) {$k$, $t$};
		\draw[prop] (I1) -- (D);
		\draw[prop] (I2) -- (D);
		\draw[prop] (I3) -- (D);
	\end{tikzpicture}\ +\
	\begin{tikzpicture}[baseline={([yshift=-.5ex]current bounding box.center)}]
		\node[vertex] (I1) at (-3.5,0.4) {$1$};
		\node[vertex] (I2) at (-2,0.4) {$2$};
		\node[vertex] (I3) at (-1.5,-0.4) {$3$};
		\node[outdens] (D) at (0,0) {$k$, $t$};
		\draw[prop] (I1) -- (I2);
		\draw[prop] (I2) -- (D);
		\draw[prop] (I3) -- (D);
	\end{tikzpicture}\ +\nonumber\\*[1ex]
	&\qquad \begin{tikzpicture}[baseline={([yshift=-.5ex]current bounding box.center)}]
		\node[vertex] (I1) at (-3.5,0.4) {$1$};
		\node[vertex] (I2) at (-1.5,0.4) {$3$};
		\node[vertex] (I3) at (-2.5,-0.4) {$2$};
		\node[outdens] (D) at (0,0) {$k$, $t$};
		\draw[prop] (I1) -- (I2);
		\draw[prop] (I2) -- (D);
		\draw[prop] (I3) -- (D);
	\end{tikzpicture}\ +\
	\begin{tikzpicture}[baseline={([yshift=-.5ex]current bounding box.center)}]
		\node[vertex] (I1) at (-3.0,0.4) {$2$};
		\node[vertex] (I2) at (-1.5,0.4) {$3$};
		\node[vertex] (I3) at (-3.5,-0.4) {$1$};
		\node[outdens] (D) at (0,0) {$k$, $t$};
		\draw[prop] (I1) -- (I2);
		\draw[prop] (I2) -- (D);
		\draw[prop] (I3) -- (D);
	\end{tikzpicture}\ +\nonumber\\*[1ex]
	&\qquad \begin{tikzpicture}[baseline={([yshift=-.5ex]current bounding box.center)}]
		\node[vertex] (I1) at (-3.5,0.4) {$1$};
		\node[vertex] (I2) at (-3,-0.4) {$2$};
		\node[vertex] (I3) at (-1.5,0) {$3$};
		\node[outdens] (D) at (0,0) {$k$, $t$};
		\draw[prop] (I1) -- (I3);
		\draw[prop] (I2) -- (I3);
		\draw[prop] (I3) -- (D);
	\end{tikzpicture}\ +\
	\begin{tikzpicture}[baseline={([yshift=-.5ex]current bounding box.center)}]
		\node[vertex] (I1) at (-4.5,0) {$1$};
		\node[vertex] (I2) at (-3.0,0) {$2$};
		\node[vertex] (I3) at (-1.5,0) {$3$};
		\node[outdens] (D) at (0,0) {$k$, $t$};
		\draw[prop] (I1) -- (I2);
		\draw[prop] (I2) -- (I3);
		\draw[prop] (I3) -- (D);
	\end{tikzpicture} \,.
\end{align}
We explain in detail why these are the correct sets of diagrams momentarily. Note that there is a certain amount of freedom in defining equivalence of diagrams which in turn induces some symmetry factors. The convention we choose here avoids symmetry factors altogether and allows for easy numerical evaluation of the resulting expressions. There is a minimal change in the Feynman rules to achieve this. If one were to use the rules as written down above, symmetry factors~$\frac{1}{2}$ and~$1$ for the diagrams of~$\tilde{\rho}_2$ and factors~$\frac{1}{6}$,~$\frac{1}{3}$,~$\frac{1}{3}$,~$\frac{1}{3}$,~$\frac{1}{2}$ and~$1$ for~$\tilde{\rho}_3$ would have to be put into the expressions.

\subsection{Symmetry Factors and Shot-Noise}
\label{sec:feynman_symmetryFactors}

As pointed out above, any discussion of symmetry factors needs to start out with defining what we mean by equivalence of diagrams. We shall use here the simple notion that two diagrams are equivalent if and only if they only differ by the labels of the vertices. In particular, we distinguish between topologically equal graphs with different horizontal ordering of the vertices (an example are diagrams two to four in~\eqref{eq:feynman_dens3}). Equivalent diagrams only need to be listed once in the sum over all diagrams. Naively, the cardinality of each such equivalence class appears to~$n!$, where $n$~is the number of vertices. This seems to conveniently cancel the factor~$\frac{1}{n!}$ coming from the Taylor expansion. However, one has to be slightly more careful.

As an example, consider the first diagram in equation~\eqref{eq:feynman_dens2}. This graph actually only appears once in the Taylor expansion, because there is only one way of connecting all interaction vertices to the density vertex. As a consequence, the cardinality of the corresponding equivalence class is less than~$n!$ in such cases. However, this can precisely be compensated for by a minor change in the Feynman rules. Namely, we demand that all vertices are time-ordered, i.e., the time coordinates should satisfy~$t_i \le t_1 \le t_2 \le \cdots \le t_n \le t$ for diagrams of order~$n$. This can easily be accomplished by adjusting the integral boundaries in the rule for the interaction vertex. The combinatorial factor from the equivalence of diagrams is exactly the same as the one from the possible time-orderings.

In summary, we obtain observables through the sum of all different time-ordered diagrams of the respective order. No symmetry factors are necessary. Note that the time ordering for vertices connected via edges is redundantly encoded by the propagator~$g_{qp}(t',t)$. Hence, we can omit the Heaviside function in~$g_{qp}(t',t)$ in the following without altering the result which can be advantageous when numerically integrating the expressions -- instead of a discontinuous function over a large domain~$t_a \in [t_i,t]$, we now integrate a continuous function over a smaller domain~$t_a \in [t_i,t_{a+1}]$.

There is a second issue which suggests another slight change to our Feynman rules. Namely, the discrete nature of the system yields shot noise terms which do not automatically vanish once we integrate over a probability distribution of initial conditions. Instead, they would require to explicitly perform the thermodynamic limit~$N \ra \infty$. This is discussed in detail in~\cite[section~5.2]{L3}. It is convenient to remove these ultimately irrelevant terms right away by demanding that density correlation functions are taken only for different particles. Hence, instead of equation~\eqref{eq:NBody_rpointDensDef}, we should use
\begin{align}
	G_{\rho \cdots \rho}(k_1, k_2,\ldots,k_r;\b{{\varphi}}^{(q)}(t)) = \sum_{\substack{\{j_1,j_2,\ldots,j_r\}\\j_s \neq j_t}} \exp\left( -i \, \sum_{s=1}^r k_s \cdot {\varphi}_{j_s}^{(q)}(t) \right) \,,
\end{align}
where the sum runs over all combinations of different indices. Note that the terms removed are precisely the shot noise terms found in~\cite{L3}.

\subsection{Particle Trajectories and Interpretation}

It is possible to obtain a very similar set of Feynman rules for the particle trajectories, i.e., the observable given by~$\mathcal{O}_j = \frac{\delta}{i \, \delta {J}_j^{(q)}(t)}$. Indeed, compared to the case of the density, the only thing to change is to replace the density vertex by the result of acting with the operator on a single factor~${J}_j^{(q)}$. It can easily be verified that the Feynman rules for the interaction vertex and the edges between them remain unchanged. The discussion of symmetry factors and time ordering remains unchanged, too. The density vertex is gone and instead all diagrams are trees ending in an outgoing edge
\begin{align}
	\label{rule:outgoing_edge}
	\begin{tikzpicture}[baseline={([yshift=-.5ex]current bounding box.center)}]
		\node[empty] (A) at (-2,0) {$a$};
		\node (B) at (0,0) {$j$,$t$};
		\draw[prop] (A) -- (B);
	\end{tikzpicture}\ &\ce - i \, g_{qp}(t,t_a') \, k_a' \, \delta_{j_aj} \,.
\end{align}
There is a special case in zeroth order, where we only have an outgoing edge. There (and only there) the operator~$\mathcal{O}_j$ can act on the free generating functional~$Z_0$ directly and hence yields the free trajectory~${\bar{\varphi}}_j^{(q)}(t;\b{x})$.

We thus obtain the contributions to the trajectory of the $j\nth$~particle up to third order as
\begin{align}
	\left({\tilde{\varphi}}_0\right)^{(q)}_j(t;\b{x}) &= {\bar{\varphi}}_j^{(q)}(t;\b{x}) \,,\\
	\label{eq:feynman_traj1st}
	\left({\tilde{\varphi}}_1\right)^{(q)}_j(t;\b{x}) &= \
	\begin{tikzpicture}[baseline={([yshift=-.5ex]current bounding box.center)}]
		\node[vertex] (I1) at (-1.5,0) {$1$};
		\node (T) at (0,0) {$j$,$t$};
		\draw[prop] (I1) -- (T);
	\end{tikzpicture}\\*
	&= - i \sum\limits_{l_1 \neq j} \int_{t_i}^t \diff t_1' \, g_{qp}(t,t_1') \int \frac{\diff^3 k_1'}{(2\pi)^3} \, m \, k_1' \, v(k_1',t_1') \, \exp\left( i \, k_1' \cdot \left( {\bar{\varphi}}_{j}^{(q)}(t_1';\b{x}) - {\bar{\varphi}}_{l_1}^{(q)}(t_1';\b{x}) \right) \right) \,, \\
	\left({\tilde{\varphi}}_2\right)_j^{(q)}(t;\b{x}) &= \
	\begin{tikzpicture}[baseline={([yshift=-.5ex]current bounding box.center)}]
		\node[vertex] (I1) at (-3.0,0) {$1$};
		\node[vertex] (I2) at (-1.5,0) {$2$};
		\node (T) at (0,0) {$j$,$t$};
		\draw[prop] (I1) -- (I2);
		\draw[prop] (I2) -- (T);
	\end{tikzpicture}\\*
	&= -i \sum\limits_{l_2 \neq j} \sum\limits_{j_1} \sum\limits_{l_1 \neq j_1} \int_{t_i}^t \diff t_2' \, g_{qp}(t,t_2') \int_{t_i}^{t_2'} \diff t_1' \, g_{qp}(t_2',t_1') \iint \frac{\diff^3 k_2'}{(2\pi)^3} \frac{\diff^3 k_1'}{(2\pi)^3} \, m^2 \nonumber \\*
	&\hspace{0.1\textwidth} k_2' \, v(k_2',t_2') \, \left( k_2'\cdot k_1' \right) \, v(k_1',t_1') \, \exp\left( i \, k_2' \cdot \left( {\bar{\varphi}}_{j}^{(q)}(t_2';\b{x}) - {\bar{\varphi}}_{l_2}^{(q)}(t_2';\b{x}) \right) \right) \nonumber\\*
	&\hspace{0.1\textwidth} \left( \delta_{j_1j} - \delta_{j_1l_2} \right) \, \exp\left( i \, k_1' \cdot \left( {\bar{\varphi}}_{j_1}^{(q)}(t_1';\b{x}) - {\bar{\varphi}}_{l_1}^{(q)}(t_1';\b{x}) \right) \right) \,, \\
	\left({\tilde{\varphi}}_3\right)^{(q)}_j(t;\b{x}) &= \
	\begin{tikzpicture}[baseline={([yshift=-.5ex]current bounding box.center)}]
		\node[vertex] (I1) at (-3.5,0.4) {$1$};
		\node[vertex] (I2) at (-3.0,-0.4) {$2$};
		\node[vertex] (I3) at (-1.5,0) {$3$};
		\node (T) at (0,0) {$j$,$t$};
		\draw[prop] (I1) -- (I3);
		\draw[prop] (I2) -- (I3);
		\draw[prop] (I3) -- (T);
	\end{tikzpicture}\ +\
	\begin{tikzpicture}[baseline={([yshift=-.5ex]current bounding box.center)}]
		\node[vertex] (I1) at (-4.5,0) {$1$};
		\node[vertex] (I2) at (-3.0,0) {$2$};
		\node[vertex] (I3) at (-1.5,0) {$3$};
		\node (T) at (0,0) {$j$,$t$};
		\draw[prop] (I1) -- (I2);
		\draw[prop] (I2) -- (I3);
		\draw[prop] (I3) -- (T);
	\end{tikzpicture} \ \,.
\end{align}
Again we only gave the diagrams for third order as the expressions become rather lengthy. The expressions up to second order are more transparent once we perform the integrations over~$k_a'$. Indeed, the first order trajectory obtained from summing the contributions of zeroth and first order,
\begin{align}
	\left({\tilde{\varphi}}_0\right)^{(q)}_j(t;\b{x}) &= {\bar{\varphi}}_j^{(q)}(t;\b{x}) \qquad \text{and} \\*
	\left({\tilde{\varphi}}_1\right)^{(q)}_j(t;\b{x}) &= - \sum\limits_{l_1 \neq j} \int_{t_i}^t \diff t_1' \, g_{qp}(t,t_1') \, m \, \nabla v \left( {\bar{\varphi}}_{j}^{(q)}(t_1';\b{x}) - {\bar{\varphi}}_{l_1}^{(q)}(t_1';\b{x}), t_1' \right) \,,
\end{align}
respectively, is just the Born approximation. This is in complete agreement to our considerations in the general case and this expression is a special case of equation~\eqref{eq:theory_phi1}. For the second order contribution we can perform the integrations, too, and arrive at
\begin{align}
	\left({\tilde{\varphi}}_2\right)^{(q)}_j(t;\b{x}) &= \sum\limits_{l_2 \neq j} \sum\limits_{j_1} \sum\limits_{l_1 \neq j_1} \int_{t_i}^t \diff t_2' \, g_{qp}(t,t_2') \int_{t_i}^{t_2'} \diff t_1' \, g_{qp}(t_2',t_1') \, m^2 \, \left( \delta_{j_1j} - \delta_{j_1l_2} \right) \nonumber\\*
	&\hspace{0.1\textwidth} H[v]\left( {\bar{\varphi}}_{j}^{(q)}(t_2';\b{x}) - {\bar{\varphi}}_{l_2}^{(q)}(t_2';\b{x}), t_2' \right) \, \nabla v\left( {\bar{\varphi}}_{j_1}^{(q)}(t_1';\b{x}) - {\bar{\varphi}}_{l_1}^{(q)}(t_1';\b{x}), t_1' \right) \,.
\end{align}
Here, $H[v]$~is the Hessian matrix of the two particle interaction potential~$v$. Again, we reproduce one of our general expressions, namely equation~\eqref{eq:theory_phi2}, specialized to an interacting $N$-body system.

For the interpretation of these expressions and in particular the associated diagrams, consider first the outgoing edge~\eqref{rule:outgoing_edge}. Here, we want to extract the trajectory of the $j\nth$~particle which is accomplished by propagating the particle~$j_a$ from the interaction vertex~$a$ at time~$t_a'$ to the final time~$t$. The fact that we use~$\delta_{jj_a}$ for this, suggests to take the particle with index~$j_a$ as the outgoing particle from the interaction vertex. Then, the form of the expression for the interaction vertex~\eqref{rule:int_vertex} becomes transparent: We consider the interaction of particle~$j_a$ with all other particles~$l_a$ at possible times~$t_a$. In order to calculate this interaction, we consider the force as evaluated along the free trajectories of the particles. Consequently, in first order we obtain the Born approximation.

Matters become more interesting once we have two interaction vertices connected together by an edge~\eqref{rule:internal_edge}. Now, the outgoing particle~$j_a$ from the left-hand interaction vertex~$a$ is identified with one of the particles taking part in the right-hand interaction vertex~$b$. More precisely, the factor~$\left( \delta_{j_aj_b} - \delta_{j_al_b} \right)$ yields two terms. The first corresponds to the case where the outgoing particle~$j_b$ is acted upon by the other particles twice at times~$t_a$ and~$t_b$. The second term takes into account the possibility that one of the particles~$l_b$ taking place in the interaction of~$j_b$ at time~$t_b$ had been acted upon already at time~$t_a$ by all other particles~$l_a$. In this case right-hand interaction vertex~$b$ determines the forces acting on particle~$j_b$ at time~$t_b$ by adding the contributions of all particles~$l_b$ as if all but one of them would be moving on a free trajectory, while this one particle~$j_a$ is moving on a Born trajectory.\footnote{\label{foot:feynman_Born}Continuing the comparison with the iterative Born approximation from footnote~\ref{foot:theory_Born}: In the second order iterative Born approximation we would have that all other particles move on (first order) Born trajectories, not just the particle~$j_a$. Again we can see that KFT is adding in these contributions slower -- order by order more particles on Born trajectories are considered for the calculation of the force acting on the outgoing particle. Diagrammatically, the second order iterative Born approximation corrects the first order trajectory by the sum over all diagrams of the form
\begin{align*}
	\begin{tikzpicture}[baseline={([yshift=-.5ex]current bounding box.center)}]
		\node[vertex] (I1) at (-2.75,1)   {$1$};
		\node[vertex] (I2) at (-2.25,0.2)   {$2$};
		\node[vertex] (I3) at (-1.5,-0.8)   {$a$};
		\node[vertex] (V) at (0,0)   {$b$};
		\node (T) at (2,0) {$j$,$t$};
		\draw[prop] (I1) -- (V);
		\draw[prop] (I2) -- (V);
		\draw[prop] (I3) -- (V);
		\draw[prop] (V) -- (T);
		\draw[prop,dotted,shorten <= 1ex,shorten >= 1ex] (I2) -- (I3);
	\end{tikzpicture}
\end{align*}
for~$a \in \N$. Conversely, KFT only adds the diagram with~$a = 1$. The remaining diagrams appear at arbitrarily high order in the KFT perturbation series. Higher-order iterative Born trajectories can likewise be expressed by infinite sums of KFT diagrams.}

While this is quite an instructive interpretation, it is slightly oversimplified. Indeed, all diagrams only give corrections to the free trajectory. Hence, interaction vertices give corrections to free particle trajectories -- such that, e.g., in the case of the interaction vertex~$b$ what we actually calculate is the difference between the force as given by all particles~$l_b$ moving freely and the force given by one of the particles moving on a Born trajectory. Only upon adding in the diagrams of lower orders we are in the situation described above.

The interpretation of diagrams involving density vertices is similar. The key difference is that the density vertex collects the density contributions of all incoming particles similarly to an interaction vertex. Depending on the precise form of the diagram, there might be one or multiple such incoming particles and they might have different orders (in the sense that the subdiagrams have a different number of interaction vertices). As such, the $n\nth$~order density calculated in KFT perturbation theory is not the density of a system of particles moving on $n\nth$~order KFT trajectories as one would naively expect. Instead, contributions are added more gradually.

\subsection{Density Correlation Functions and Power Spectra}

The sections above contain all necessary ingredients to derive expressions for density $r$-point correlation functions. However, since these are the key observables in the cosmological applications we are aiming at, we want to discuss them explicitly here. As stated above, the diagrams to consider for the $r$-point correlation function in $n\nth$~order KFT perturbation theory are the ones with $n$~internal vertices and $r$~density vertices. In accordance with our discussion in section~\ref{sec:feynman_symmetryFactors} we avoid symmetry factors and shot-noise terms by enforcing a time ordering~$t_a \le t_{a+1} \all a$ and remove tuples with common indices from the sums in the density vertices.

Then, the contributions to the density $2$-point correlation functions up to second order are given by
\begin{align}
	\left(\tilde{G}_{\rho\rho}\right)_0(k_1,k_2,t;\b{x}) &=\ 
	\begin{tikzpicture}[baseline={([yshift=-.5ex]current bounding box.center)}]
		\node[outdens] (D1) at (0,0) {$k_1$, $t$};
		\node[outdens] (D2) at (0,-0.8) {$k_2$, $t$};
	\end{tikzpicture}\
	= \sum_{m_1} \sum_{m_2 \neq m_1}  \exp\left( -i \, \left( k_1 \cdot {\bar{\varphi}}_{m_1}^{(q)}(t;\b{x}) + k_2 \cdot {\bar{\varphi}}_{m_2}^{(q)}(t;\b{x}) \right) \right) \,,\\[1ex]
	\left(\tilde{G}_{\rho\rho}\right)_1(k_1,k_2,t;\b{x}) &=\ 
	\begin{tikzpicture}[baseline={([yshift=-.5ex]current bounding box.center)}]
		\node[vertex] (I1) at (-1.5,0) {$1$};
		\node[outdens] (D1) at (0,0) {$k_1$, $t$};
		\node[outdens] (D2) at (0,-0.8) {$k_2$, $t$};
		\draw[prop] (I1) -- (D1);
	\end{tikzpicture} \ + \
	\begin{tikzpicture}[baseline={([yshift=-.5ex]current bounding box.center)}]
		\node[vertex] (I1) at (-1.5,-0.8) {$1$};
		\node[outdens] (D1) at (0,0) {$k_1$, $t$};
		\node[outdens] (D2) at (0,-0.8) {$k_2$, $t$};
		\draw[prop] (I1) -- (D2);
	\end{tikzpicture}\ \,,\\[1ex]
	\left(\tilde{G}_{\rho\rho}\right)_2(k_1,k_2,t;\b{x}) &=\ 
	\begin{tikzpicture}[baseline={([yshift=-.5ex]current bounding box.center)}]
		\node[vertex] (I1) at (-2,0.2) {$1$};
		\node[vertex] (I2) at (-1.5,-0.6) {$2$};
		\node[outdens] (D1) at (0,0) {$k_1$, $t$};
		\node[outdens] (D2) at (0,-0.8) {$k_2$, $t$};
		\draw[prop] (I1) -- (D1);
		\draw[prop] (I2) -- (D1);
	\end{tikzpicture} \ + \
	\begin{tikzpicture}[baseline={([yshift=-.5ex]current bounding box.center)}]
		\node[vertex] (I1) at (-3,0) {$1$};
		\node[vertex] (I2) at (-1.5,0) {$2$};
		\node[outdens] (D1) at (0,0) {$k_1$, $t$};
		\node[outdens] (D2) at (0,-0.8) {$k_2$, $t$};
		\draw[prop] (I1) -- (I2);
		\draw[prop] (I2) -- (D1);
	\end{tikzpicture} \ + \nonumber\\*[2ex]
	&\qquad\begin{tikzpicture}[baseline={([yshift=-.5ex]current bounding box.center)}]
		\node[vertex] (I1) at (-2,0) {$1$};
		\node[vertex] (I2) at (-1.5,-0.8) {$2$};
		\node[outdens] (D1) at (0,0) {$k_1$, $t$};
		\node[outdens] (D2) at (0,-0.8) {$k_2$, $t$};
		\draw[prop] (I1) -- (D1);
		\draw[prop] (I2) -- (D2);
	\end{tikzpicture} \ + \
	\begin{tikzpicture}[baseline={([yshift=-.5ex]current bounding box.center)}]
		\node[vertex] (I1) at (-1.5,0) {$2$};
		\node[vertex] (I2) at (-2,-0.8) {$1$};
		\node[outdens] (D1) at (0,0) {$k_1$, $t$};
		\node[outdens] (D2) at (0,-0.8) {$k_2$, $t$};
		\draw[prop] (I1) -- (D1);
		\draw[prop] (I2) -- (D2);
	\end{tikzpicture} \ + \nonumber\\*[2ex]
	&\qquad\begin{tikzpicture}[baseline={([yshift=-.5ex]current bounding box.center)}]
		\node[vertex] (I1) at (-2,-0.2) {$1$};
		\node[vertex] (I2) at (-1.5,-1) {$2$};
		\node[outdens] (D1) at (0,0) {$k_1$, $t$};
		\node[outdens] (D2) at (0,-0.8) {$k_2$, $t$};
		\draw[prop] (I1) -- (D2);
		\draw[prop] (I2) -- (D2);
	\end{tikzpicture} \ + \
	\begin{tikzpicture}[baseline={([yshift=-.5ex]current bounding box.center)}]
		\node[vertex] (I1) at (-3,-0.8) {$1$};
		\node[vertex] (I2) at (-1.5,-0.8) {$2$};
		\node[outdens] (D1) at (0,0) {$k_1$, $t$};
		\node[outdens] (D2) at (0,-0.8) {$k_2$, $t$};
		\draw[prop] (I1) -- (I2);
		\draw[prop] (I2) -- (D2);
	\end{tikzpicture} \,.
\end{align}
Let us write down the mathematical expression encoded by one of the diagrams to provide a further example of the usage of our Feynman rules. It is
\begin{align}
	\begin{tikzpicture}[baseline={([yshift=-.5ex]current bounding box.center)}]
		\node[vertex] (I1) at (-1.5,0) {$1$};
		\node[outdens] (D1) at (0,0) {$k_1$, $t$};
		\node[outdens] (D2) at (0,-0.8) {$k_2$, $t$};
		\draw[prop] (I1) -- (D1);
	\end{tikzpicture} &= \sum_{m_1} \sum_{m_2 \neq m_1} \exp\left( -i \, \left( k_1 \cdot {\bar{\varphi}}_{m_1}^{(q)}(t;\b{x}) + k_2 \cdot {\bar{\varphi}}_{m_2}^{(q)}(t;\b{x}) \right) \right) \sum\limits_{l_1 \neq m_1} \int_{t_i}^t \diff t_1' \, g_{qp}(t,t_1') \nonumber \\*
	&\hspace{0.1\textwidth} \int \frac{\diff^3 k_1'}{(2\pi)^3} \, m \, v(k_1',t_1') \, k_1' \cdot k_1 \, \exp\left( i \, k_1' \cdot \left( {\bar{\varphi}}_{m_1}^{(q)}(t_1';\b{x}) - {\bar{\varphi}}_{l_1}^{(q)}(t_1';\b{x}) \right) \right) \,.
\end{align}

For our applications we are specifically interested in the density fluctuation power spectrum~$\tilde{P}_\delta$. Given a probability distribution of initial conditions~$P(\b{x})$ which is spatially homogeneous, the power spectrum is defined implicitly via
\begin{align}
	\label{eq:feynman_PowerSpectrumDef}
	\int \diff^{6N} \b{x} \, P(\b{x}) \, \tilde{G}_{\rho\rho}(k_1,k_2,t;\b{x}) = \left( 2 \pi \right)^6 \, \bar{\rho}^2 \, \delta_D\left(k_1\right) \, \delta_D\left(k_2\right) + \left( 2 \pi \right)^3 \, \bar{\rho}^2 \, \delta_D\left(k_1 + k_2\right) \, \tilde{P}_\delta(k_1,t) \,,
\end{align}
where $\bar{\rho}$~is the mean density. As evident from this equation, the power spectrum fully describes the density two-point correlation function. If the distribution~$P(\b{x})$ is also statistically isotropic, the power spectrum~$\tilde{P}_\delta(k_1,t)$ only depends on the norm~$\norm{k_1}$. We discuss the process of integrating observables over initial conditions in the following section.

For $3$-point functions we would consider three density vertices and hence obtain three diagrams in first and twelve diagrams in second order. In the corresponding expressions, the sums over indices~$m_1, m_2, m_3$ would be such that~$m_2 \neq m_1$ and~$m_3 \neq m_1,m_2$. For $4$-point functions the diagrams become even more numerous, particularly at higher-orders. As such, the manual usage of the Feynman rules becomes quite tedious rather quickly. Therefore we automated the generation of diagrams and the transcription into mathematical expressions as well as their numerical evaluation.


\section{Integration over Initial Conditions}

The KFT formalism and the associated perturbation theory as presented in the previous sections can be used to study classical Hamiltonian systems with explicit initial conditions~$\b{x} \in \b{X}$. However, in applications we rarely have access to, nor are interested in the specific initial conditions of the system. Rather, we often prefer to deduce statistical properties of a system. KFT allows for this generalization in a natural way. Indeed, instead of having our generating function depend explicitly on initial phase space coordinates~$\b{x}$, we can integrate it over a probability distribution~$P(\b{x})$ of them. We define the averaged generating functional
\begin{align}
	\label{eq:ini_averagedZ}
	\langle Z[\b{J}] \rangle \ce \int_{\b{X}} \diff \b{x} \, Z[\b{J}; \b{x}] \, P(\b{x}) \,.
\end{align}
It has been common practice to use this object as the starting point in KFT~\cite{L3,bartelmann2019cosmic,bartelmann2020kinetic}. If one performs the integration in equation~\eqref{eq:ini_averagedZ}, the observables obtained via functional derivatives with respect to~$\b{J}$ are averaged over initial conditions, i.e.,
\begin{align}
	\langle \tilde{\mathcal{O}} \rangle(t) \ce \int_{\b{X}} \diff \b{x} \, \tilde{\mathcal{O}}(t;\b{x}) \, P(\b{x}) = \mathcal{O}\left(\frac{\delta}{i \, \delta \b{J}(t)}\right) \, \restr{\langle Z[\b{J}] \rangle}{\b{J} = \b{0}} \,.
\end{align}

Without doubt it is convenient to have a single object~$\langle Z[\b{J}] \rangle$ describing the averaged system and being able to deduce observables from it. However, it is quite difficult to perform the integration in equation~\eqref{eq:ini_averagedZ} for a non-trivial probability distribution~$P(\b{x})$ -- after all, the generating functional~$Z[\b{J}; \b{x}]$ contains all information of our system and hence is the most difficult object of the theory. In comparison, it is significantly easier to perform this integration over an observable obtained via the perturbative approach presented above.

Mathematically, of course, we obtain the same result regardless of whether we perform the integration or the functional derivatives first. More precisely, if we have an expression for the $n\nth$~order generating functional~$Z_n[\b{J};\b{x}]$ and intend to find the $n\nth$~order expectation value of an observable~$\mathcal{O}$, we can use either
\begin{align}
	\label{eq:ini_DoingKFTBad}
	\langle \tilde{\mathcal{O}}_n \rangle (t) &= \restr{\mathcal{O}\left(\frac{\delta}{i \, \delta \b{J}(t)}\right) \langle Z_n[\b{J}] \rangle}{\b{J} = \b{0}} = \restr{\mathcal{O}\left(\frac{\delta}{i \, \delta \b{J}(t)}\right) \left( \int_{\b{X}} \diff \b{x} \, Z[\b{J}; \b{x}] \, P(\b{x}) \right)}{\b{J} = \b{0}} \quad \text{or} \\
	\label{eq:ini_DoingKFTGood}
	\langle \tilde{\mathcal{O}}_n \rangle (t) &= \langle \tilde{\mathcal{O}}_n(t;\b{x}) \rangle = \int_{\b{X}} \diff \b{x} \, \left( \restr{\mathcal{O}\left(\frac{\delta}{i \, \delta \b{J}(t)}\right) Z_n[\b{J};\b{x}]}{\b{J} = \b{0}} \right) \, P(\b{x}) \,.
\end{align}
We opt for the latter option here which in our opinion is advantageous for the perturbative treatment. In this section we apply this strategy to increasingly complicated physical systems.

\subsection{Toy Model: Gaussian Distribution}
\label{sec:ini_toyGauss}

In describing physical systems with KFT there are two independent aspects of complexity: Firstly, if we have a complicated Hamiltonian with non-trivial interactions, obtaining perturbative expressions for observables may be laborious. Secondly, if we have complicated initial conditions, the final integration in equation~\eqref{eq:ini_DoingKFTGood} may become rather challenging. While the first aspect generally is less of a problem in itself (the Feynman rules only require derivatives), the more complicated the pertrubative expressions become, the more difficult is the integration over initial conditions. Hence, even if we have a relatively simple probability distribution~$P(\b{x})$, it might be infeasible to perform this integration. The second aspect is more direct: If~$P(\b{x})$ is a difficult distribution, averaging any non-trivial function over it requires some effort.

In order to get used to the procedure and to introduce some calculational techniques, we consider a series of toy models. The physical system we want to investigate is an ensemble of particles of mass~$m$ moving in one dimension and interacting via a two-particle interaction potential~$v(k)$ without time-dependence. We can regard this as a one-dimensional interacting $N$-body system on a static background and use the one-dimensional analogues of the expressions derived above. For the initial values at time~$t_i = 0$ let us use the probability distribution
\begin{align}
	\label{eq:ini_toyModelP}
	P(\b{x}) = \prod_{j=1}^N P(x_j) \qquad \text{with} \qquad P(x_j) = \frac{1}{2\pi\sigma\xi}\exp\left( - \frac{\left({x}_j^{(q)} - \mu \right)^2}{2 \sigma^2}\right) \exp\left( - \frac{\left({x}_j^{(p)} - \lambda \right)^2}{2 \xi^2}\right) \,,
\end{align}
i.e., a Gaussian distribution both in position and momentum. An elementary calculation shows that the expectation value for the zeroth order KFT trajectory for any particle~$j$ reproduces the center of mass trajectory
\begin{align}
	\left\langle \left({\tilde{\varphi}}_0\right)^{(q)}_j \right\rangle (t) = \int \diff \b{x} \, {\bar{\varphi}}_j^{(q)}(t;\b{x}) \, P(\b{x}) = \int \diff x_j \, \left( {x}_j^{(q)} + g_{qp}(t,0) \, {x}_j^{(p)} \right) \, P(x_j) = \mu + g_{qp}(t,0) \, \lambda
\end{align}
as it should. None of higher-order perturbative contributions to the trajectory alter that result. We show this explicitly for the diagram coming from first order KFT perturbation theory, taken from equation~\eqref{eq:feynman_traj1st}:
\begin{align}
	\left\langle \ \begin{tikzpicture}[baseline={([yshift=-.5ex]current bounding box.center)}]
		\node[vertex] (I1) at (-1.5,0) {$1$};
		\node (T) at (0,0) {$j$,$t$};
		\draw[prop] (I1) -- (T);
	\end{tikzpicture} \right\rangle(t)
	&= - i \int \diff \b{x} \, P(\b{x}) \sum\limits_{l_1 \neq j} \int_0^t \diff t_1' \, g_{qp}(t,t_1') \int \frac{\diff k_1'}{2\pi} \, m \nonumber\\*
	&\hspace{0.1\textwidth} k_1 \, v(k_1') \, \exp\left( i \, k_1' \cdot \left( {\bar{\varphi}}_{j}^{(q)}(t_1';\b{x}) - {\bar{\varphi}}_{l_1}^{(q)}(t_1';\b{x}) \right) \right) \\
	\label{eq:feynman_avg1stTrajStep1}
	&= - i \sum\limits_{l_1 \neq j} \int_0^t \diff t_1' \, g_{qp}(t,t_1') \int \frac{\diff k_1'}{2\pi} \, m \iint \diff x_j \, \diff x_{l_1} \, P(x_j) \, P(x_{l_1})  \nonumber\\*
	&\hspace{0.1\textwidth}  k_1 \, v(k_1') \, \exp\left( i \, k_1' \cdot \left( {\bar{\varphi}}_{j}^{(q)}(t_1';\b{x}) - {\bar{\varphi}}_{l_1}^{(q)}(t_1';\b{x}) \right) \right).
\end{align}
In the last line we already performed the integration over all components of~$\b{x}$ except~$x_j$ and~$x_{l_1}$. This is one key simplification of performing the integration of perturbative observables compared to integrating the generating functional directly -- the latter necessarily contains all particle indices which prevents this step. Inserting the expression of the free trajectory for~${\bar{\varphi}}_{j}^{(q)}(t_1';\b{x})$ and~${\bar{\varphi}}_{l_1}^{(q)}(t_1';\b{x})$ in equation~\eqref{eq:feynman_avg1stTrajStep1} we obtain a term
\begin{align}
	\int \diff x_j \, P(x_j) \, \exp\left( i \, k_1' \cdot {x}_j^{(q)} \right) \, \exp\left( i \, g_{qp}(t_1',0) \, k_1' \cdot {x}_j^{(p)} \right),
\end{align}
and a similar factor with indices~$l_1$. This clearly yields Fourier transforms of the position and momentum distributions. In the present case, these Fourier transforms are again Gaussians multiplied with some conveniently cancelling phase factors and therefore their product is an even function. Together with the tacit assumption that the interaction potential~$v$ is symmetric, the integration over~$k_1$ yields zero.

Above we have encountered a generic feature of the averaging process for perturbative observables. Because the free trajectory and hence the initial values only appear inside Fourier phase factors, the integration over~$\b{x}$ always yields Fourier transforms of the initial probability distribution~$P(\b{x})$. In probability theory, such Fourier transforms of a probability distribution~$P(x)$ of a random variable~$x$ is known by the name of \emph{characteristic function} and usually denoted by~$\Phi_x$. It can be defined in terms of expectation values via
\begin{align}
	\Phi_x(f) = \left\langle \exp\left(i\, f \cdot x \right) \right\rangle = \int \diff x \, \exp\left(i\, f \cdot x \right) \, P(x) \,,
\end{align}
where we point out the unusual sign convention for the Fourier transform. We use the letter~$f$ for the conjugate variable for~$x$ instead of the in this context usually used letter~$t$ to avoid confusion with time coordinates.

In terms of characteristic functions, the expression in equation~\eqref{eq:feynman_avg1stTrajStep1} can be compactly written as
\begin{align}
	\left\langle \ \begin{tikzpicture}[baseline={([yshift=-.5ex]current bounding box.center)}]
		\node[vertex] (I1) at (-1.5,0) {$1$};
		\node (T) at (0,0) {$j$,$t$};
		\draw[prop] (I1) -- (T);
	\end{tikzpicture} \right\rangle(t)
	&= - i \sum\limits_{l_1 \neq j} \int_0^t \diff t_1' \, g_{qp}(t,t_1') \int \frac{\diff k_1'}{2\pi} \, m \, k_1 \, v(k_1') \, \Phi_{x_j}(f) \, \Phi_{x_{l_1}}(-f) \,,
\end{align}
where we abbreviated~$f \ce \left( k_1', g_{qp}(t_1',0) \, k_1' \right)^\tp$. Given that in our toy model the distributions for position and momentum are uncorrelated, the characteristic functions factorize into characteristic functions for position and momentum, respectively. In our applications this is generally not possible and we need to work with the characteristic functions of~$x$ or even~$\b{x}$.

As a more quantitative example, we consider the same Gaussian initial distributions, but determine the expectation value of the density. In zeroth order perturbation theory, the expectation value of the density is simply
\begin{align}
	\label{eq:ini_toyModelrho0}
	\left\langle \tilde{\rho}_0 \right\rangle(k,t) &= \left\langle \ \begin{tikzpicture}[baseline={([yshift=-.5ex]current bounding box.center)}]
		\node[outdens] (D) at (0,0) {$k$,$t$};
	\end{tikzpicture} \  \right\rangle =
	\sum_{j} \int \diff x_j \, P(x_j) \, \exp\left( - i \, k \cdot \left( {x}_j^{(q)} + g_{qp}(t,0) \, {x}_j^{(p)} \right) \right) \,.
\end{align}
Let us immediately perform a Fourier transform to obtain the density as a function of position~$q$:
\begin{align}
	\left\langle \tilde{\rho}_0 \right\rangle(q,t) &= \int \frac{\diff k}{2\pi} \, \left\langle \tilde{\rho}_0 \right\rangle(k,t) \, \exp(i \, k \cdot q)
	= \sum_j \int \diff {x}_j^{(p)} P\left( q - g_{qp}(t,0) \, {x}_j^{(p)}, {x}_j^{(p)} \right) \,.
\end{align}
Using the definition of the inital probability distribution from equation~\eqref{eq:ini_toyModelP} we obtain
\begin{align}
	\left\langle \tilde{\rho}_0 \right\rangle(q,t) &= \frac{N}{\sqrt{2\pi \left(\sigma^2 + g_{qp}(t,0) \, \xi^2\right)}} \exp\left(- \frac{\left(q - \left(\mu + g_{qp}(t,0) \, \lambda\right)\right)^2}{2 \left(\sigma^2 + g_{qp}(t,0) \, \xi^2\right)}\right) \,,
\end{align}
i.e., a Gaussian with mean given by the center of mass position~$\left\langle \left({\tilde{\varphi}}_0^{(q)}\right)_j \right\rangle (t)$ and variance~$\sigma^2 + g_{qp}(t,0) \, \xi^2$ which is monotonically increasing (for standard choices for the propagator $g_{qp}$). While this is the expected result, the free density could have been obtained directly without using KFT perturbation theory.

\begin{figure}[ht]
    \centering
    \includegraphics[width=0.8\textwidth]{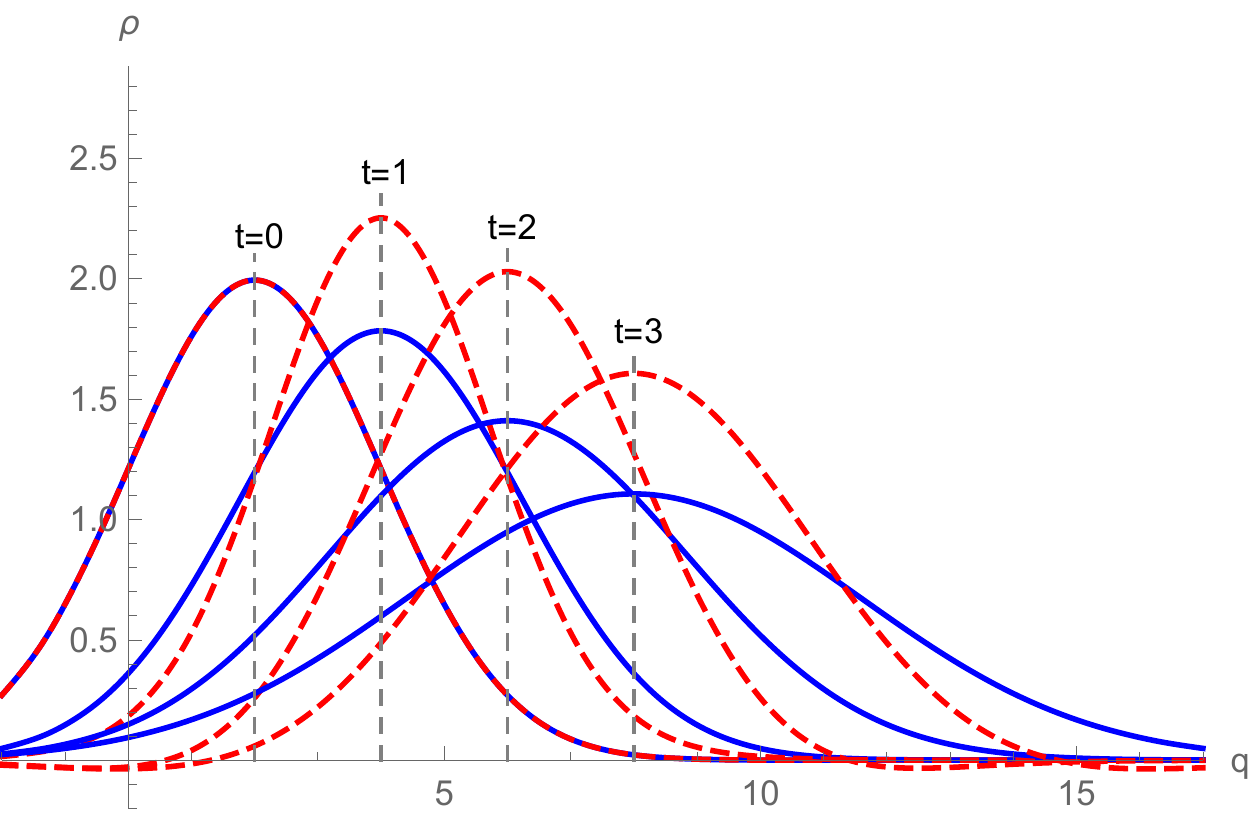} 
    \caption{This figure shows the density evolution for zeroth and first order perturbation theory in blue and red, respectively. The initial spatial distribution at $t=0$ has mean value~$\mu = 2$ and standard deviation~$\sigma = 2$ (leftmost Gaussian). The momentum distribution has mean value~$\lambda = 2$ and standard deviation~$\xi = 1$, such that in every unit of time the density distribution moves to the right by two units and broadens. This precisely is the behaviour in zeroth order perturbation theory (blue). The first order correction counteracts the diffusion due to the encoded attractive interaction (red).~\cite{Mathematica}}
    \label{fig:gaussian_toy_model}
\end{figure}

For an actual test of the predictions of KFT perturbation theory, let us consider the first order contribution to the density given by
\begin{align}
	\left\langle \tilde{\rho}_1 \right\rangle(k,t) &= \left\langle\ \begin{tikzpicture}[baseline={([yshift=-.5ex]current bounding box.center)}]
		\node[vertex] (I1) at (-1.5,0) {$1$};
		\node[outdens] (D) at (0,0) {$k$,$t$};
		\draw[prop] (I1) -- (D);
	\end{tikzpicture}\ \right\rangle \,.
\end{align}
The expression encoded in the diagram can again be simplified and integrated mostly analytically. For simplicity and transparency, we assume a static background, for which the propagator takes the simple form~$g_{qp}(t,t') = \frac{1}{m} (t-t') \, \theta(t-t')$ as derived in equation~\eqref{eq:Nbody_GreensFunction}. Then, the resulting evolution of the density for zeroth and first order is shown in figure~\ref{fig:gaussian_toy_model}. Clearly, attractive interaction encoded by the first order correction counteracts the diffusion. Going to higher-orders, this behaviour is enhanced and prolonged.\footnote{Since the interaction is evaluated along free trajectories, at no finite order a bound system can be modelled, i.e., diffusion always eventually wins in this example. With respect to our intended application to cosmic structure formation, this means that at any finite order of perturbation theory we are unable to accurately describe objects below a certain size. Specifically, we expect that for any finite order the resulting non-linear density fluctuation power spectrum falls below the observed one above a certain wave number.}

Note that not all physical properties of observables are preserved at finite interaction orders. In the specific example, there can be regions where the density is negative in the first order approximation. This is to be expected because the first order correction needs to be negative in some regions in order not to change the normalization. By increasing the interaction strength, the negative terms can then dominate over the positive zeroth order contribution. However, higher-order terms correct for this and restore positivity of the density.

\subsection{Toy Model: Shell-Crossing}

In the previous toy model, the distributions for position and momentum were completely independent. Here we study an example where the momenta and positions of the particles are correlated in the sense that the momentum assigned to particles depends on their position. This is a preparation for the upcoming section on correlated initial conditions which generalizes this. However, the toy model calculations are also of interest in their own right because they show that KFT is not affected by the notorious shell-crossing problem one encounters in cosmic structure formation. Common hydrodynamic approaches break down when streams of dark matter particles cross due to the assumption of the existence of a (single-valued) velocity field. KFT considers the particle dynamics in phase space where this conceptual limitation is circumvented.

Let us consider an interacting $N$-body system in one dimension subject to the probability distribution
\begin{align}
	P(\b{x}) = \prod_{j=1}^N \frac{1}{\sqrt{2\pi} \, \sigma} \exp\left(- \frac{\left({{x}_j^{(q)}}\right)^2}{2\sigma^2} \right) \, \delta_D\left({x}_j^{(p)} + {x}_j^{(q)}\right)
\end{align}
of initial values~$\b{x}$ at time~$t_i = 0$. Evidently, the particle positions initially form a Gaussian distribution, but their momenta are assigned such that a particle at initial position~${x}_j^{(q)}$ has momentum~${x}_j^{(p)} = - {x}_j^{(q)}$. This linear relationship implies that there is a time~$t_c$, where simultaneously the free trajectory of each particle
\begin{align}
	{\bar{\varphi}}_j^{(q)}(t_c;\b{x}) = {x}_j^{(q)} + g_{qp}(t_c,0) \, {x}_j^{(p)} = \left( 1 - g_{qp}(t_c,0) \right) \, {x}_j^{(q)}
\end{align}
is zero. Hence, ignoring interactions, at time~$t_c$ all particles are at position~$q = 0$ as the streams of left-moving particles starting out at~${x}_j^{(q)} > 0$ and right-moving particles starting out at~${x}_j^{(q)} < 0$ cross. In this instance of time, the density is a $\delta_D$-function making a description using conventional hydrodynamic methods impossible.

\begin{figure}
    \centering
    \includegraphics[width=0.19\textwidth]{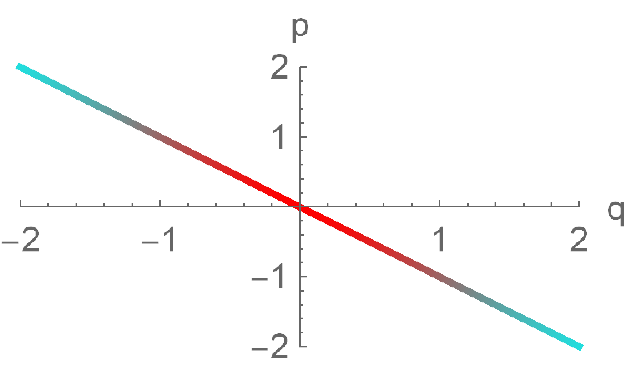}
    \includegraphics[width=0.19\textwidth]{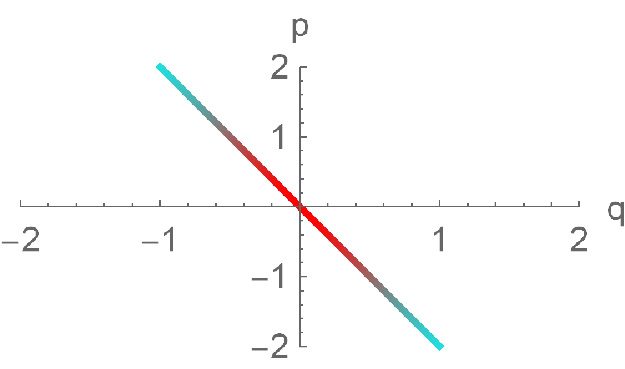}
    \includegraphics[width=0.19\textwidth]{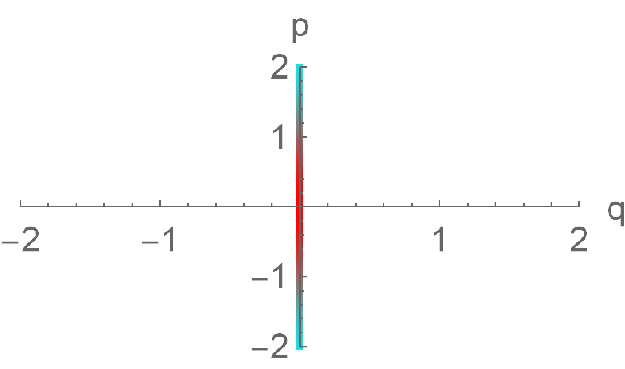}
    \includegraphics[width=0.19\textwidth]{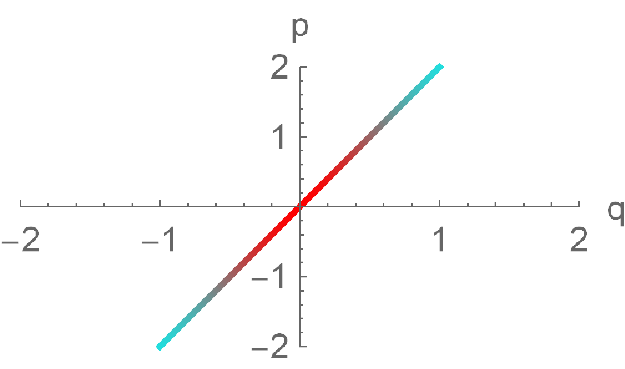}
    \includegraphics[width=0.19\textwidth]{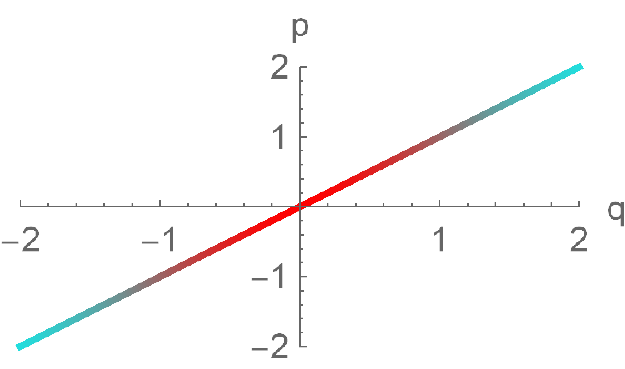}\\[6pt]
    \includegraphics[width=0.19\textwidth]{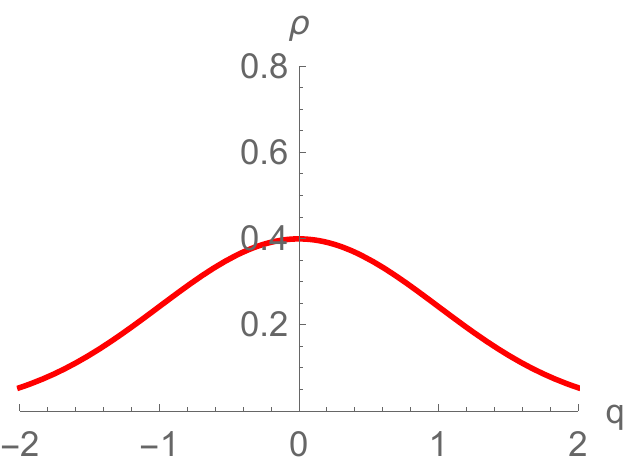}
    \includegraphics[width=0.19\textwidth]{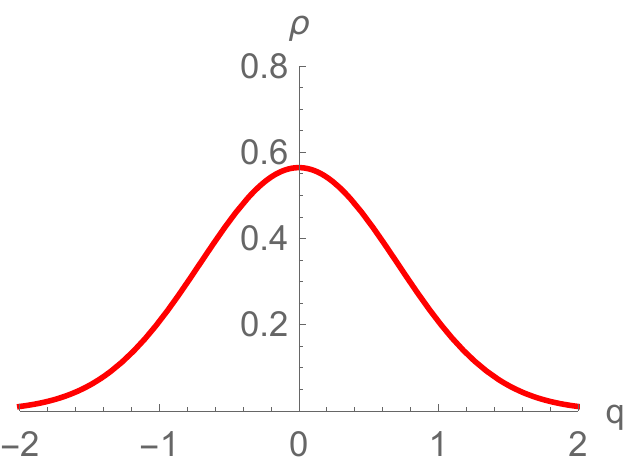}
    \includegraphics[width=0.19\textwidth]{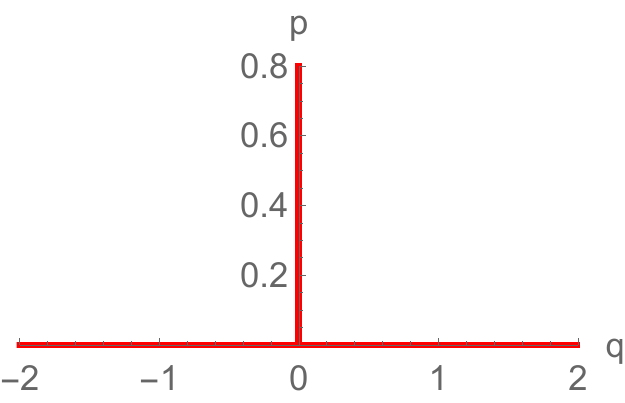}
    \includegraphics[width=0.19\textwidth]{plots/fig2-12.pdf}
    \includegraphics[width=0.19\textwidth]{plots/fig2-11.pdf}
    \caption{Evolution of the toy model system over time (left to right). The upper row shows the distribution in phase space (density indicated by colour), the lower row the corresponding spatial density function. The singular density function at $t_c$ (middle column) is a result of the spatial projection rather than an inherent singularity of the system.~\cite{Mathematica}}
    \label{fig:shell-crossing_toy_model}
\end{figure}

Starting from equation~\eqref{eq:ini_toyModelrho0}, an elementary calculation yields
\begin{align}
	\left\langle \tilde{\rho}_0 \right\rangle(q,t) 
	&= \frac{N}{\sqrt{2\pi} \, \abs{1 - g_{qp}(t,0)} \, \sigma} \exp\left(- \frac{q^2}{2\left( 1 - g_{qp}(t,0) \right)^2\sigma^2} \right)  \,.
\end{align}
This is a Gaussian distribution with variance~$\left( 1 - g_{qp}(t,0) \right) \, \sigma$ which transitions through a $\delta_D$-distribution as~$t \ra t_c$. Note that the density evolution for~$t > t_c$ is completely unaffected by the singularity at~$t_c$ in the KFT formalism. As can be seen in figure~\ref{fig:shell-crossing_toy_model} (again evaluated for static background), the singularity in the density is a result of the projection of the phase space distribution of the system onto the spatial axis and therefore does not constitute a problem when working in phase space.

\subsection{Correlated Initial Conditions}

As our next step, we want to consider a case of initial conditions where all particle positions and momenta are correlated with each other. More precisely, we sample a (statistically homogeneous and isotropic) density field and assign velocities according to an associated velocity field. This is no longer a mere toy model, but rather is the way in which standard cosmological $N$-body simulations set up their initial conditions~\cite{initialConditions}. Indeed, standard cosmological perturbation theory is applicable in the early stages of cosmic structure formation in good accuracy.\footnote{Note that we can set the initial conditions very early without major downsides (unlike, e.g., $N$-body simulations, where this increases computational cost and accumulated errors). Hence, some of the criticism of this method in~\cite{initialConditions} can be sidestepped.} Stopping the evolution prior to shell-crossing we obtain a density and a velocity field which we can then discretize and follow the particle dynamics as structure formation becomes non-linear. This way, we use the hydrodynamic picture in its range of validity and switch to a description via $N$-body dynamics as this becomes necessary. We follow Appendix~A of~\cite{L3} here, but adapt it somewhat.

Let us attempt to formalize the ideas of the previous section. We want that the initial positions of our particle ensemble sample a (number) density field~$\rho(q)$ at initial time~$t_i$. This can be encoded in the equation
\begin{align}
\label{eq:ini_probabilityOfXqpgivenRho}
    P\left( {x}_j^{(q)} \mid \rho \right) = \frac{1}{N} \, \rho\left( {x}_j^{(q)} \right) \,.
\end{align}
The meaning of this equation can be read off: Given the density field, the probability of finding a particle~$j$ at the position~${x}_j^{(q)}$ should be given by the density at that precise point. The prefactor~$\frac{1}{N}$ accounts for the fact that in a given volume~$V$ we deposit a total particle number~$N$. Having assigned a position to a particle, we want to equip it with a momentum. Again, we can write this down in a rather transparent equation
\begin{align}
\label{eq:ini_probabilityOfXpgivenV}
    P\left( {x}_j^{(p)} \mid v \right) = \delta_D \left( {x}_j^{(p)} - m \, v\left( {x}_j^{(q)} \right) \right) \,.
\end{align}
Here, $v(q)$ is the velocity field at the initial time~$t_i$. It can clearly be seen from this equation, that individually for each particle its momentum is correlated with its position.

So far, we have treated the density and velocity fields completely independently. However, in our case of cosmic structure formation, we can regard them as the result of a hydrodynamic description valid for times~$t < t_i$. Then, the equations of standard cosmic perturbation theory yield a relationship between the two. As a first step, we assert that the velocity field is irrotational. Even if the primordial perturbations have a vector component in the usual scalar-vector-tensor decomposition (which for standard inflationary scenarios they have not), it decays during cosmic expansion. Therefore there exists a velocity potential~$\psi$ such that the velocity field is its gradient, $v(q) = \nabla \psi(q)$.

Assuming that the density and velocity fields indeed arise via a hydrodynamic description, the are related via the equations of standard cosmic perturbation theory. The relevant equation is
\begin{align}
\label{eq:ini_densContrast}
    - \kappa \,\nabla^2 \psi(q) = \delta(q) \ce \frac{\rho(q) - \bar{\rho}}{\bar{\rho}}
\end{align}
which is the final result of the discussion in~\ref{sec:app_PoissonInitial}. Here, $\delta(q)$ is the density contrast, $\bar{\rho}$ the mean density and $\kappa$ a constant which depends on the choice of time variable. Using this equation as well as the definition of~$\psi$ as the velocity potential, we can replace the density and velocity fields in equations~\eqref{eq:ini_probabilityOfXqpgivenRho} and~\eqref{eq:ini_probabilityOfXpgivenV}. This way, we promote the velocity potential~$\psi$ to the fundamental object controlling the statistics of our initial particle distribution. Indeed, the initial positions and momenta of our particles are subject to the conditional probability distributions
\begin{align}
\label{eq:ini_probabilityOfXgivenPsi}
	P\left( {x}_j^{(q)} \mid \psi \right) = \frac{\bar{\rho}}{N} \left( 1 - \kappa \,\nabla^2 \psi\left( {x}_j^{(q)} \right) \right) \qquad \text{and} \qquad P\left( {x}_j^{(p)} \mid \psi \right) = \delta_D \left( {x}_j^{(p)} - m \,\nabla \psi\left( {x}_j^{(q)} \right) \right) \,.
\end{align}
Formally, in order to combine these conditional probabilities into a probability distribution~$P(\b{x})$, we have to average them over~$P(\psi)$. Written as an equation, it is
\begin{align}
\label{eq:ini_probx1}
	P(\b{x}) = \int \diff \psi \, \left( \prod_{j=1}^N P\left( {x}_j^{(q)} \mid \psi \right) \, P\left( {x}_j^{(p)} \mid \psi \right)  \right) \, P(\psi) \,,
\end{align}
where the integration runs over all possible smooth field configurations~$\psi$.

On the surface, the integration in equation~\eqref{eq:ini_probx1} seems rather difficult. However, our task is simplified substantially by the fact that the velocity potential~$\psi$ is a Gaussian random field if we choose our initial time~$t_i$ sufficiently early. A good choice for~$t_i$ is the release of the Cosmic Microwave Background (CMB) at redshift~$z \approx 1090$. This is well in the linear regime of cosmic structure formation and CMB measurements give us very precise statistics for the density contrast~$\delta(q)$. In particular, we know that the density contrast at time~$t_i$ is very well described by a Gaussian random field~\cite{Planck2018gaussianity}. Hence, the velocity potential~$\psi(q)$ likewise is a Gaussian random field and therefore its statistics are solely described by its power spectrum~$P_\psi(k) = \kappa^{-2} \norm{k}^{-4} P_\delta(k)$.

In addition to this, the conditional probabilities only depend on the derivatives of~$\psi$ at the particle positions~${x}_j^{(q)}$. Since~$\psi$ is a Gaussian random field, whenever we pick a set of positions~$\b{{x}^{(q)}} = \left( {x}_1^{(q)}, {x}_2^{(q)}, \ldots, {x}_N^{(q)} \right)^\tp$, the values~$\b{\psi} \ce \left( \psi\left( {x}_1^{(q)} \right), \psi\left( {x}_2^{(q)} \right), \ldots, \psi\left( {x}_N^{(q)} \right) \right)^\tp$ follow a multivariate Gaussian distribution. More precisely,
\begin{align}
	P\left( \b{\psi} \right) &= \frac{1}{\sqrt{(2\pi)^N \, \det\left( \b{C^{\psi\psi}}\left( \b{{x}}^{(q)} \right) \right)}} \exp\left( - \frac{1}{2} \, \b{\psi}^\tp \! \left(\b{C^{\psi\psi}}\right)^{-1}\left( \b{{x}}^{(q)} \right) \, \b{\psi} \right) \qquad \text{with} \\
	C^{\psi\psi}_{jl}\left( \b{{x}}^{(q)} \right) &= \int \frac{\diff^3 k''}{(2\pi)^3} P_\psi(k'') \, \exp\left( -i k'' \cdot \left( {x}_j^{(q)} - {x}_l^{(q)} \right) \right) \,.
\end{align}
It is a standard result that the statistics of the derivatives of a Gaussian random field are related to the statistics of the field itself in a simple manner. Indeed, it is
\begin{align}
\label{eq:ini_probPsi}
	P\left( \b{\nabla\psi}, \b{\nabla^2\psi} \right) &= \frac{1}{\sqrt{(2\pi)^{4N} \, \det\left( \b{C}\left( \b{{x}}^{(q)} \right) \right)}} \exp\left( - \frac{1}{2} \, \begin{pmatrix} \b{\nabla^2 \psi} \\ \b{\nabla \psi} \end{pmatrix}^\tp \! \b{C}^{-1}\left( \b{{x}^{(q)}} \right) \, \begin{pmatrix} \b{\nabla^2 \psi} \\ \b{\nabla \psi} \end{pmatrix} \right) \quad \text{with} \\
	C_{jl}\left( \b{{x}^{(q)}} \right) &= \begin{pmatrix} C^{\delta\delta}_{jl}\left( \b{{x}^{(q)}} \right) & C^{\delta p}_{jl}\left( \b{{x}^{(q)}} \right) \\ \left(C^{\delta p}_{jl}\right)^\tp\left( \b{{x}^{(q)}} \right) & C^{pp}_{jl}\left( \b{{x}^{(q)}} \right) \end{pmatrix} \label{eq:ini_corrMatrix} \\
	&= \int \frac{\diff^3 k''}{(2\pi)^3} \begin{pmatrix} - \norm{k''}^4 & i \, k''^\tp \, \norm{k''}^2 \\ i \, k'' \, \norm{k''}^2 & k'' k''^\tp \end{pmatrix} P_\psi(k'') \, \exp\left( -i k'' \cdot \left( {x}_j^{(q)} - {x}_l^{(q)} \right) \right) \,.
	\label{eq:ini_corrMatrixExpressions}
\end{align}
Here we abbreviated the velocity field~$\b{\nabla \psi} \ce \left( \nabla \psi\left( {x}_1^{(q)} \right), \nabla \psi\left( {x}_2^{(q)} \right), \ldots, \nabla \psi\left( {x}_N^{(q)} \right) \right)^\tp$ and the (negative) density contrast~$\b{\nabla^2 \psi} \ce \left( \nabla^2 \psi\left( {x}_1^{(q)} \right), \nabla^2 \psi\left( {x}_2^{(q)} \right), \ldots, \nabla^2 \psi\left( {x}_N^{(q)} \right) \right)^\tp$. The correlation matrix~$\b{C}$ is of size $4N \times 4N$ and contains initial density-density, density-momentum and momentum-momentum correlations. Using the joint probability $P\left( \b{\nabla\psi}, \b{\nabla^2\psi} \right)$, we can rewrite equation~\eqref{eq:ini_probx1} into
\begin{align}
	\label{eq:ini_probx2}
	P(\b{x}) &= \iint \diff^{3N} \b{\nabla \psi} \, \diff^N \b{\nabla^2\psi} \, \left( \prod_{j=1}^N P\left( {x}_j^{(q)} \mid \b{\nabla^2 \psi} \right) \, P\left( {x}_j^{(p)} \mid \b{\nabla \psi} \right) \right) \, P\left( \b{\nabla\psi}, \b{\nabla^2\psi} \right) \,.
\end{align}

Equation~\eqref{eq:ini_probx2} is the initial phase space probability distribution we use for the cosmic structure formation. The initial positions and momenta of the particles are correlated via density and momentum correlations. Instead of the probability distribution~$P(\b{x})$ we work with its characteristic function~$\Phi_{\b{x}}(\b{f})$ below. Therefore, let us derive a compact expression for it:
\begin{align}
	\Phi_{\b{x}}(\b{f}) &= \int \diff^{6N} \b{x} \, \exp(i \, \b{f} \cdot \b{x}) \, P(\b{x}) \\
	&= \int \diff^{6N} \b{x} \, \int \diff^N \b{\nabla^2\psi} \, \left(\prod_{j=1}^N \frac{\bar{\rho}}{N} \left( 1 - \kappa \left(\nabla^2 \psi \right)_j \right)\right) \frac{m^{-3N}}{\sqrt{(2\pi)^{4N} \, \det\left( \b{C}\left( \b{{x}^{(q)}} \right) \right)}} \nonumber\\*
	&\hspace{0.1\textwidth} \left( \restr{\exp\left(i \, m \, \b{\xi} \cdot \b{\nabla^2 \psi} \right)}{\b{\xi} = \b{0}}\right) \, \exp\left( - \frac{1}{2} \, \begin{pmatrix} \b{\nabla^2 \psi} \\ m^{-1} \b{{x}^{(p)}} \end{pmatrix}^\tp \! \b{C}^{-1}\left( \b{{x}^{(q)}} \right) \, \begin{pmatrix} \b{\nabla^2 \psi} \\ m^{-1} \b{{x}^{(p)}} \end{pmatrix} + i \, \b{f} \cdot \b{x} \right) \\
	&= \int \diff^{6N} \b{x} \, \int \diff^N \!\left( m \, \b{\nabla^2\psi} \right) \, \left(\prod_{j=1}^N \frac{\bar{\rho}}{N} \left( 1 - \frac{\kappa}{m} \frac{\partial}{i \, \partial \xi_j} \right)\right) \frac{m^{-4N}}{\sqrt{(2\pi)^{4N} \, \det\left( \b{C}\left( \b{{x}^{(q)}} \right) \right)}}  \, \exp\left(i \, \b{{f}^{(q)}} \cdot \b{{x}^{(q)}} \right) \nonumber\\*
	&\hspace{0.1\textwidth} \restr{\exp\left( - \frac{1}{2m^2} \, \begin{pmatrix} m \b{\nabla^2 \psi} \\ \b{{x}^{(p)}} \end{pmatrix}^\tp \! \b{C}^{-1}\left( \b{{x}^{(q)}} \right) \, \begin{pmatrix} m \b{\nabla^2 \psi} \\ \b{{x}^{(p)}} \end{pmatrix} + i \, \b{{f}^{(p)}} \cdot \b{{x}^{(p)}} + i \, m\, \b{\xi} \cdot \b{\nabla^2 \psi} \right)}{\b{\xi} = \b{0}} \\
	\label{eq:ini_characteristicFunction_fin}
	&= \int \diff^{3N} \b{{x}^{(q)}} \, \left(\prod_{j=1}^N \frac{\bar{\rho}}{N} \left( 1 - \frac{\kappa}{m} \frac{\partial}{i \, \partial \xi_j} \right)\right) \restr{\exp\left( - \frac{m^2}{2} \, \begin{pmatrix} \b{\xi} \\ \b{{f}^{(p)}} \end{pmatrix}^\tp \! \b{C}\left( \b{{x}^{(q)}} \right) \, \begin{pmatrix} \b{\xi} \\ \b{{f}^{(p)}} \end{pmatrix} \right) \, \exp\left(i \, \b{{f}^{(q)}} \cdot \b{{x}^{(q)}} \right)}{\b{\xi} = \b{0}} \,.
\end{align}
In the second equality we inserted equations~\eqref{eq:ini_probx2}, \eqref{eq:ini_probPsi} and~\eqref{eq:ini_probabilityOfXgivenPsi} and immediately used the Dirac $\delta$-function to cancel the integral over~$\b{\nabla \psi}$. We also inserted a factor equal to one which allows to perform the Gaussian integral by introducing an auxiliary Fourier conjugate~$\b{\xi}$ for~$m \, \b{\nabla^2 \psi}$. Note that in practice the form of~$\b{f}$ can be read off from the exponential factors in our expressions for observables. A simple example is the expectation value of the zeroth order density
\begin{align}
	\left\langle \tilde{\rho}_0 \right\rangle(k,t) &=
	\int \diff^{6N}\b{x} \, P(\b{x}) \, \sum_{m} \exp\left( - i \, k \cdot \left( {x}_m^{(q)} + g_{qp}(t,t_i) \, {x}_m^{(p)} \right) \right) \,.
\end{align}
After exchanging the sum with the integration, each term with index~$m$ is precisely given by the characteristic function~$\Phi_{\b{x}}(\b{f})$, where the $m\nth$~component of~$\b{f}$ is
\begin{align}
	{f}_m^{(q)} = -k \qquad \text{and} \qquad {f}_m^{(p)} = -g_{qp}(t,t_i) \, k \,,
\end{align}
while all other components are zero. Inserting this into equation~\eqref{eq:ini_characteristicFunction_fin}, we obtain
\begin{align}
	\left\langle \tilde{\rho}_0 \right\rangle(k,t) = \sum_m \Phi_{\b{x}}(\b{f})
	&= \sum_m \int \diff^{3N} \b{{x}^{(q)}} \, \left(\prod_{j=1}^N \frac{\bar{\rho}}{N} \left( 1 - \frac{\kappa}{m} \frac{\partial}{i \, \partial \xi_j} \right)\right) \nonumber\\*
	\label{eq:ini_Density_Zeroth}
	&\hspace{0.1\textwidth} \restr{\exp\left( - \frac{m^2}{2} \, \begin{pmatrix} \b{\xi} \\ \b{{f}^{(p)}} \end{pmatrix}^\tp \! \b{C}\left( \b{{x}^{(q)}} \right) \, \begin{pmatrix} \b{\xi} \\ \b{{f}^{(p)}} \end{pmatrix} \right) \, \exp\left(- i \, k \cdot {x}_m^{(q)} \right)}{\b{\xi} = \b{0}} \,.
\end{align}
The Gaussian exponential factor remains difficult even in this simple example.

\subsection{Correlation Hierarchy}

We have seen in the previous section that already the zeroth order density expectation value yields a complicated expression~\eqref{eq:ini_Density_Zeroth}. If we go to higher-order perturbation theory or consider higher-point density correlation functions, the expressions become even more intimidating. However, structurally they are quite similar. Therefore, in this section, we study in detail how the expression in equation~\eqref{eq:ini_Density_Zeroth} can be simplified. Afterwards, we show how to deal with the general case of $r$-point density correlation functions in higher-order perturbation theory. Ultimately, we are interested in the density fluctuation power spectrum.

We start our analysis by simplifying the derivatives with respect to $\b{\xi}$ equation~\eqref{eq:ini_Density_Zeroth}. Using the decomposition of the correlation matrix~$\b{C}$ in equation~\eqref{eq:ini_corrMatrix} into density-density, density-momentum and momentum-momentum correlations, we have
\begin{align}
   & \prod_{j=1}^N \left( 1 - \frac{\kappa}{m}\frac{\partial}{i \, \partial \xi_j}\right)\exp\left( - \frac{m^2}{2} \, \begin{pmatrix} \b{\xi} \\ \b{{f}^{(p)}} \end{pmatrix}^\tp \! \b{C} \, \begin{pmatrix} \b{\xi} \\ \b{{f}^{(p)}} \end{pmatrix} \right)=\nonumber\\
   \label{eq:ini_splitCorrelations}
   &\hspace{0.1\textwidth} \exp{\left(-\frac{m^2}{2} \left(\b{{f}^{(p)}}\right)^\tp \b{C^{pp}} \b{{f}^{(p)}}\right)} \prod_{j=1}^N \left( 1 + \frac{i \, \kappa}{m} \frac{\partial}{\partial \xi_j}\right) \, \exp{\left(-\frac{m^2}{2} \b{{\xi}}^\tp \b{C^{\delta\delta}} \b{{\xi}} - m^2 \b{{\xi}}^\tp \b{C^{\delta p}} \b{{f}^{(p)}} \right)} \,.
\end{align}
Here and in the following we refrain from explicitly writing the dependence of~$\b{C}$ and its components on~$\b{x^{(q)}}$. The products of $\left( 1 +  \frac{i \, \kappa}{m} \frac{\partial}{\partial \xi_j}\right)$ applied to the exponential can be expressed as 
\begin{align}
   \label{eq:ini_derivativeXi}
   &\left( 1+\frac{i \, \kappa}{m}\sum_j\frac{\partial}{\partial \xi_j} + \frac{1}{2!}\left(\frac{i \, \kappa}{m}\right)^2 \! \sum_{\substack{j, k\\j\neq k}}\frac{\partial^2}{\partial \xi_j \partial \xi_k} + \frac{1}{3!} \left(\frac{i \, \kappa}{m}\right)^3 \!\! \sum_{\substack{j, k, l\\j\neq k\neq l}}\frac{\partial^3}{\partial \xi_j \partial \xi_k \partial \xi_l} + \dots\right) \, \exp{\left(-\frac{m^2}{2} \b{{\xi}}^\tp \b{C^{\delta\delta}} \b{{\xi}} - m^2 \b{{\xi}}^\tp \b{C^{\delta p}} \b{{f}^{(p)}} \right)} \,.
\end{align}
Since we have a quadratic expression in the exponential in equation~\eqref{eq:ini_derivativeXi} and $\b{\xi}$ is set to zero after taking the derivative, derivatives act either alone
\begin{align}
\label{eq:ini_derivativeXi1}
    \frac{i \, \kappa}{m} \frac{\partial}{\partial \xi_j} \, \restr{\exp{\left(-\frac{m^2}{2} \b{{\xi}}^\tp \b{C^{\delta\delta}} \b{{\xi}} - m^2 \b{{\xi}}^\tp \b{C^{\delta p}} \b{{f}^{(p)}} \right)}}{\b{\xi} = \b{0}} &= \sum_n  \left( - i \, m \, \kappa \, C^{\delta p}_{j n} f^{(p)}_n \right)
\intertext{or in pairs}
\label{eq:ini_derivativeXi2}
    \left(\frac{i \, \kappa}{m}\right)^2 \! \frac{\partial^2}{\partial \xi_j \partial \xi_k} \, \restr{\exp{\left(-\frac{m^2}{2} \b{{\xi}}^\tp \b{C^{\delta\delta}} \b{{\xi}} - m^2 \b{{\xi}}^\tp \b{C^{\delta p}} \b{{f}^{(p)}} \right)}}{\b{\xi} = \b{0}} &= \kappa^2 C^{\delta\delta}_{jk} + \sum_{n_1,n_2} \left(- i \, m \, \kappa \,C^{\delta p}_{j n_1} f^{(p)}_{n_1}\right) \left(- i \, m \, \kappa \,C^{\delta p}_{k n_2} f^{(p)}_{n_2}\right) \,.
\end{align}
We used~$C^{\delta\delta}_{jl} = C^{\delta\delta}_{lj}$ to cancel the factor of~$\frac12$ (we implicitly assume that $P_\psi(-k) = P_\psi(k)$ which is justified in a statistically isotropic setting). In the second term, we wrote the products on the right-hand side in brackets to remind ourselves that they are actually scalar products since~$C^{\delta p}_{jk}$ are $1\times3$-matrices for all particle indices~$j,k$. By combining the terms from equations~\eqref{eq:ini_derivativeXi1} and~\eqref{eq:ini_derivativeXi2} appropriately, one can construct the expressions for higher derivatives with respect to~$\b{\xi}$ in a systematic manner. As an example, for the third order derivative in equation~\eqref{eq:ini_derivativeXi} we obtain
\begin{align}
     &\sum_{n} \kappa^2 C^{\delta \delta}_{j k} \left( - i \, m \, \kappa \, C^{\delta p}_{l n} f^{(p)}_{n} \right) + \sum_{n} \kappa^2 \, C^{\delta \delta}_{k l} \left( - i \, m \, \kappa \, C^{\delta p}_{j n} f^{(p)}_{n} \right) + \sum_{n} \kappa^2 \, C^{\delta \delta}_{l j} \left( - i \, m \, \kappa \, C^{\delta p}_{k n} f^{(p)}_{n} \right) \nonumber\\*
     &\hspace{0.1\textwidth} + \sum_{n_1,n_2,n_3} \left( - i \, m \, \kappa \, C^{\delta p}_{j n_1} f^{(p)}_{n_1} \right) \left( - i \, m \, \kappa \, C^{\delta p}_{k n_2} f^{(p)}_{n_2} \right) \left( - i \, m \, \kappa \, C^{\delta p}_{l n_3} f^{(p)}_{n_3} \right) \,.
\end{align}

In order to treat the momentum-momentum correlations~$\b{C^{pp}}$ on the same footing as~$\b{C^{\delta\delta}}$ and~$\b{C^{\delta p}}$, we can perform a Taylor expansion of the exponential containing them in equation~\eqref{eq:ini_splitCorrelations}. This way we obtain a product of two infinite sums in which we order terms in powers of the initial power spectrum or, equivalently, the number of~$\b{C^{\bullet \bullet}}$ for~$\bullet \in \{\delta,p\}$. This is in contrast to the treatment in~\cite{L3}, where the different kinds of correlations are treated differently.\footnote{In particular, in~\cite{L3} the authors treat the auto- and cross-correlations between the particles differently. This leads to a damping term~$Q_D = \frac12 \sum_j (f_j^{(p)})^\tp C_{jj}^{pp} f_j^{(p)}$ which subsequently needs to be treated with special care. We do not see a strong reason for treating particle auto-correlations differently and therefore do not obtain a separate damping term. Instead, the contribution~$Q_D$ is kept as part of the momentum-momentum correlation matrix~$\b{C^{pp}}$ and automatically appears at the appropriate orders of our correlation hierarchy.} In our case, the correlation hierarchy is build up from the three types of terms
\begin{align}
\label{eq:ini_correlation-factors}
    D^{\delta\delta}_{jl} \ce \kappa^2 C^{\delta \delta}_{jl} \,, \qquad D^{\delta p}_{jl} \ce - i \, m \, \kappa \, C^{\delta p}_{jl} f^{(p)}_{l} \,, \qquad D^{pp}_{jl} \ce -\frac{m^2}{2} \left(f^{(p)}_j\right)^\tp C^{pp}_{jl} f^{(p)}_l \,.
\end{align}
At~$m\nth$ order in the correlation expansion, we take a product of $m$~of these terms. Each such product acquires a prefactor~$\frac{1}{r!}$, where $r$~is the number of momentum-momentum correlation factors, coming from the Taylor expansion of the exponential. Moreover, there is a prefactor~$\frac{1}{s!}$, where~$s$ is the number of indices not paired with a factor~$f^{(p)}_\cdot$. Finally, we sum over all appearing indices (of which there are~$2m$), demanding that the set formed by both indices of factors~$D^{\delta \delta}_{jl}$ and first indices of factors~$D^{\delta p}_{jl}$ has no duplicates (the cardinality of this set is~$s$). This is due to the way the derivatives appear in equation~\eqref{eq:ini_derivativeXi} with distinct indices in each sum. Equivalently, all indices which are not contracted with a factor of~$f^{(p)}_l$ must not be equal.

A priori, all indices in the correlation terms constructed above are summed over the total number of particles~$N$. Once again, if we were to integrate the generating functional~$Z$ over the initial probability distribution~$P(\b{x})$, this would be the best we could do. However, given that we are considering observables, we can reduce the number of terms substantially. Let us consider the expectation value~$\langle \tilde{\rho}_0\rangle(k,t)$ which has been our example throughout. There we know that~$\b{f^{(p)}}$ has exactly one non-vanishing component~$f^{(p)}_m$, where $m$~is the index being summed over to obtain the contributions of each particle to the density. Thus, we immediately see that all the sums in our correlation expansions over indices appearing in~$D^{pp}_{jl}$ or as second index in~$D^{\delta p}_{jl}$ collapse into a single term. Generally, for any density $r$-point correlation functions, these sums receive contributions only for indices which already appear in the expression prior to integration over the initial conditions.

Perhaps surprisingly, the same statement is true for the remaining sums. Note that generally components of~$\b{f^{(q)}}$ are non-zero if and only if the corresponding component of~$\b{f^{(p)}}$ is non-zero. This is because they originate from Fourier factors of the form
\begin{align}
    \exp\left( - i \, k \cdot \left( {x}_m^{(q)} + g_{qp}(t,0) \, {x}_m^{(p)} \right) \right) \,.
\end{align}
Thus, the general form for density $r$-point correlation functions in equation~\eqref{eq:ini_characteristicFunction_fin} depends on the position variables~$x^{(q)}_l$ with indices other than the ones with non-zero~$f^{(p)}_l$ only via the correlation factors~$\b{C^{\bullet\bullet}}\left(\b{x^{(q)}}\right)$. However, this dependence can only be via any index of~$D^{\delta\delta}_{jl}$ or via the first index of~$D^{\delta\delta}_{jl}$ due to the reasoning in the previous paragraph.

Crucially, the set formed by the indices of~$D^{\delta\delta}_{jl}$ and the first indices of~$D^{\delta\delta}_{jl}$ must not contain duplicates. Therefore, in any product of correlation terms, position variables~$x^{(q)}_l$ with~$f^{(p)}_l = 0$ can appear at most once. This, however, implies that these variables can not appear at all. To see this, recall that the correlation factors have the form
\begin{align}
    C_{j l}^{\bullet \bullet}=\int \frac{\diff^3 k''}{\left(2\pi\right)^3} \, \alpha^{\bullet\bullet}(k'') \, \exp \left(-i k'' \cdot \left(x_{j}^{(q)}-x_{l}^{(q)}\right)\right)
\end{align}
for certain functions~$\alpha^{\bullet\bullet}(k'')$ and~$\bullet \in \{\delta,p\}$. If the variable~$x_{l}^{(q)}$ does not appear elsewhere in the expectation value, then the integral over it results in
\begin{align}
    \int \frac{\diff^3 k''}{\left(2\pi\right)^3} \, \alpha^{\bullet\bullet}(k'') \, \delta_{D}(k'') \, \exp\left(-i k'' \cdot x_{j}^{(q)}\right) = \alpha^{\bullet\bullet}(0)
\end{align}
which vanishes for usual choices for the initial density fluctuation power spectrum~$P_\delta(k)$.

Summarizing this discussion, we cannot have any indices in the correlation expansion which are not present in the expression for the density $r$-point correlation function prior to integration over the initial conditions. This is a substantial simplification, because it means that we can perform the integration over almost all components of~$\b{x^{(q)}}$ in equation~\eqref{eq:ini_characteristicFunction_fin} immediately. Specifically for the zeroth order density expectation value in equation~\eqref{eq:ini_Density_Zeroth} we find that only the zeroth order correlation gives a contribution:
\begin{align}
	\left\langle \tilde{\rho}_0 \right\rangle(k,t) &= \frac{\bar{\rho}}{N} \sum_j \int \diff^3 x^{(q)}_j \, \exp\left( - ik \cdot x_j^{(q)} \right) = (2\pi)^3 \, \delta_D(k) \, \bar{\rho} \,.
\end{align}

\subsection{Correlations vs.\ Interactions}
\label{sec:ini_corrInt}

We have seen above that for the expectation value of a general density $r$-point correlation function all the integrals over initial phase space coordinates except for the position variables~$x^{(q)}_l$ which explicitly appear in the expression for the diagrams are trivial. In the following, we consider the remaining integrals in more detail. This will lead us to the surprising observation that initial correlations between particles are much more difficult to treat computationally than the interactions during the evolution of the system.

Once again, we use the density function as an example in order to make the discussion more concrete. This time, we consider the first order correction
\begin{align}
	\tilde{\rho}_1(k,t;\b{x}) &= \
	\begin{tikzpicture}[baseline={([yshift=-.5ex]current bounding box.center)}]
		\node[vertex] (I1) at (-2,0) {$1$};
		\node[outdens] (D) at (0,0) {$k$, $t$};
		\draw[prop] (I1) -- (D);
	\end{tikzpicture}\\*
		&= - \sum_j \exp\left( - i \, k \cdot {\bar{\varphi}}_{m}^{(q)}(t;\b{x}) \right) \sum\limits_{l_1 \neq j} \int_{t_i}^\infty \diff t_1' \, g_{qp}(t,t_1') \nonumber \\*
		&\hspace{0.1\textwidth} \int \frac{\diff^3 k_1'}{(2\pi)^3} \, m \, v(k_1',t_1') \, k_1' \cdot k \, \exp\left( i \, k_1' \cdot \left( {\bar{\varphi}}_{j}^{(q)}(t_1';\b{x}) - {\bar{\varphi}}_{l_1}^{(q)}(t_1';\b{x}) \right) \right)
\end{align}
and focus on the contribution coming from the density-momentum correlation~$D^{\delta p}_{l_1j}$. Recall that the free trajectory is given by~${\bar{\varphi}}_{j}^{(q)}(t;\b{x}) = x^{(q)}_j + g_{qp}(t,t_i) \, x^{(p)}_j$, such that once we combine the exponential factors, we obtain
\begin{align}
    \tilde{\rho}_1(k,t;\b{x}) = - \sum_{\substack{j,l_1\\l_1\neq j}} \int_{t_i}^\infty \diff t_1' \, g_{qp}(t,t_1') \int \frac{\diff^3 k_1'}{(2\pi)^3} \, m \, v(k_1',t_1') \, k_1' \cdot k \, \exp\left( i \, \b{x} \cdot \b{f} \right) \,
\end{align}
where the only non-vanishing components of~$\b{f}$ are the position and momentum parts with indices~$j$ and~$l_1$. Hence, the expectation value  of the first order correction to the density can be written compactly as
\begin{align}
    \langle \tilde{\rho}_1 \rangle(k,t) = - \sum_{\substack{j,l_1\\l_1\neq j}} \int_{t_i}^\infty \diff t_1' \, g_{qp}(t,t_1') \int \frac{\diff^3 k_1'}{(2\pi)^3} \, m \, v(k_1',t_1') \, k_1' \cdot k \, \Phi_{\b{x}}(\b{f}) \,,
\end{align}
where we used the characteristic function~$\Phi_{\b{x}}(\b{f})$. We remark that this procedure is generally possible for all density $r$-point correlation functions for any perturbation order.

As the next step, we rewrite the characteristic function as in equation~\eqref{eq:ini_characteristicFunction_fin} and expand in the correlations. Leveraging the discussion in the previous subsections, we know that we obtain an infinite sum of contributions with certain correlation factors~\eqref{eq:ini_correlation-factors}. Crucially, the possible particle indices for the correlation factors are severely restricted, as they must be picked from the set~$\{l_1,j\}$. Therefore, the possible contributions are proportional to $D^{\delta \delta}_{l_1j}, D^{\delta p}_{l_1j}, D^{\delta p}_{jl_1}, D^{\delta p}_{l_1l_1}, D^{\delta p}_{jj}, D^{p p}_{l_1m}, D^{p p}_{l_1l_1}$ or $D^{p p}_{jj}$ in first correlation order. Here, we used symmetry of~$D^{\delta\delta}_{\cdot\cdot}$ and~$D^{pp}_{\cdot\cdot}$ to combine contributions.\footnote{Note that while~$C^{\delta p}_{\cdot\cdot}$ itself is symmetric in the indices, only the second index is contracted with a factor~$f^{(p)}_\cdot$, as can be seen in~\eqref{eq:ini_correlation-factors}. Thus the contribution for exchanged indices is generally different.} In second correlation order, the list is already quite lengthy, consisting of $D^{\delta p}_{l_1j} D^{\delta p}_{jl_1}$, $D^{\delta p}_{l_1j} D^{\delta p}_{jl_1}$, $D^{\delta p}_{jl_1} D^{\delta p}_{l_1l_1}$, $D^{\delta p}_{l_1l_1} D^{\delta p}_{jj}$ as well as all products of two terms of $1\st$~order involving at least one $D^{pp}_{\cdot\cdot}$ factor. The restriction that the set formed by the indices of~$D^{\delta\delta}_{\cdot\cdot}$ and the first indices of~$D^{\delta p}_{\cdot\cdot}$ must not have duplicates, removes two contributions in first and roughly halves the number of contributions in second correlation order.

Let us return to our example, where we specifically wanted to analyze the contributions coming from~$D^{\delta p}_{l_1j}$ which is given by
\begin{align}
\label{eq:ini_dens1-C}
        - \left( \frac{\bar{\rho}}{N} \right)^2 \sum_{\substack{j,l_1\\l_1\neq j}} \int_0^\infty \diff t_1' \, g_{qp}(t,t_1') \int \frac{\diff^3 k_1'}{(2\pi)^3} \, m \, v(k_1',t_1') \, k_1' \cdot k \int \diff^3x^{(q)}_{l_1} \, \diff^3x^{(q)}_{j} \, \left( - i \, m \, \kappa \, C^{\delta p}_{l_1j} f^{(p)}_{j} \right) \, \exp\left( i \, \b{x^{(q)}} \cdot \b{f^{(q)}} \right) \,.
\end{align}
Note that the position variables only appear in Fourier phase factors and thus the integrals over them can be performed analytically. Using the definition
\begin{align}
    C^{\delta p}_{l_1j} = \int \frac{\diff^3 k_1''}{(2\pi)^3} \norm{k_1''}^2 \, \left(i \, k_1''\right)^\tp \, P_\psi(k_1'') \, \exp\left( -i \, k_1'' \cdot \left( {x}_{l_1}^{(q)} - {x}_j^{(q)} \right) \right)
\end{align}
each of the two integrals yields a Dirac $\delta$-function. These encode the linear system of equations
\begin{align}
    \begin{matrix} f^{(q)}_{l_1} - k_1'' &= 0 \\ f^{(q)}_{j} + k_1'' &= 0 \end{matrix} \qquad \LRa \qquad \begin{matrix} \quad\ - k_1' - k_1'' &= 0 \\ -k + k_1' + k_1'' &= 0 \end{matrix}
\end{align}
Adding these two equations, we immediately see that the condition~$k = 0$ is implied. However, the entire contribution~\eqref{eq:ini_dens1-C} is proportional to~$k$ and thus vanishes. In fact, it can be shown that for any perturbation and correlation orders the expectation value of the density function does not receive corrections.\footnote{For zeroth correlation order, one needs to employ equation~\eqref{eq:app_interactionPotentialZero} to see that the contributions from positive perturbation orders are zero. Depending on the choice of initial density-fluctuation power spectrum~$P_\delta$, this subtlety might also come to the rescue at higher correlation orders, especially if one considers contribution involving exclusively momentum-momentum-correlations.} Consequently, the expectation value of the density~$\langle \tilde{\rho} \rangle(k,t) = \langle \tilde{\rho}_0 \rangle(k,t) = (2\pi)^3 \, \delta_D(k) \, \bar{\rho}$ is the (Fourier transform of the) mean density as it should be for statistically homogeneous and isotropic initial conditions.

The appearance of such a system of linear equations is a generic feature. In fact, its form can be read off quite easily from the diagram and the indices of the correlations. Let us consider the general case of a density $r$-point correlation function~$G^{(r)}_\rho(k_1,\ldots, k_r)$ at $n\nth$~order perturbation theory and at $m\nth$~order in the expansion in the initial correlations. We temporarily ignore any initial particle auto-correlations, i.e., terms~$C^{\bullet \bullet}_{jj}$ with repeated indices~$j$. For those we do not have any exponential factor such that they do not contribute to the system of linear equations and the corresponding integrals over wave vectors can be evaluated independently.

Generally, the number of equations is given by $r+n$, as this is the number of particle indices appearing in the expression for the diagrams. Note that while any interaction vertex has two indices~$l_a$ and~$j_a$, the latter is always identified with another index by the Kronecker deltas in the edges. The number of variables in the equations is equal to $r+n+m$. Specifically, these are the arguments~$k_1,\ldots, k_r$ of the correlation function, the wave vectors integrated over in the interactions~$k_1',\ldots, k_n'$ and those coming from the initial correlations~$k_1'',\ldots, k_m''$. Because our diagrams are acyclic, the system of linear equations always has full rank. Indeed, the $(r+n)$~columns corresponding to~$(k_1,\ldots, k_r, k_1',\ldots, k_n')$ are linearly independent.\footnote{One easy way of proving this is to order the variables as they appear from right to left in the diagram. The columns corresponding to~$(k_1,\ldots, k_r)$ have exactly one non-vanishing entry and thus form the negative of the unit matrix. The remaining $n$~columns also have entries~$-1$ on the diagonal and in addition a single other entry $1$~above it. The determinant of the $(r+n)\times(r+n)$ matrix thus is~$(-1)^{r+n}$ which implies the linear independence of the rows of the entire system.} Consequently, we have $r+n$ independent Dirac $\delta$-functions which can be used to cancel $r+n$~integrals over wave numbers~$k_j'$ and~$k_l''$. Of these integrals there are~$n+m$ such that naively the remaining number of integrals is~$m-r$. In practice, however, it is possible that after solving the linear system of equations some of the Dirac $\delta$-functions only contain the arguments~$(k_1,\ldots, k_r)$. This is the case, e.g., for the zeroth order density expectation value~$\langle \tilde{\rho}_0 \rangle(k,t)$ as we saw above. In these cases an equal number of additional integrals remain. Thus, denoting the number of remaining Dirac $\delta$-functions by~$p$, the actual number of remaining integrals is~$m - r + p$. As an explicit example, the density fluctuation power spectrum~$P_\delta$ has~$r = 2$ and~$p = 1$ (corresponding to a factor $\delta_D(k_1 + k_2)$, where~$k_1$ and~$k_2$ are the wave vectors coming from the density vertices), and hence there are~$m-1$ integrals remaining. In particular, there is no contribution coming from the zeroth order term in the correlation expansion.

Even in the case of a maximum of~$r$ remaining Dirac $\delta$-functions, there is only~$m$ integrals over wave vectors remaining. In particular, it is always the case that all integrals coming from the interactions can be cancelled. Conversely, in general the integrals over wave vectors coming from the correlations remain and need to be solved numerically. In practice, this makes the expansion in the initial correlations significantly more difficult than the expansion in the interactions. Indeed, generically we expect one additional simple time integral for any interaction order, but one additional difficult multidimensional wave vector integral for any correlation order.


\section{Reproducing Linear Cosmic Structure Growth}

We have seen above that going to high orders in the expansion in the correlations is difficult in practice. Generically, in the resulting expressions there are multidimensional integrals over wave vectors remaining which are numerically challenging to solve. It is beyond the scope of this paper to discuss these computational issues in detail. Instead, we restrict ourselves to the first order in the correlation expansion in the following. We will see momentarily that this way we reproduce exactly the linear growth of perturbations as predicted by standard hydrodynamic cosmic perturbation theory. The non-linear growth of structures is exclusively encoded in the higher-order initial correlations and will be studied in detail in a follow-up paper~\cite{we2022}. Below we study the density two-point correlation function from which we extract the density fluctuation power spectrum -- an important observable in the context of cosmic structure formation. In an analogous fashion, one can derive higher correlation functions and obtain, e.g., the density fluctuation bi- and trispectra.

Before we can interpret the results we obtain for the power spectrum, we need to lay the groundwork by specifying the form of the propagator~$g_{qp}(t,t')$ and the two-particle interaction potential~$v(k,t)$ in the context of cosmic structure formation. In particular, this requires us to transfer our construction of KFT on an expanding background. Similarly to numerical simulations of cosmic structure formation, Einstein's theory of General Relativity is only incorporated on the background level, while the particle interaction is given by Newtonian gravity, ignoring baryonic, radiative and relativistic effects at~$t > t_i$. This is a well-established approximate treatment on sub-horizon scales and can be justified both theoretically and numerically (cf.~e.g.~\cite{PhysRevD.83.123505}).

\subsection{Free Linear Power Spectrum}

In our quest for the linear power spectrum, we start with the expression for the density two-point correlation function in zeroth order perturbation theory given by
\begin{align}
\label{eq:cos_Grho0}
    \left(\tilde{G}_{\rho\rho}\right)_0(k_1,k_2,t;\b{x}) &=\ 
	\begin{tikzpicture}[baseline={([yshift=-.5ex]current bounding box.center)}]
		\node[outdens] (D1) at (0,0) {$k_1$, $t$};
		\node[outdens] (D2) at (0,-0.8) {$k_2$, $t$};
	\end{tikzpicture}\
	= \sum_{\substack{m_1,m_2\\m_1\neq m_2}} \exp\left( -i \, \left( k_1 \cdot {\bar{\varphi}}_{m_1}^{(q)}(t;\b{x}) + k_2 \cdot {\bar{\varphi}}_{m_2}^{(q)}(t;\b{x}) \right) \right) \,.
\end{align}
The aim of this subsection is to obtain the expectation value of this observable and to extract the density fluctuation power spectrum from it. In order to do so, we integrate the expression in equation~\eqref{eq:cos_Grho0} over the probability distribution given in equation~\eqref{eq:ini_probx2}. As explained above, we limit ourselves to the contributions up to first order in the expansion in the initial particle correlations. Specifically, these contributions are proportional to~$1$, $D^{\delta\delta}_{m_1m_2}$, $D^{\delta p}_{m_1m_2}$ and $D^{p p}_{m_1m_2}$ as well as the corresponding (and here in all cases equal) contributions for swapped indices. Contributions with repeated indices like~$D^{p p}_{m_1m_1}$ vanish due to the appearance of factors like~$k_1 \, \delta_D(k_1)$.

The calculations of these contributions are analogous to the ones in the previous section. Firstly, we rewrite the integral over~$\left(\tilde{G}_{\rho\rho}\right)_0(k_1,k_2,t;\b{x})$ in terms of the characteristic function~$\varphi_{\b{x}}(\b{f})$ and utilize its expansions in the correlations. Then we consider each contribution individually. As an example, for the correlation contribution~$D^{\delta p}_{m_1m_2}$ it is\footnote{\label{foot:cos_notation}Our notation in the first line is somewhat cavalier. In particular, we use the summation indices within~$\tilde{G}_{\rho\rho}$ also outside of it. Moreover, we cancel by hand the exponential factors involving~$\b{x^{(p)}}$ even though technically they go away through integration.}
\begin{align}
    &~\left(\frac{\bar{\rho}}{N}\right)^2 \int \diff^3 x^{(q)}_{m_1} \, \diff^3 x^{(q)}_{m_2} \, \left(\tilde{G}_{\rho\rho}\right)_0(k_1,k_2,t;\b{x}) \, \exp\left( -i \, \b{f^{(p)}} \cdot \b{x^{(p)}} \right) \, \left( - i \, m\, \kappa \, C^{\delta p}_{m_1m_2} \, f^{(p)}_{m_2} \right) \nonumber\\*
    =&~- i \, m\, \kappa \, \left(\frac{\bar{\rho}}{N}\right)^2 \sum_{\substack{m_1,m_2\\m_1\neq m_2}} \int \diff^3 x^{(q)}_{m_1} \, \diff^3 x^{(q)}_{m_2} \, \exp\left( -i \, \left( k_1 \cdot x^{(q)}_{m_1} + k_2 \cdot x^{(q)}_{m_2} \right) \right) \nonumber\\*
        &\hspace{0.1\textwidth} \int \frac{\diff^3 k_1''}{(2\pi)^3} \, \left( i \, k_1'' \, \norm{k_1''}^2 \right) \cdot \left( - g_{qp}(t,t_i) \, k_2 \right) \, P_\psi(k_1'') \, \exp\left( -i \, k_1'' \cdot \left( x^{(q)}_{m_1} - x^{(q)}_{m_2} \right) \right) \\*
    =&~m\, \kappa \, \left(\frac{\bar{\rho}}{N}\right)^2 g_{qp}(t,t_i) \sum_{\substack{m_1,m_2\\m_1\neq m_2}} \int \frac{\diff^3 k_1''}{(2\pi)^3} \, \norm{k_1''}^2 \, k_1'' \cdot k_2 \, P_\psi(k_1'') \, \delta_D(k_1 + k_1'') \, \delta_D(k_2 - k_1'') \\*
    =&~(2\pi)^3 \, \bar{\rho}^2 \, m \, \kappa \, g_{qp}(t,t_i) \, \frac{N(N-1)}{N^2} \, \norm{k_1}^4 \, P_\psi(k_1) \, \delta_D(k_1 + k_2) \,.
\end{align}
In the limit of large particle numbers $N \ra \infty$, the fraction in the last line approaches unity. Moreover, we can replace the initial power spectrum of the velocity potential~$P_\psi(k)$ by the initial density fluctuation power spectrum~$P_\delta(k) = \kappa^{2} \norm{k}^4 P_\psi(k)$.

The full list of non-zero contributions is given by
\begin{align}
    1~~\qquad \quad&:\quad (2\pi)^6 \, \bar{\rho}^2 \, \delta_D(k_1) \, \delta_D(k_2) \,, \\
    D^{\delta\delta}_{m_1m_2}, D^{\delta\delta}_{m_2m_1} \quad&:\quad (2\pi)^3 \, \frac{\bar{\rho}^2}{2} \, P_\delta(k_1) \, \delta_D(k_1 + k_2) \,, \\
    D^{\delta p}_{m_1m_2}, D^{\delta p}_{m_2m_1} \quad&:\quad (2\pi)^3 \, \bar{\rho}^2 \, \frac{m \, g_{qp}(t,t_i)}{\kappa} \, P_\delta(k_1) \, \delta_D(k_1 + k_2) \,, \\
    D^{p p}_{m_1m_2}, D^{p p}_{m_2m_1} \quad&:\quad (2\pi)^3 \, \frac{\bar{\rho}^2}{2} \, \left(\frac{m \, g_{qp}(t,t_i)}{\kappa}\right)^2 \, P_\delta(k_1) \, \delta_D(k_1 + k_2) \,.
\end{align}
We remark that the factor of~$\frac{1}{2}$ in the expression from contributions proportional to~$D^{\delta\delta}_{\cdot\cdot}$ is the prefactor of the second order term in the expansion in equation~\eqref{eq:ini_derivativeXi}. Combining these four contributions, the expectation value for the density two-point correlation function is given by
\begin{align}
    \left\langle \left(\tilde{G}_{\rho\rho}\right)_0 \right\rangle(k_1,k_2,t) = (2\pi)^6 \, \bar{\rho}^2 \, \delta_D(k_1) \, \delta_D(k_2) + (2\pi)^3 \, \bar{\rho}^2 \, \delta_D(k_1 + k_2) \, \left(1 + \frac{m \, g_{qp}(t,t_i)}{\kappa}\right)^2 P_\delta(k_1) + \mathcal{O}\left(\b{C}^2\right) \,.
\end{align}
This is precisely the form of a general two-point correlation function of statistically homogeneous density field as given in equation~\eqref{eq:feynman_PowerSpectrumDef}. In particular, we can extract the density fluctuation power spectrum
\begin{align}
\label{eq:cos_PowerSpectrum01}
    \left(\tilde{P}_\delta\right)_0(k_1,t) = \left(1 + \frac{m \, g_{qp}(t,t_i)}{\kappa} \right)^2 P_\delta(k_1) + \mathcal{O}\left(\b{C}^2\right) \,.
\end{align}
The subscript~$0$ in~$\left(\tilde{P}_\delta\right)_0(k_1,t)$ indicates that we are working in zeroth order perturbation theory for the interactions between the particles. The term~$\mathcal{O}\left(\b{C}^2\right)$ symbolizes that we are working up to first order in the expansions in the initial correlations between the particles. We observe that the power spectrum in equation~\eqref{eq:cos_PowerSpectrum01} is proportional to the initial power spectrum. This should come as no surprise given that we argued earlier that the first order correlation contributions to the power spectrum do not come with any integrals over~$k$.

\subsection{Higher-Order Linear Power Spectrum}

Our next step is to compute the first order correction to the two-point density correlation function. It is given by
\begin{align}
\label{eq:cos_Grho1}
    \left(\tilde{G}_{\rho\rho}\right)_1(k_1,k_2,t;\b{x}) &=\ 
	\begin{tikzpicture}[baseline={([yshift=-.5ex]current bounding box.center)}]
		\node[vertex] (I1) at (-1.5,0) {$1$};
		\node[outdens] (D1) at (0,0) {$k_1$, $t$};
		\node[outdens] (D2) at (0,-0.8) {$k_2$, $t$};
		\draw[prop] (I1) -- (D1);
	\end{tikzpicture} \ + \ 
	\begin{tikzpicture}[baseline={([yshift=-.5ex]current bounding box.center)}]
		\node[vertex] (I1) at (-1.5,-0.8) {$1$};
		\node[outdens] (D1) at (0,0) {$k_1$, $t$};
		\node[outdens] (D2) at (0,-0.8) {$k_2$, $t$};
		\draw[prop] (I1) -- (D2);
	\end{tikzpicture} \\
	&= \sum_{m_1} \sum_{m_2 \neq m_1} \exp\left( -i \, \left( k_1 \cdot {\bar{\varphi}}_{m_1}^{(q)}(t;\b{x}) + k_2 \cdot {\bar{\varphi}}_{m_2}^{(q)}(t;\b{x}) \right) \right) \sum\limits_{l_1 \neq m_1} \int_{t_i}^t \diff t_1' \, g_{qp}(t,t_1') \nonumber \\*
	&\hspace{0.05\textwidth} \int \frac{\diff^3 k_1'}{(2\pi)^3} \, m \, v(k_1',t_1') \, k_1' \cdot k_1 \, \exp\left( i \, k_1' \cdot \left( {\bar{\varphi}}_{m_1}^{(q)}(t_1';\b{x}) - {\bar{\varphi}}_{l_1}^{(q)}(t_1';\b{x}) \right) \right) + \Big( 1 \leftrightarrow 2 \Big)
\end{align}
for explicit initial conditions~$\b{x}$. The expression for the second diagram is the same as for the first diagram, but the wave vectors~$k_1$ and~$k_2$ as well as the indices~$m_1$ and~$m_2$ are swapped. We have already familiarized ourselves with the procedure of integrating such expressions over our probability distribution of initial values~$P(\b{x})$. As a first step, we read off the non-zero components of~$\b{f^{(p)}}$ which are given by
\begin{align}
	{f}_{m_1}^{(p)} &= -g_{qp}(t,t_i) \, k_1 + g_{qp}(t_1',t_i) \, k_1' \,, \nonumber\\*
	{f}_{m_2}^{(p)} &= -g_{qp}(t,t_i) \, k_2 \,,\nonumber\\*
	{f}_{l_1}^{(p)} &= -g_{qp}(t_1',t_i) \, k_1' \,. 	
\end{align}
for the first diagram. Then, we rewrite the entire expression for~$\left\langle \left(\tilde{G}_{\rho\rho}\right)_1 \right\rangle$ in terms of the characteristic function and expand to first order in the correlations.

We know already that the only contributions~$D_{jl}^{\bullet\bullet}$ are those where indices~$j$ and~$l$ already appear in the expression of~$\left(\tilde{G}_{\rho\rho}\right)_1$. However, there is a very useful further condition. Namely, any index~$m$ corresponding to the left-most vertices of the diagram must appear as an index of some correlation factor~$D_{jl}^{\bullet\bullet}$. This is because these left-most vertices always come with a single exponential phase factor containing the particle position~$x^{(q)}_m$ which upon integration yield a Dirac $\delta$-function containing the corresponding wave vector. Since this wave vector always appears as an overall factor (or as the argument of the initial power spectrum), the expression results in zero. Examples are the indices~$l_1$ and~$m_2$ for the first diagram in equation~\eqref{eq:cos_Grho1}. Since we have only two indices of the correlation factors available in our first order expansion, the only non-zero contributions come from~$D^{\delta\delta}_{m_2l_1}$, $D^{\delta p}_{m_2l_1}$ and~$D^{pp}_{m_2l_1}$ as well as their counterparts with exchanged indices.

To provide an example for the calculation, let us consider the contribution of~$D^{\delta\delta}_{m_2l_1}$ from the first diagram. It is given by (cf.~footnote~\ref{foot:cos_notation})
\begin{align}
    &~\frac{1}{2} \left(\frac{\bar{\rho}}{N}\right)^3 \int \diff^3 x^{(q)}_{m_1} \, \diff^3 x^{(q)}_{m_2} \, \diff^3 x^{(q)}_{l_1} \,  \left(\tilde{G}_{\rho\rho}\right)_1(k_1,k_2,t;\b{x}) \, \exp\left( i \, \b{f^{(p)}} \cdot \b{x^{(p)}} \right) \, \left( \kappa^2 C^{\delta\delta}_{m_2l_1} \right) \nonumber\\*
    =&~\frac{\kappa^2}{2} \left(\frac{\bar{\rho}}{N}\right)^3 \sum_{\substack{m_1, m_2 \\ m_1 \neq m_2}} \sum\limits_{l_1 \neq m_1} \int \diff^3 x^{(q)}_{m_1} \, \diff^3 x^{(q)}_{m_2} \, \diff^3 x^{(q)}_{l_1} \, \exp\left( -i \, \left( k_1 \cdot {x}_{m_1}^{(q)} + k_2 \cdot {x}_{m_2}^{(q)} \right) \right) \int_{t_i}^t \diff t_1' \, g_{qp}(t,t_1') \int \frac{\diff^3 k_1'}{(2\pi)^3} \, m \nonumber \\* 
    &\hspace{0.05\textwidth} v(k_1',t_1') \, k_1' \cdot k_1 \, \exp\left( i \, k_1' \cdot \left( {x}_{m_1}^{(q)} - {x}_{l_1}^{(q)} \right) \right) \int \frac{\diff^3 k_1''}{(2\pi)^3} \, \left( -\norm{k_1''}^4 \right) \, P_\psi(k_1'') \, \exp\left( -i \, k_1'' \cdot \left( x^{(q)}_{m_2} - x^{(q)}_{l_1} \right) \right) \\*
    =&~-\frac{\kappa^2}{2} \left(\frac{\bar{\rho}}{N}\right)^3 \sum_{\substack{m_1, m_2 \\ m_1 \neq m_2}} \sum\limits_{l_1 \neq m_1} \int_{t_i}^t \diff t_1' \, g_{qp}(t,t_1') \int \frac{\diff^3 k_1'}{(2\pi)^3} \, m \, v(k_1',t_1') \, k_1' \cdot k_1 \nonumber \\* 
    &\hspace{0.05\textwidth} \int \frac{\diff^3 k_1''}{(2\pi)^3} \, \norm{k_1''}^4 \, P_\psi(k_1'') \, (2\pi)^9 \, \delta_D\left(k_1 - k_1'\right) \, \delta_D\left(k_2 + k_1''\right) \, \delta_D\left(k_1' - k_1''\right) \\*
    =&~-(2\pi)^3 \, \frac{\kappa^2}{2} \left(\frac{\bar{\rho}}{N}\right)^3 N(N-1)^2 \int_{t_i}^t \diff t_1' \, g_{qp}(t,t_1') \, m \, v(k_1,t_1') \, \norm{k_1}^6 \, P_\psi(k_1) \, \delta_D\left(k_2 + k_1\right) \,.
\end{align}
Like for the zeroth order case, we take the limit of the particle number~$N$ going to infinity and replace the power spectrum for the initial velocity potential~$P_\psi$ by the initial density fluctuation power spectrum~$P_\delta$. Then, the non-zero contributions are given by
\begin{align}
    D^{\delta\delta}_{m_2l_1}, D^{\delta\delta}_{l_1m_2} ~~&:~~ - (2\pi)^3 \, \frac{\bar{\rho}^3}{2} \, P_\delta(k_1) \, \delta_D(k_1 + k_2) \, \int_0^t \diff t_1' \, g_{qp}(t,t_1') \, m \, \norm{k_1}^2 v(k_1,t_1') \,, \\
    D^{\delta p}_{m_2l_1}\quad ~~&:~~ - (2\pi)^3 \, \bar{\rho}^3 \, P_\delta(k_1) \, \delta_D(k_1 + k_2) \, \int_0^t \diff t_1' \, g_{qp}(t,t_1') \, \frac{m^2 \, g_{qp}(t_1',t_i)}{\kappa}  \, \norm{k_1}^2 v(k_1,t_1') \,, \\
    D^{\delta p}_{l_1m_2}\quad ~~&:~~ - (2\pi)^3 \, \bar{\rho}^3 \, P_\delta(k_1) \, \delta_D(k_1 + k_2) \, \int_0^t \diff t_1' \, g_{qp}(t,t_1') \, \frac{m^2 \, g_{qp}(t,t_i)}{\kappa} \, \norm{k_1}^2 v(k_1,t_1') \,, \\
    D^{p p}_{m_2l_1}, D^{p p}_{l_1m_2} ~~&:~~ - (2\pi)^3 \, \frac{\bar{\rho}^3}{2} \, P_\delta(k_1) \, \delta_D(k_1 + k_2) \, \int_0^t \diff t_1' \, g_{qp}(t,t_1') \, \frac{m^3 \, g_{qp}(t_1',t_i) \, g_{qp}(t,t_i)}{\kappa^2} \, \norm{k_1}^2 v(k_1,t_1') \,.
\end{align}

Combining the contributions listed above, we obtain the first order correction to the expectation value to the density fluctuation power spectrum
\begin{align}
\label{eq:cos_PowerSpectrum11}
    \left(\tilde{P}_\delta\right)_1(k_1,t) = -2 \int_{t_i}^t \diff t_1' \, g_{qp}(t,t_1') \, \left(1 + \frac{m \, g_{qp}(t_1',t_i)}{\kappa} \right) \, \left(1 + \frac{m \, g_{qp}(t,t_i)}{\kappa} \right) \, m\, \bar{\rho} \, \norm{k_1}^2 v(k_1,t_1') \, P_\delta(k_1) + \mathcal{O}\left(\b{C}^2\right) \,,
\end{align}
where the factor~$2$ in front accounts for the contribution of the second diagram. In order to identify the two contributions we assumed that the two-particle interaction potential~$v(k_1,t_1')$ is symmetric in the wavevector, i.e., it is~$v(-k_1,t_1') = v(k_1,t_1')$. We observe that if the potential is of Newtonian form, i.e., $v(k_1,t_1') \propto \norm{k_1}^{-2}$, we again only obtain a rescaling of the initial density fluctuation power spectrum.

Going to higher-orders, there is a crucial simplification when we only consider the first order contributions in the correlation expansion. Let us consider the diagram
\begin{align}
    \begin{tikzpicture}[baseline={([yshift=-.5ex]current bounding box.center)}]
		\node[vertex] (I1) at (-2,0.2) {$1$};
		\node[vertex] (I2) at (-1.5,-0.6) {$2$};
		\node[outdens] (D1) at (0,0) {$k_1$, $t$};
		\node[outdens] (D2) at (0,-0.8) {$k_2$, $t$};
		\draw[prop] (I1) -- (D1);
		\draw[prop] (I2) -- (D1);
	\end{tikzpicture}
\end{align}
which is part of the second order two-point density correlation function~$\left(\tilde{G}_{\rho\rho}\right)_2(k_1,k_2,t;\b{x})$. We argued above that all left-most vertices must have their respective particle indices appear as indices of the correlation contributions~$D^{\bullet\bullet}_{\cdot\cdot}$. This is because these vertices -- here both interaction vertices and the lower density vertex -- feature a Fourier phase factor of a particle position (with indices~$l_1$, $l_2$ and~$m_2$, respectively) which does not appear elsewhere in the expression. However, in linear order of the correlation expansion, we have only two slots for indices, hence there always is one of these phase factors left which subsequently sets the corresponding wavevector to zero and leads to a vanishing result for the entire term. The same is even true if we consider the contributions proportional to~$\delta_{j_aj_b}$ in the Feynman rule for the propagator given in equation~\eqref{rule:internal_edge}, because then a Fourier phase factor involving~$x^{(q)}_{l_b}$ remains. Hence, there are yet more contributions which may be ignored when considering linear initial particle correlations only.

Since all diagrams for the density two point function are forests consisting of two trees with density vertices as their roots, these trees must never branch out to make sure that there are only two left-most vertices. Hence, the $n\nth$~order contribution to the density two-point correlation function is given by
\begin{align}
    \left\langle \left(\tilde{G}_{\rho\rho}\right)_n \right\rangle(k_1,k_2,t) = \sum_{r=0}^n \sum_{\text{orderings}} \left\langle \ 
    \begin{tikzpicture}[baseline={([yshift=-.5ex]current bounding box.center)}]
		\node[vertex] (I1) at (-4.5,0) {$1$};
		\node[vertex] (Ik) at (-1.5,0) {$r$};
		\node[vertex] (Ik1) at (-4.5,-0.8) {\resizebox{1em}{!}{$r\hspace{-0.4ex}+\hspace{-0.6ex}1$}};
		\node[vertex] (In) at (-1.5,-0.8) {$n$};
		\node[outtraj] (dotsu) at (-3,0) {\raisebox{-5.5pt}{$\,\b{\cdots}$}};
		\node[outtraj] (dotsd) at (-3,-0.8) {\raisebox{-5.5pt}{$\,\b{\cdots}$}};
		\node[outdens] (D1) at (0,0) {$k_1$, $t$};
		\node[outdens] (D2) at (0,-0.8) {$k_2$, $t$};
		\draw[prop] (I1) -- (dotsu);
		\draw[prop] (dotsu) -- (Ik);
		\draw[prop] (Ik) -- (D1);
		\draw[prop] (Ik1) -- (dotsd);
		\draw[prop] (dotsd) -- (In);
		\draw[prop] (In) -- (D2);
	\end{tikzpicture} \ 
	\right\rangle \,,
\end{align}
where the sum over the orderings includes all the possible ways the interaction vertices in the upper and lower tree can be horizontally positioned relative to each other (which might require relabeling to abide by time-ordering). However, recalling our discussion of time-ordering in section~\ref{sec:feynman_symmetryFactors}, this sum can be completely avoided by demanding time-ordering only within each tree instead of for the entire diagram. As an example, the two diagrams
\begin{align}
    \begin{tikzpicture}[baseline={([yshift=-.5ex]current bounding box.center)}]
		\node[vertex] (I1) at (-2,0) {$1$};
		\node[vertex] (I2) at (-1.5,-0.8) {$2$};
		\node[outdens] (D1) at (0,0) {$k_1$, $t$};
		\node[outdens] (D2) at (0,-0.8) {$k_2$, $t$};
		\draw[prop] (I1) -- (D1);
		\draw[prop] (I2) -- (D2);
	\end{tikzpicture} \ \propto \int_{t_i}^t \diff t_2' \int_0^{t_2'} \diff t_1' \cdots \qquad \text{and} \qquad
    \begin{tikzpicture}[baseline={([yshift=-.5ex]current bounding box.center)}]
		\node[vertex] (I1) at (-2,-0.8) {$1$};
		\node[vertex] (I2) at (-1.5,0) {$2$};
		\node[outdens] (D1) at (0,0) {$k_1$, $t$};
		\node[outdens] (D2) at (0,-0.8) {$k_2$, $t$};
		\draw[prop] (I2) -- (D1);
		\draw[prop] (I1) -- (D2);
	\end{tikzpicture} \ \propto \int_{t_i}^t \diff t_2' \int_0^{t_2'} \diff t_1' \cdots
\end{align}
combine (after relabeling~$1 \leftrightarrow 2$ in, e.g., the second diagram) into a single expression in which both~$t_1'$ and~$t_2'$ are integrated from~$t_i$ to~$t$.

Let us examine the contributions to the power spectrum in zeroth and first interaction order given in equations~\eqref{eq:cos_PowerSpectrum01} and~\eqref{eq:cos_PowerSpectrum11} once more. In both cases, the combined contributions from~$D^{\delta\delta}_{\cdot\cdot}$, $D^{\delta p}_{\cdot\cdot}$ and~$D^{pp}_{\cdot\cdot}$ can be written as a product of two terms of which one depends on the time variables of the upper and the other on the time variables of the lower tree.\footnote{This is somewhat concealed in these cases by the fact that both density vertices are evaluated at the same time~$t$. We can write these contributions from the different kinds of initial correlations as a product because it happens to hold that~$2 D^{\delta\delta}_{jl} D^{pp}_{jl} = D^{\delta p}_{jl} D^{\delta p}_{lj}$ upon integrating over~$x^{(q)}_j$ and~$x^{(q)}_l$ (the integrals identify the wave vectors and cancel the integrals in the correlation factors~$C^{\bullet\bullet}_{\cdot\cdot}$).} This remains valid at higher-orders and allows us to completely decouple the two trees in the diagrams contributing to the power spectrum. Indeed, we arrive at
\begin{align}
\label{eq:cos_PowerSpectrumn1}
    \left(\tilde{P}_\delta\right)_n(k_1,t) &= \sum_{r=0}^n f_r(k_1,t) \, f_{n-r}(k_1,t) \, P_\delta(k_1) + \mathcal{O}\left(\b{C}^2\right) \,, \qquad \text{where} \\*
\label{eq:cos_PowerSpectrumnfr}
    f_r(k,t) &= \begin{cases} \left(1 + \frac{m \, g_{qp}(t,t_i)}{\kappa} \right) &, r = 0\,, \\ -\int_{t_i}^t \diff t_r' \, g_{qp}(t,t_r') \, f_{r-1}(k,t_r') \, m \, \bar{\rho} \norm{k}^2 v(k,t_r') &, r > 0 \,.\end{cases}
\end{align}
Note that the quantity~$f_r(k,t)$ is defined iteratively. As an example, it is
\begin{align}
    f_2(k,t) &= \int_{t_i}^t \diff t_2' \int_{t_i}^{t_2'} \diff t_1' \, g_{qp}(t,t_2') \, g_{qp}(t_2',t_1') \, \left(1 + \frac{m \, g_{qp}(t_1',t_i)}{\kappa} \right)\, m^2 \, \bar{\rho}^2 \norm{k}^4 v(k,t_2') \, v(k,t_1') \,.
\end{align}

For the calculation of~$f_r(k,t)$ on an expanding background, we use the expressions derived in~\ref{sec:app_ParticlesExpandingBackground}. In this case, it is
\begin{align}
    f_0(k,a) &=1 + \int_{a_i}^{a} \diff a'' \frac{m \, a_i^2 \, H(a_i) \, f(a_i)}{m_{\text{eff}}(a'')} \qquad \text{and} \\*
    f_r(k,a) &= - \int_{a_i}^{a} \diff a_r' \, f_{r-1}(k,a_r') \int_{a_r'}^{a} \diff a'' \frac{1}{m_{\text{eff}}(a'')} \theta(a - a_r') \, m_{\text{eff}}(a_r') \, \bar{\rho} \, \norm{k}^2 \, \frac{3 \, \Omega_m(a_r')}{2 \, (a_r')^2 \, \bar{\rho} \, \norm{k}^2} \\
    &= - \frac{3}{2} \int_{a_i}^{a} \diff a_r' \, f_{r-1}(k,a_r') \int_{a_r'}^{a} \diff a'' \frac{m_{\text{eff}}(a_r') \, \Omega_m(a_r')}{(a_r')^2 \, m_{\text{eff}}(a'')}
\end{align}
for~$r > 0$ and~$k \neq 0$.\footnote{Note that on an expanding background, the factor~$m$ in equation~\eqref{eq:cos_PowerSpectrumnfr} needs to replaced by~$m_{\text{eff}}(a_r')$, as is evident from comparing the equations of motion in equation~\eqref{eq:Nbody_eom} with the Hamiltonian in equation~\eqref{eq:app_Hamiltonian}.} Aside from exposing that~$f_r(k,a)$ is dimensionless and scale-invariant for the case of Newtonian gravitational interaction, this also shows that the particle mass~$m$ and the mean density~$\bar{\rho}$ do not affect the result of this calculation. The numerical evaluation of this expression can be done iteratively, such that the computational complexity is linear in the perturbation order. Depending on the desired numerical accuracy, the calculation takes about two minutes per perturbation order on a laptop.

\begin{figure}
    \centering
    \includegraphics[width=0.9\textwidth]{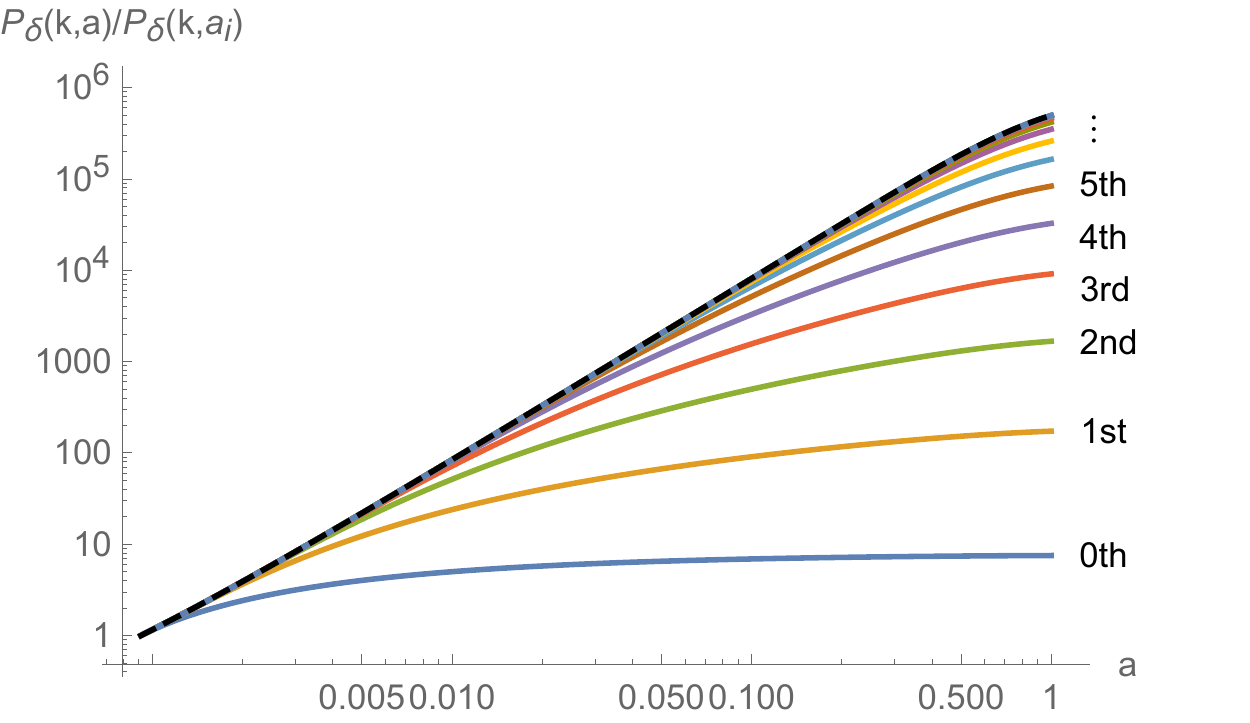}
    \caption{Evolution of the amplitude of the power spectrum against scale factor~$a$ for the cosmological standard model. The lowest curve is the zeroth order approximation and the curves above add incrementally the higher-order corrections. Shown are the curves up to $15\nth$~order in perturbation theory. The highest-order curves are visibly indistinguishable from the expected linear growth given by the dashed black line.~\cite{Mathematica}}
    \label{fig:cosmo_1}
\end{figure}

\begin{figure}
    \centering
    \includegraphics[width=0.48\textwidth]{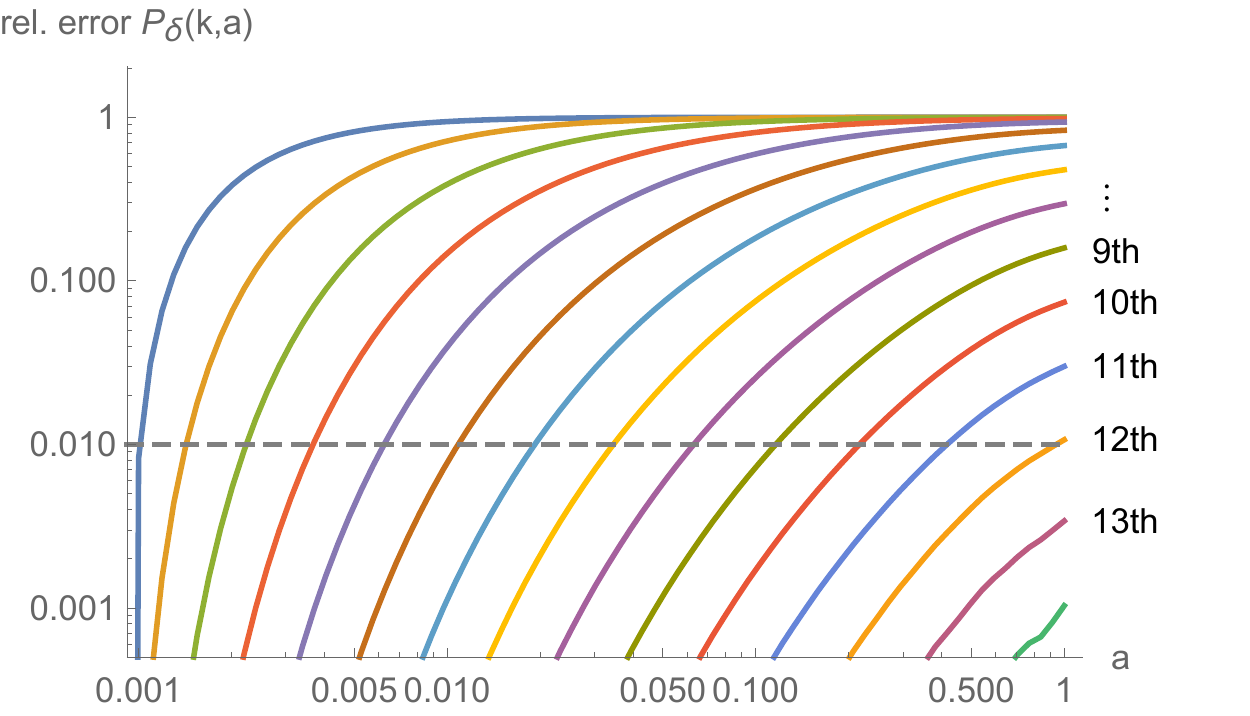}
    \includegraphics[width=0.48\textwidth]{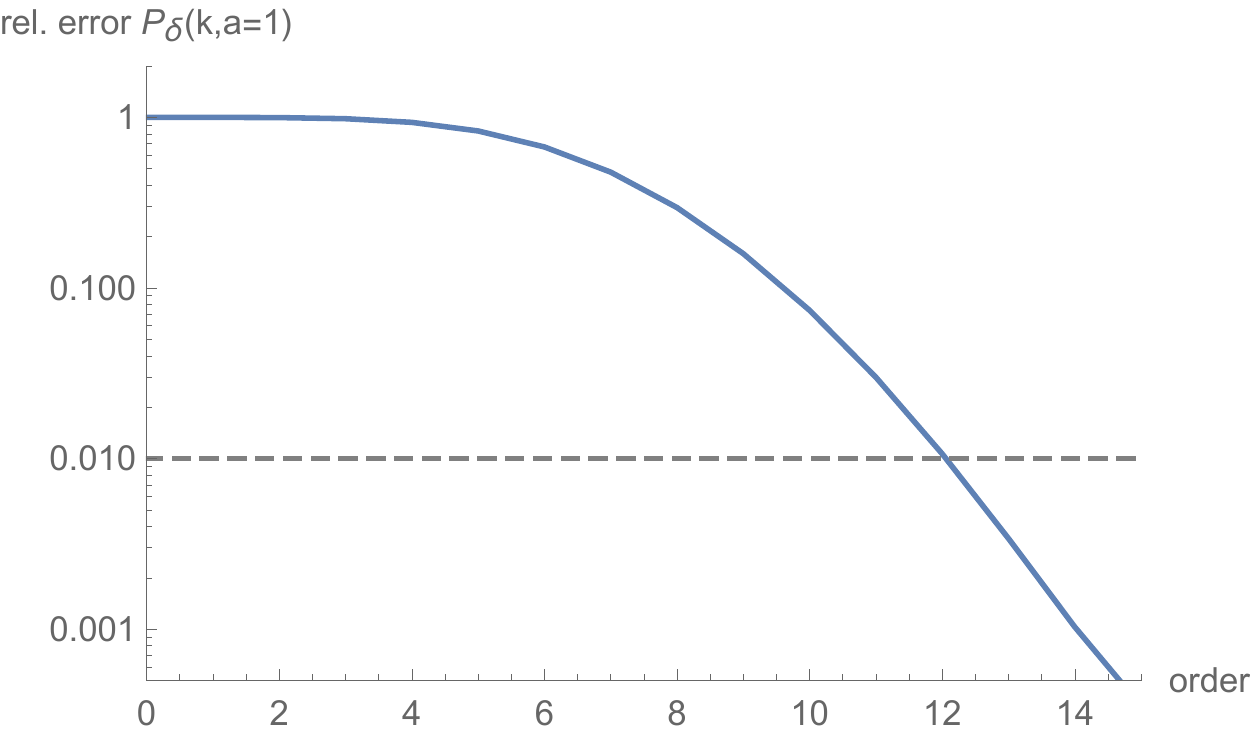}
    \caption{Left panel: Relative difference between the expected linear growth of the power spectrum and the perturbative approximations for up to $15\nth$~order. The x-axis is scale factor~$a$. Hence this panel effectively shows the residuals of figure~\ref{fig:cosmo_1}. Right panel: Like the left panel, but the relative difference is plotted against the perturbation order at final scale factor~$a = 1$. Percent-level accuracy (indicated by the grey dashed curve) is achieved by going to at least~$12\nth$ perturbation order.~\cite{Mathematica}}
    \label{fig:cosmo_2}
\end{figure}

In figure~\ref{fig:cosmo_1} we plot the growth of the amplitude of the power spectrum using equation~\eqref{eq:cos_PowerSpectrumn1} up to $15\nth$~order in perturbation theory as a function of scale factor for a standard $\Lambda$CDM-model. The colorful curves correspond to the orders in perturbation theory, starting from zeroth order with the lowest blue curve. The orange curve above adds the first order correction, the green curve additionally the second order correction, and so forth. Visually, the curves converge against the dashed black line which is simply the square of the linear growth factor. Going to high enough orders in KFT perturbation theory for linear initial correlations, we fully recover the linear growth of the power spectrum.

This is quantitatively verified by figure~\ref{fig:cosmo_2}. Its left panel shows the residuals of the previous plot, i.e., the relative difference between the expected linear growth and the perturbative approximations. The zeroth order curve (leftmost blue) is accurate to percent-level only up to scale factor~$a \approx 0.0010$ which is almost immediately after the scale factor at which we set our initial conditions~$a(t_i) \approx 0.0009$. However, as we include more and more orders in perturbation theory, our approximation remains accurate for longer. If we go up to about $12\nth$~order, we retain percent-level accuracy for the entire evolution. The right panel of figure~\ref{fig:cosmo_2} shows the relative mismatch at scale factor~$a = 1$, i.e., at the end of evolution, plotted against the order of perturbation. It can clearly be seen that the perturbation series converges towards the correct value.

\section{Conclusions and Outlook}

Based on the path integral approach to classical mechanics, Kinetic Field Theory (KFT) provides a framework to obtain observables of a classical physical system subject to probabilistic initial conditions. In this work we have disentangled the deterministic evolution and the averaging over the initial conditions. In particular, we have written down the generating functional with explicit dependence on the initial conditions in equations~\eqref{eq:theory_Z} and~\eqref{eq:theory_Z0}. These general expressions -- valid for any Hamiltonian system which allows for a split into free evolution and interactions -- can be taken as a starting point for obtaining perturbative expressions for observables. Indeed, it is possible to generalize the Feynman rules we derived in section~\ref{sec:feynman_rules} to this general setting.

For the sake of concreteness, we presented the perturbative approach to KFT specialized to interacting $N$-body systems. Conceptually, the perturbative approach expands in the force between particles and thus, effectively, in the deviation from the free trajectories. We have argued that the perturbative expansion of KFT adds the interactions in a relatively slow fashion. Therefore, in systems for which the free evolution yields a good approximation to an observable, we expect that the perturbation expansion of KFT provides stable convergence to this observable. We point out that our main application, cosmic structure formation, has precisely this property -- free evolution already provides a reasonable estimate of the density fluctuation power spectrum.

The main result of our work is the systematic perturbation scheme presented in section~\ref{sec:feynman}. Using our Feynman rules we can write down the mathematical expressions for any density $r$-point correlation functions for arbitrary order in the perturbative expansion. They also have a clear physical interpretation: At interaction vertices the outgoing particle is perturbed by the entire system via the two-particle interaction potential. If there are any ingoing edges, some of the particles participating in this interaction might already have been perturbed by earlier interactions. The expansion parameter is the interaction potential such that the $n\nth$~order contribution to an observable accounts for all combinations of $n$-fold perturbation of particles of the system.

Up to this point, the entire scheme was explicitly dependent on a specific initial phase space position for the system~$\b{x}$. The true potential of KFT lies in the averaging process of observables over a probability distribution of the initial values~$P(\b{x})$. Crucially, this averaging can be done in large part analytically. In particular, even in the limit of the particle number becoming infinite, all integrals over initial particle positions and momenta can be performed analytically for the observables of interest. This is due to the fact that the initial values~$\b{x}$ only appear inside Fourier phase factors in our Feynman rules which allows us to express expectation values for observables in terms of the characteristic function~$\Phi_{\b{x}}(\b{f})$ of the probability distribution~$P(\b{x})$. We explained this using a toy model in section~\ref{sec:ini_toyGauss}.

The initial conditions for cosmic structure formation feature a Gaussian random field~$\psi$ which acts as the velocity potential and determines the density contrast~$\delta$ via its Laplacian. This gives rise to density-density, density-momentum and momentum-momentum correlations~$\b{C^{\bullet\bullet}}$ between different particles. The complicated nature of this probability distribution and its characteristic function (given in equation~\eqref{eq:ini_characteristicFunction_fin}) suggests an expansion in the initial correlations~$\b{C^{\bullet\bullet}}$. Indeed, without this expansion the momentum-momentum correlations appear in an exponential and thus the initial particle positions appear as a Fourier phase factor within this exponential.

The initial conditions for cosmic structure formation feature a Gaussian random field~$\psi$ which acts as the velocity potential and determines the density contrast~$\delta$ via its Laplacian. This gives rise to density-density, density-momentum and momentum-momentum correlations~$\b{C^{\bullet\bullet}}$ between different particles. The complicated nature of this probability distribution and its characteristic function (given in equation~\eqref{eq:ini_characteristicFunction_fin}) suggests an expansion in the initial correlations~$\b{C^{\bullet\bullet}}$. In doing so, the particle positions appear exclusively inside Fourier phase factors and thus can be integrated analytically. In fact, these integrations yield Dirac $\delta$-functions which cancel the wave vector integrals from the expansion in the interactions.

We have given a general recipe to deduce the form of any $r$-point density correlation function in $m\nth$~order in the expansions in initial correlations and in $n\nth$~order in the expansion in the interactions. Combining the expressions for explicit initial conditions obtained from the Feynman rules with appropriate combinations of the correlation factors given in equation~\eqref{eq:ini_correlation-factors} we obtain the relevant terms. We have automated these process and are able to produce all terms of any order in our two expansions. The remaining difficulties are in, firstly, the rapidly growing number of diagrams and correlation combinations at higher orders and, secondly, the numerical evaluation of the remaining integrals in their mathematical expressions. In section~\ref{sec:ini_corrInt} we provided a precise count of the remaining integrals in our expressions for expectation values of density $r$-point functions in cosmology.

From this count we observe that for any interaction order there remains one time-integral, while for any correlation order we have one leftover integral over a wave vector. The precise count of the wave vectors we need to integrate over is lower if we calculate higher-point correlation functions, but higher if there are leftover Dirac $\delta$-functions. Some simple algebra reveals that if we consider linear correlations, then the terms contributing to the density fluctuation power spectrum do not feature any integrals over wave vectors. In fact, we have seen that the power spectrum can be calculated via an iterative method in this case which makes the numerical evaluation rather efficient.

One key feature of the case of linear initial correlations is that the density fluctuation power spectrum receives an enhancement which is independent of wave number, i.e., we obtain linear growth. By computing the relevant contributions, we show that KFT fully reproduces the expected linear growth of the power spectrum. Setting up initial conditions at redshift~$z = 1090$, expanding these up to linear order in the initial density- and momentum-correlations and solving the Hamiltonian dynamics perturbatively up to $12th$~order in the particle interactions reproduces the amplitude of the power spectrum to percent accuracy. In figure~\ref{fig:cosmo_1} we show how the various perturbation orders approximate the expected linear growth~$P_{\delta}(k,a) \propto D_+(a)^2$ up to increasingly low redshift. Given that the perturbative expansion is essentially in the number of interactions per particle, this means that surprisingly few gravitational encounters are sufficient to accurately reproduce linear structure growth.

The fact that linear initial correlations fully reproduce linear growth of the power spectrum also implies that the entire non-linear growth of the power spectrum is due to non-linear initial correlations. In particular, this implies that the momentum-autocorrelations appearing at higher correlation orders are exactly cancelled, most likely order-by-order in the expansion. We therefore caution against treating these autocorrelations separately as a damping term. The formalism we have presented can be used to study non-linear correlations, too, and we will present the results in a future paper~\cite{we2022}. We also intend to study the relationship of our perturbative approach to the Resummed-KFT (RKFT) formalism~\cite{L3c}.

	
	\section*{Acknowledgements}
	\addcontentsline{toc}{section}{Acknowledgements}
	
	We thank Matthias Bartelmann for valuable discussions. We are grateful to Andrej Brandalik for pointing out the subtlety discussed in the final paragraph of~\ref{sec:app_ParticlesExpandingBackground}. SZ also thanks Fabio D'Ambrosio, Andrea Giusti, Robert Lilow, Isabel Oldengott and Ricardo Waibel for helpful comments.
	Parts of this work were discussed in the first Annual KFT Workshop in Heidelberg in April 2022 and we are grateful for the many interesting discussions and useful comments.
	LH is supported by funding from the European Research Council (ERC) under the European Unions Horizon 2020 research and innovation programme grant agreement No 801781 and by the Swiss National Science Foundation grant 179740.

	
	\appendix
	\include{Components}
	

\section{Particles on an Expanding Background}
\label{sec:app_ParticlesExpandingBackground}

When incorporating an expanding background into KFT, the first decision to make is whether our coordinates~$\b{x} \in \b{X}$ are comoving or absolute. While both choices are viable, there are a couple of advantages of working with comoving coordinates. Firstly, in comoving coordinates free motion is along straight paths and, secondly, we immediately obtain the commonly used comoving wave vectors via Fourier transform. Therefore, let~$\b{x}$ be comoving coordinates and~$\b{\varphi}(t)$ the comoving phase space trajectory, i.e., the physical coordinates of the $j\nth$~particle are given by $r_j(t) = a(t) \, \varphi^{(q)}_j(t)$.

We remark that by using an uniformly expanding universe described by a scale factor~$a(t)$, we implicitly assumed homogeneity and isotropy of our $N$-particle system when averaged over large enough volumes. Since we are working in a flat and thus infinite universe, this necessitates taking the limit of the particle number~$N$ going to infinity. Preferring to work with a finite dimensional phase space for the moment, we postpone taking this limit. We note, however, that working with finite~$N$ can also be viewed as considering a finite volume~$V$ of an infinite universe filled with statistically homogeneously and isotropically distributed particles. When we later send~$N$ to infinity for constant average density~$\bar{\rho}$, we implicitly take simultaneously the limit~$V \ra \infty$. 

Labelled by physical coordinates, the particles are subject to the Lagrangian
\begin{align}
\label{eq:app_LagrangianPhysicalCoordinates}
	\mathcal{L}\left(\b{r},\partial_t \b{r},t\right) = \frac{m}{2} \, \partial_t \b{r}(t) \cdot \partial_t \b{r}(t) - \frac{m}{2} \sum_{j} \sum_{k \neq j} \nu_E\left( r_k(t) - r_j(t), t \right) \,,
\end{align}
where we introduced the Eulerian two-particle interaction potential~$\nu_E$. We can rewrite the second term of~$\mathcal{L}$ as
\begin{align}
	\frac{m}{2} \sum_{j} \sum_{k \neq j} \nu_E\left( r_k(t) - r_j(t), t \right) = m \, V_E\left( t; \b{r} \right) \,,
\end{align}
where the Eulerian potential~$V_E\left( t; \b{r} \right)$ is defined in analogy to equation~\eqref{eq:Nbody_potential}. This expression should be thought of as the potential energy of the $N$-particle system. If we regard one of the particle positions~$r \ce r_l(t)$ as a free variable, we can write down the cosmological Poisson equation
\begin{align}
\label{eq:app_PoissonEquationPhys}
    \nabla_r^2 V_E\left( t; \b{r} \right) \ce \restr{\left( \frac{\delta}{\delta r_l(t)} \cdot \frac{\delta}{\delta r_l(t)} \right) V_E\left( t; \b{r} \right)}{r_l(t) \ec r} = 4 \pi G \, m \, \rho_E\left( r; \b{r}(t) \right) - \Lambda \,.
\end{align}
Here, $G$ is the gravitational constant and $\Lambda$ the cosmological constant. The (Eulerian) particle number density field~$\rho_E\left( r, \b{r}(t) \right)$ is simply given by a sum of Dirac $\delta$-functions,
\begin{align}
\label{eq:app_DensityDecomp}
	\rho_E\left(r; \b{r}(t)\right) = \sum_j \delta_D\left(r - r_j(t)\right) \,.
\end{align}

Proceeding similarly to~\cite{L3d} and section~7 of~\cite{fabis2015statistical} we translate the above to comoving coordinates~$\b{\varphi}$. Up to a total derivative, the above Lagrangian is equal to
\begin{align}
\label{eq:app_LagrangianCom}
	\mathcal{L}\left(\b{\varphi^{(q)}},\partial_t \b{\varphi^{(q)}},t\right) = \frac{m}{2} \, a(t)^2 \, \partial_t \b{\varphi^{(q)}}(t) \cdot \partial_t \b{\varphi^{(q)}}(t) - m \, V_L\left( t; \b{\varphi^{(q)}} \right) \,,
\end{align}
where the peculiar Lagrangian potential
\begin{align}
    V_L\left( t; \b{\varphi^{(q)}} \right) \ce V_E \left( t; a \, \b{\varphi^{(q)}} \right) + \frac{1}{2} a(t) \left(\partial_t^2 a(t)\right) \, \b{\varphi^{(q)}}(t) \cdot \b{\varphi^{(q)}}(t)
\end{align}
appears. Using equation~\eqref{eq:app_PoissonEquationPhys} as well as the cosmic acceleration equation, one can show that the Lagrangian potential~$V_L$ satisfies
\begin{align}
\label{eq:app_PoissonEquationCom}
	\nabla_q^2 V_L\left( t; \b{\varphi^{(q)}} \right) \ce \restr{\left( \frac{\delta}{\delta \varphi^{(q)}_l(t)} \cdot \frac{\delta}{\delta \varphi^{(q)}_l(t)} \right) V_L\left( t; \b{\varphi^{(q)}} \right)}{\varphi^{(q)}_l(t) \ec q} = \frac{4 \pi G \, m}{a(t)} \, \left( \rho\left(q;\b{\varphi^{(q)}}(t)\right) - \bar{\rho} \right) \,.
\end{align}
In this equation we introduced the comoving number density~$\rho\left(q;\b{\varphi^{(q)}}(t)\right) \ce a(t)^3 \rho_E\left(a(t) \, q; a(t) \, \b{\varphi^{(q)}}(t)\right)$ which henceforth, as well as in the main document, is used exclusively. Note that the mean comoving number density~$\bar{\rho}$ is time-independent.

Returning to the Lagrangian~\eqref{eq:app_LagrangianCom} a central goal of this section is to perform a Legendre transformation to obtain the Hamiltonian. However, it is convenient to first change our time coordinate from~$t$ to scale factor~$a$. Let us rewrite the comoving Lagrangian~\eqref{eq:app_LagrangianCom} in terms of this time parameter. First, observe that
\begin{align}
	\diff a = \frac{\diff a}{\diff t} \, \diff t = a \, H(a) \, \diff t
\end{align}
such that~$\mathcal{L}(t) = \left( a \, H(a) \right) \mathcal{L}(a)$ to ensure invariance of the action. As usual, $H = a^{-1} \, \partial_t a$ is the Hubble function. In addition, we need to replace the time derivatives of~$\b{\varphi^{(q)}}$ with derivatives with respect to~$a$. Then the Lagrangian becomes
\begin{align}
\label{eq:4_Langrangian_comoving_tau}
	\mathcal{L}\left(\b{\varphi^{(q)}},\partial_a \b{\varphi^{(q)}},a\right) = \frac{m_{\text{eff}}(a)}{2} \, \partial_a \b{\varphi^{(q)}}(a) \cdot \partial_a \b{\varphi^{(q)}}(a) - \frac{m}{a \, H(a)} \, V_L\left( a; \b{\varphi^{(q)}} \right) \,,
\end{align}
where we defined the time-dependent effective particle mass
\begin{align}
	m_{\text{eff}}(a) \ce m \, a^3 \, H(a) \,.
\end{align}

Next, we perform the Legendre transform of the Lagrangian~\eqref{eq:app_LagrangianCom} and arrive at the Hamiltonian
\begin{align}
\label{eq:app_Hamiltonian}
    \mathcal{H}\left(\b{\varphi^{(q)}},\b{\varphi^{(p)}},a\right) = \frac{\b{\varphi^{(p)}}(a) \cdot \b{\varphi^{(p)}}(a)}{2 \, m_{\text{eff}}(a)} + m_{\text{eff}}(a) \, V\left( a; \b{\varphi^{(q)}} \right)
\end{align}
featuring the conjugate momentum
\begin{align}
    \b{\varphi^{(p)}}(a) = \frac{\partial \mathcal{L}\left(\b{\varphi^{(q)}},\partial_a \b{\varphi^{(q)}},a\right)}{\partial \b{\dot{\varphi}^{(q)}}(a)} = m_{\text{eff}}(a) \, \partial_a \b{\varphi}^{(q)}(a) \,.
\end{align}
Moreover, we redefined the potential
\begin{align}
    V\left( a; \b{\varphi^{(q)}} \right) \ce \frac{1}{a^4 \, H(a)^2} \, V_L\left( a; \b{\varphi^{(q)}} \right)
\end{align}
such that the Hamiltonian equations of motion are precisely given by equations~\eqref{eq:Nbody_eom} upon replacing~$t \mapsto a$ and~$m \mapsto m_{\text{eff}}(a)$. Next, we want to derive the form of the Green's function and the two-particle interaction potential for this Hamiltonian.

The free equations of motion take the form
\begin{align}
    \partial_a \varphi^{(q)}_j(a) = \frac{1}{m_{\text{eff}}(a)} \, {\varphi}_j^{(p)}(a) \qquad \text{and} \qquad \partial_a \varphi^{(p)}_j(a) = 0 \qquad \text{for all $j$.}
\end{align}
One can easily verify, that a suitable Green's function for these equations is given by
\begin{align}
	\mathcal{G}(a,a') &= \begin{pmatrix} \theta(a-a') \, \mathcal{I}_3 & g_{qp}(a,a') \, \mathcal{I}_3 \\ 0 & \theta(a-a') \, \mathcal{I}_3 \end{pmatrix} \qquad \text{with} \\
	g_{qp}(a,a') &= \int_{a'}^{a} \frac{\diff a''}{m_{\text{eff}}(a'')} \, \theta(a-a') \,.
\end{align}
Note, that the expression~$g_{qp}(t,t') = \frac{1}{m} (t-t') \, \theta(t-t')$ derived in the special case of a static background in equation~\eqref{eq:Nbody_GreensFunction} is recovered upon replacing~$a \mapsto t$ and~$m_{\text{eff}}(a) \mapsto m$.

The second object we are interested in is the two-particle interaction potential. The potential~$V$ satisfies the Poisson equation
\begin{align}
\label{eq:app_PoissonEquationMod}
	\nabla_q^2 V\left( a; \b{\varphi^{(q)}} \right) = \frac{4 \pi G \, m}{a^5 \, H(a)^2} \, \left( \rho\left(q;\b{\varphi^{(q)}}(a)\right) - \bar{\rho} \right) \,.
\end{align}
We can define the (comoving) two-particle interaction potential~$\nu$ implicitly via a decomposition of the potential~$V$ as
\begin{align}
	V\left( a; \b{\varphi^{(q)}} \right) \ec \frac{1}{2} \sum_{j} \sum_{k \neq j} \nu\left( \varphi^{(q)}_k(a) - \varphi^{(q)}_j(a), a \right) \,.
\end{align}
Similarly, the density field can be written as a sum of Dirac $\delta$-functions, analogously to equation~\eqref{eq:app_DensityDecomp}. Using these, the Poisson equation decomposes into contributions for individual particles
\begin{align}
	\nabla_q^2 \nu\left( q, a \right) = \frac{4 \pi G \, m}{a^5 \, H(a)^2} \, \left( \delta_D\left(q\right) - \frac{\bar{\rho}}{N} \right) \,.
\end{align}

Performing a Fourier transform~$q \mapsto k$, this equation becomes
\begin{align}
    -\norm{k}^2 \, \nu\left(k, a\right) = \frac{4 \pi G \, m}{a^5 \, H(a)^2} \, \left( 1 - \frac{(2\pi)^3  \delta_D(k)}{V} \right) \,,
\end{align}
where we used that~$\bar{\rho} = \frac{N}{V}$. For~$k \neq 0$, this simply yields the familiar Newtonian scaling as~$\nu(k) \propto \norm{k}^{-2}$. Specifically, it is
\begin{align}
\label{eq:app_interactionPotential}
    \nu(k,a) = - \frac{4 \pi G \, m}{a^5 \, H(a)^2 \, \norm{k}^2} = - \frac{3 \, \Omega_m(a)}{2 \, \bar{\rho} \, a^2 \, \norm{k}^2} \qquad \text{for $k \neq 0$} \,,
\end{align}
where in the last step we introduced the matter density parameter~$\Omega_m(a)$. Interestingly, however, the Dirac $\delta$-function originating from the mean density does have an effect for~$k = 0$. In that case, we observe that
\begin{align}
\label{eq:app_interactionPotentialZero}
    \delta_D(k) \, \norm{k}^2 \, \nu\left(k, \tau\right) = - \frac{3 \, \Omega_m(a)}{2 \, \bar{\rho} \, a^2} \, \left( \delta_D(k) - \frac{(2\pi)^3 \, \delta_D(k) \, \delta_D(k)}{V} \right) = 0
\end{align}
using the identity~$(2\pi)^3 \, \delta_D(k) \, \delta_D(k) = V \, \delta_D(k)$. This subtlety is actually of importance when calculating certain expectation values.\footnote{We remark that in section~$7.3$ of~\cite{fabis2015statistical} this issue is dealt with by demanding that the Fourier transform of the potential should be well-defined, obtaining a more general condition than equation~\eqref{eq:app_interactionPotentialZero}.} Indeed, naively using equation~\eqref{eq:app_interactionPotential} to calculate the expectation value for the density would yield
\begin{align}
    \langle \rho \rangle(q,a) = \bar{\rho} \, \exp\left( \int_{a_i}^{a} \diff a' \, g_{qp}(a, a') \right)
\end{align}
instead of the correct result~$\langle \rho \rangle(q,a) = \bar{\rho}$.

\section{Poisson Equation for the Initial Density Contrast}
\label{sec:app_PoissonInitial}

In order to set up cosmological initial conditions, we need a relationship between the density and the velocity potential. This can be achieved by using the continuity equation of hydrodynamics which is given by
\begin{align}
\label{eq:app_continuityEqEul}
    \frac{\partial \rho_E(r,t)}{\partial t} + \nabla_r \cdot \left( \rho_E(r,t) \, v_E(r,t) \right)=0 \,.
\end{align}
Here, the subscript~$E$ for the number density field~$\rho_E$ and the velocity field~$v_E$ indicates that this equation is valid in the Eulerian description of hydrodynamics which relies on the physical coordinates~$r$. We can go over to the Lagrangian description by splitting the velocity and density fields into a background and a perturbative component. Specifically, we write
\begin{align}
    v_E(r,t) &= H(t) \, r + v(r,t) \qquad \text{and} \\
    \rho_E(r,t) &= \bar{\rho}_E(t) \left(1 + \delta(r,t) \right) \,,
\end{align}
where~$H(t) = a(t)^{-1} \, \partial_t a(t)$ is the Hubble function. We assume that the Lagrangian peculiar velocity field~$v(r,t)$ and the density contrast~$\delta(r,t)$ are small compared to the Hubble flow and unity, respectively.

Defining the total time derivative
\begin{align}
    \frac{\diff}{\diff t} = \frac{\partial}{\partial t} + H(t) \, r \cdot \nabla_r \,,
\end{align}
the continuity equation~\eqref{eq:app_continuityEqEul} assumes the form
\begin{align}
    \left( \frac{\diff}{\diff t} - H(t) \, r \cdot \nabla_r \right) \left( \bar{\rho}_E(t) \left(1 + \delta(r,t) \right) \right) + \nabla_r \cdot \left( \left( \bar{\rho}_E(t) \left(1 + \delta(r,t) \right) \right) \, \left( H(t) \, r + v(r,t) \right) \right) &= 0 \\*
    \LRa \qquad \left[ \frac{1}{\bar{\rho}_E(t)} \frac{\diff \bar{\rho}_E}{\diff t}(t) + 3 H(t) \right] \left(1 + \delta(r,t) \right) + \frac{\diff \delta}{\diff t}(r,t) + \nabla_r \cdot v(r,t) + \mathcal{O}\left( \delta(r,t) \, v(r,t) \right) &= 0 \,.
\end{align}
In the last term in the second line we collected all terms quadratic in the perturbation. In accordance with the usual treatment in linear hydrodynamic perturbation theory -- which is valid in the very early universe where we set up our initial conditions -- we will neglect these terms in the following. Furthermore, the first term in the second line yields zero by means of the Friedmann equation. Indeed, it is
\begin{align}
    \frac{\diff \bar{\rho}_E}{\diff t}(t) + 3 H(t) \, \bar{\rho}_E(t) = 0
\end{align}
for pressureless cold dark matter even in a universe filled by a mixture of non-interacting fluids. Finally, we introduce comoving coordinates by~$r(t) = a(t) \, q(t)$ and obtain the Lagrangian version of the continuity equation
\begin{align}
\label{eq:app_continuityEqLagr}
    \frac{\diff \delta}{\diff t}(q,t) + \frac{1}{a(t)} \, \nabla_q \cdot v(q,t) = 0 \,.
\end{align}

As is common in linear hydrodynamic perturbation theory, we choose a product ansatz for the density contrast, i.e., we assume that~$\delta(q,t) = D_{+}(t) \, \delta(q)$. The object~$D_{+}(t)$ is the so-called linear growth factor which we normalize via the condition~$D_{+}(t_i) = 1$. This implies that~$\delta(q) = \delta(q,t_i)$ is simply the initial density contrast. Moreover, we recall that we made the assumption that the velocity field is irrotational at initial time~$t_i$, i.e., we can write $v(q,t_i) = \nabla_r \psi(q) = a(t_i)^{-1} \nabla_q \psi(q)$, where~$\psi$ is the velocity potential. Lastly, we change our time parameter from cosmic time~$t$ to scale factor~$a$ and obtain
\begin{align}
\label{eq:app_Poisson}
    \delta(q) = - \kappa \, \nabla_q^2 \psi(q) \qquad \text{with} \qquad \kappa = \left( a(t_i)^2 \, \frac{\diff D_+}{\diff t}(t_i) \right)^{-1} = \frac{1}{a_i^2 \, H(a_i) \, f(a_i)} \,.
\end{align}
Here we introduced the logarithmic derivative~$f(a) = \frac{\diff \ln D_+}{\diff \ln a}(a)$ and used~$D_+(a_i) = 1$.

The derivation of the Poisson equation for the initial density contrast relied on a number of assumptions which we want to collect here. Firstly, we assumed the validity of linear hydrodynamic perturbation theory at (and only at) the initial time~$t_i$. This includes the assumption of the existence of a velocity field~$v$. Moreover, we assumed that this velocity field is irrotational at initial time~$t_i$ such that there exists a velocity potential~$\psi(q)$. For the density contrast we assumed that we can at initial time factorize the spatial and temporal dependencies, making use of the linear growth factor~$D_+(a)$. Quantitatively, we only need the inputs~$f(a_i)$ and~$H(a_i)$. We stress that we only need to know these quantities at initial scale factor~$a_i$.


	\addcontentsline{toc}{section}{References}

	\bibliography{res/9-bib.bib}
	
\end{document}